\def\Di{27 Novembre 1998}
\headline{\hss \ottorm Draft \#20}

\newcount\mgnf\newcount\tipi\newcount\tipoformule
\newcount\aux\newcount\piepagina\newcount\xdata
\mgnf=0
\aux=1           
\tipoformule=1   
\piepagina=1     
\xdata=1         

\ifnum\mgnf=1 \aux=0 \tipoformule =1 \piepagina=1 \xdata=1\fi
\newcount\bibl
\ifnum\mgnf=0\bibl=0\else\bibl=1\fi
\bibl=0

%
%
%
%
\ifnum\bibl=0
\def\ref#1#2#3{[#1#2]\write8{#1@#2}}
\def\rif#1#2#3#4{\item{[#1#2]} #3}
\fi

\ifnum\bibl=1
\openout8=ref.b
\def\ref#1#2#3{[#3]\write8{#1@#2}}
\def\rif#1#2#3#4{}

\fi

\def\9#1{\ifnum\aux=1#1\else\relax\fi}
\ifnum\piepagina=0 \footline={\rlap{\hbox{\copy200}\
$\st[\number\pageno]$}\hss\tenrm \foglio\hss}\fi \ifnum\piepagina=1
\footline={\rlap{\hbox{\copy200}} \hss\tenrm \folio\hss}\fi
\ifnum\piepagina=2\footline{\hss\tenrm\folio\hss}\fi

\ifnum\mgnf=0 \magnification=\magstep0
\hsize=16.4truecm\vsize=21truecm \parindent=4.pt\fi
\ifnum\mgnf=1 \magnification=\magstep1
\hsize=16.0truecm\vsize=22.5truecm\baselineskip14pt\vglue5.0truecm
\overfullrule=0pt \parindent=4.pt\fi

\let\a=\alpha\let\b=\beta \let\g=\gamma \let\d=\delta
\let\e=\varepsilon \let\z=\zeta \let\h=\eta
\let\th=\vartheta\let\k=\kappa \let\l=\lambda \let\m=\mu \let\n=\nu
\let\x=\xi \let\p=\pi \let\r=\rho \let\s=\sigma \let\t=\tau
 \let\f=\varphi\let\ch=\chi  \let\o=\omega
 \let\G=\Gamma \let\D=\Delta \let\Th=\Theta
\let\X=\Xi   \let\F=\Phi
 \let\O=\Omega 
{\count255=\time\divide\count255 by 60 \xdef\oramin{\number\count255}
\multiply\count255 by-60\advance\count255 by\time
\xdef\oramin{\oramin:\ifnum\count255<10 0\fi\the\count255}}
\def\ora{\oramin }

\ifnum\xdata=0
\def\data{\number\day/\ifcase\month\or gennaio \or
febbraio \or marzo \or aprile \or maggio \or giugno \or luglio \or
agosto \or settembre \or ottobre \or novembre \or dicembre
\fi/\number\year;\ \ora}
\else
\def\data{\Di}
\fi

\setbox200\hbox{$\scriptscriptstyle \data $}
\newcount\pgn \pgn=1
\def\foglio{\number\numsec:\number\pgn
\global\advance\pgn by 1} \def\foglioa{A\number\numsec:\number\pgn
\global\advance\pgn by 1}
\global\newcount\numsec\global\newcount\numfor \global\newcount\numfig
\gdef\profonditastruttura{\dp\strutbox}
\def\senondefinito#1{\expandafter\ifx\csname#1\endcsname\relax}
\def\SIA #1,#2,#3 {\senondefinito{#1#2} \expandafter\xdef\csname
#1#2\endcsname{#3} \else \write16{???? ma #1,#2 e' gia' stato definito
!!!!} \fi} \def\etichetta(#1){(\veroparagrafo.\veraformula) \SIA
e,#1,(\veroparagrafo.\veraformula) \global\advance\numfor by 1
\9{\write15{\string\FU (#1){\equ(#1)}}} \9{ \write16{ EQ \equ(#1) == #1
}}} \def \FU(#1)#2{\SIA fu,#1,#2 }
\def\etichettaa(#1){(A\veroparagrafo.\veraformula) \SIA
e,#1,(A\veroparagrafo.\veraformula) \global\advance\numfor by 1
\9{\write15{\string\FU (#1){\equ(#1)}}} \9{ \write16{ EQ \equ(#1) == #1
}}} \def\getichetta(#1){Fig.  \verafigura \SIA e,#1,{\verafigura}
\global\advance\numfig by 1 \9{\write15{\string\FU (#1){\equ(#1)}}} \9{
\write16{ Fig.  \equ(#1) ha simbolo #1 }}} \newdimen\gwidth \def\BOZZA{
\def\alato(##1){ {\vtop to \profonditastruttura{\baselineskip
\profonditastruttura\vss
\rlap{\kern-\hsize\kern-1.2truecm{$\scriptstyle##1$}}}}}
\def\galato(##1){ \gwidth=\hsize \divide\gwidth by 2 {\vtop to
\profonditastruttura{\baselineskip \profonditastruttura\vss
\rlap{\kern-\gwidth\kern-1.2truecm{$\scriptstyle##1$}}}}} }
\def\alato(#1){} \def\galato(#1){}
\def\veroparagrafo{\number\numsec}\def\veraformula{\number\numfor}
\def\verafigura{\number\numfig}
\def\geq(#1){\getichetta(#1)\galato(#1)}
\def\Eq(#1){\eqno{\etichetta(#1)\alato(#1)}}
\def\eq(#1){\etichetta(#1)\alato(#1)}
\def\Eqa(#1){\eqno{\etichettaa(#1)\alato(#1)}}
\def\eqa(#1){\etichettaa(#1)\alato(#1)}
\def\eqv(#1){\senondefinito{fu#1}$\clubsuit$#1\write16{No translation
for #1} \else\csname fu#1\endcsname\fi}
\def\equ(#1){\senondefinito{e#1}\eqv(#1)\else\csname e#1\endcsname\fi}

\openin13=#1.aux \ifeof13 \relax \else \input #1.aux \closein13\fi
\openin14=\jobname.aux \ifeof14 \relax \else \input \jobname.aux
\closein14 \fi \9{\openout15=\jobname.aux} \newskip\ttglue

\font\titolone=cmbx12 scaled \magstep1
\font\titolo=cmbx10 scaled \magstep1

\font\ottorm=cmr8\font\ottoi=cmmi7\font\ottosy=cmsy7
\font\ottobf=cmbx7\font\ottott=cmtt8\font\ottosl=cmsl8\font\ottoit=cmti7
\font\sixrm=cmr6\font\sixbf=cmbx7\font\sixi=cmmi7\font\sixsy=cmsy7

\font\tenmib=cmmib10\font\sevenmib=cmmib10 scaled 700
\font\fivemib=cmmib10 scaled 500
\textfont5=\tenmib  \scriptfont5=\sevenmib  \scriptscriptfont5=\fivemib
\font\fiverm=cmr5\font\fivesy=cmsy5\font\fivei=cmmi5\font\fivebf=cmbx5

\def\ottopunti{\def\rm{\fam0\ottorm}\textfont0=\ottorm%
\scriptfont0=\sixrm\scriptscriptfont0=\fiverm\textfont1=\ottoi%
\scriptfont1=\sixi\scriptscriptfont1=\fivei\textfont2=\ottosy%
\scriptfont2=\sixsy\scriptscriptfont2=\fivesy\textfont3=\tenex%
\scriptfont3=\tenex\scriptscriptfont3=\tenex\textfont\itfam=\ottoit%
\def\it{\fam\itfam\ottoit}\textfont\slfam=\ottosl%
\def\sl{\fam\slfam\ottosl}\textfont\ttfam=\ottott%
\def\tt{\fam\ttfam\ottott}\textfont\bffam=\ottobf%
\scriptfont\bffam=\sixbf\scriptscriptfont\bffam=\fivebf%
\textfont5\sevenmib\scriptfont5=\fivei
\def\bf{\fam\bffam\ottobf}
\tt\ttglue=.5em plus.25em minus.15em%

\setbox\strutbox=\hbox{\vrule height7pt depth2pt width0pt}%
\normalbaselineskip=9pt\let\sc=\sixrm\normalbaselines\rm}
\catcode`@=11
\def\footnote#1{\edef\@sf{\spacefactor\the\spacefactor}#1\@sf
\insert\footins\bgroup\ottopunti\interlinepenalty100\let\par=\endgraf
\leftskip=0pt \rightskip=0pt \splittopskip=10pt plus 1pt minus 1pt
\floatingpenalty=20000
\smallskip\item{#1}\bgroup\strut\aftergroup\@foot\let\next}
\skip\footins=12pt plus 2pt minus 4pt\dimen\footins=30pc\catcode`@=12
\let\nota=\ottopunti\newdimen\xshift \newdimen\xwidth \newdimen\yshift
\def\ins#1#2#3{\vbox to0pt{\kern-#2 \hbox{\kern#1
#3}\vss}\nointerlineskip} \def\eqfig#1#2#3#4#5{ \par\xwidth=#1
\xshift=\hsize \advance\xshift by-\xwidth \divide\xshift by 2
\yshift=#2 \divide\yshift by 2 \line{\hglue\xshift \vbox to #2{\vfil #3
\includegraphics{#4.ps} }\hfill\raise\yshift\hbox{#5}}} \def\8{\write13}

\def\didascalia#1{\vbox{\nota\0#1\hfill}\vskip0.3truecm}

\def\T#1{#1\kern-4pt\lower9pt\hbox{$\widetilde{}$}\kern4pt{}}
\let\dpr=\partial \let\io=\infty\let\ig=\int
\def\fra#1#2{{#1\over#2}}\let\0=\noindent
\def\guida{\leaders\hbox to 1em{\hss.\hss}\hfill}
\def\tende#1{\vtop{\ialign{##\crcr\rightarrowfill\crcr
\noalign{\kern-1pt\nointerlineskip} \hglue3.pt${\scriptstyle
#1}$\hglue3.pt\crcr}}} \def\otto{\
{\kern-1.truept\leftarrow\kern-5.truept\to\kern-1.truept}\ }

\def\pagina{\vfill\eject}\let\ciao=\bye

\def\tst{\textstyle}\def\st{\scriptscriptstyle}
\def\*{\vskip0.3truecm}

\def\lis#1{{\overline #1}}\def\etc{\hbox{\it etc}}\def\eg{\hbox{\it e.g.\ }}
\def\ap{\hbox{\it a priori\ }}
\def\ie{\hbox{\it i.e.\ }}

\def\fiat{{}}
\def\\{\hfill\break} \def\={{ \; \equiv \; }}
\def\Re{{\rm\,Re\,}} 

\def\annota#1{\footnote{${}^#1$}}
\ifnum\aux=1\BOZZA\else\relax\fi
\ifnum\tipoformule=1\let\Eq=\eqno\def\eq{}\let\Eqa=\eqno\def\eqa{}
\def\equ{{}}\fi
\def\defi{\,{\buildrel def \over =}\,}

\def\pallino{{\0$\bullet\;$}}\def\1{\ifnum\mgnf=0\pagina\else\relax\fi}
\def\W#1{#1_{\kern-3pt\lower6.6truept\hbox to 1.1truemm
{$\widetilde{}$\hfill}}\kern2pt\,}
\def\Re{{\rm Re}\,}

\def\igb{
\mathop{\raise4.pt\hbox{\vrule height0.2pt depth0.2pt width6.pt}
\kern0.3pt\kern-9pt\int}}

\def\FINE{
\*
\0{\it Internet:
Authors' preprints at:  {\tt http://ipparco.roma1.infn.it}

\0\sl e-mail: {\it users:} giovanni, gentile, vieri,
{\it address}: {\tt @ipparco.roma1.infn.it}
}}
\def\V#1{{\,\underline#1\,}}

\def\RR{{\cal R}}
\def\NN{{\cal N}}\def\CC{{\cal C}}
\def\FFFF{{\cal F}}\def\UU{{\cal U}}
\def\MM{{\cal M}}\def\HH{{\cal H}}\def\PP{{\cal P}}
\def\II{{\cal I}}\def\TT{{\cal T}}
\def\OO{{\cal O}}\def\LL{{\cal L}}\def\VV{{\cal V}}
 \def\WW{{\cal W}}

\def\giu{{\downarrow}}\def\su{{\uparrow}}

\def\Val{{\,{\rm Val}\,}}

\def\AA{{\bf A}}\def\XX{{\bf X}}\def\FF{{\bf F}}
\def\UU{{\bf U}}\def\QQ{{\bf Q}}

\def\V0{{\bf 0}}

\def\nvec{{\underline{\nu}}}\def\mmm{{\underline{m}}}\def\Zz{{\underline{Z}}}

\mathchardef\aa = "050B
\mathchardef\bb = "050C
\mathchardef\xxx= "0518
\mathchardef\zz = "0510
\mathchardef\oo = "0521
\mathchardef\ll = "0515
\mathchardef\mm = "0516
\mathchardef\Dp = "0540
\mathchardef\H  = "0548
\mathchardef\ppp= "0570
\mathchardef\nn = "0517
\mathchardef\pps= "0520
\mathchardef\XXX= "0504
\mathchardef\FFF= "0508

\def\ndpr{{\kern1pt\raise 1pt\hbox{$\not$}\kern1pt\dpr\kern1pt}}
\def\Ndpr{{\kern1pt\raise 1pt\hbox{$\not$}\kern0.3pt\dpr\kern1pt}}
\def\hdp{{\h^{\fra12}}}\def\hdm{{\h^{-\fra12}}}

\font\cs=cmcsc10

\font\msytw=msbm10 scaled\magstep1

\font\msytwww=msbm7 scaled\magstep1

\def\RRR{\hbox{\msytw R}}

 \def\ZZZ{\hbox{\msytw Z}}
 \def\zzz{\hbox{\msytwww Z}}
\def\TTT{\hbox{\msytw T}}


\fiat

\null
\hskip1.truecm

\centerline{\titolone Lindstedt series and Hamilton--Jacobi equation}
\centerline{\titolone for hyperbolic tori in three time scales problems}
\*\*

\centerline{\titolo G. Gallavotti, G. Gentile, V. Mastropietro}
\*
\centerline{Universit\`a di Roma 1,2,3 }
\centerline{\Di}
\vskip.8truecm
\line{\vtop{
\line{\hskip1.5truecm\vbox{\advance \hsize by -3.1 truecm
\\{\cs Abstract.}
{\it Interacting systems consisting of two rotators and a pendulum are
considered, in a case in which the uncoupled systems have three very
different characteristic time scales.  The abundance of unstable quasi
periodic motions in phase space is studied via Lindstedt series.  The
result is a strong improvement, compared to our previous results, on
the domain of validity of bounds that imply existence of invariant
tori, large homoclinic angles, long heteroclinic chains and
drift--diffusion in phase space.}}\hfill} }}

\vskip1.5truecm

\0{\titolo \S 1. The Hamiltonian system.}
\numsec=1\numfor=1\*

\0{\bf 1.1.}
Let $(\f,\a_1,\a_2)=(\f,\aa)\in \TTT^3$ be three angles (\ie
positions on circles); let $(I,A_1,A_2)=(I,\AA)\in \RRR^3$ be their
conjugate momenta (or ``{\it actions}''). We consider the Hamiltonian
function, depending on two parameters $\e,\h>0$, defined by
$$ \HH = \hdp \O_{1}A_1  + \h \fra{A_1^2}{2J} + \hdm \O_{2} A_2
+ \fra{I^2}{2J_0} +J_0 g_0^2 (\cos\f-1)+\e f(\f,\a_1,\a_2)
\; , \Eq(1.1) $$
with $f$ an {\it even} trigonometric polynomial of degree $N,N_0$ in
$\aa,\f$ respectively; $\O_{1},\O_{2},J,J_0,g_0$ are positive constants.

This system describes two rotators (one anisochronous, labeled $\#1$,
and one isochronous, labeled $\#2$) interacting with a pendulum which
has its free (\ie with $\e=0$) unstable equilibrium position at
$I=0,\f=0$ and the stable one at $I=0,\f=\p$.
The scale of frequency of the pendulum is $O(1)$ in $\h$;
at the same time the two rotators rotate at
constant speed $O(\hdp)$ and $O(\hdm)$ respectively. Hence the system has
three time scales: we assume $\h<1$ so that the {\it slow} rotator is
the \#1 rotator.

The free motion admits invariant tori of dimension $2$ (namely
parameterized by $\AA$, a constant vector, by $\aa$ arbitrary, and
with $I=0$, $\f=0$) which are unstable and possess stable (labeled
$+$) and unstable (labeled $-$) $3$--dimensional manifolds
(parameterized by $\AA$, the same constant vector, by $\aa,\f$ arbitrary,
and with $I=\pm J_0g_0\sqrt{2(1-\cos\f)}$).

We shall study properties that eventually hold when $\h\to0$.  It is
well known ([HMS,CG] for instance) that if $\e$ is small most of the
unperturbed tori and their manifolds still exist, just a little
deformed.  This means that (under the condition stated below) there
exist functions $\UU^\pm_{\AA'}(\f,\aa)$ and $V^\pm_{\AA'}(\f,\aa)$
which are divisible by $\e$ and analytic in $\aa,\f,\e$, for $\aa\in
\TTT^2$, $|\f|<2\p$, $|\e|<\e_0$, with $\e_0$ small enough, such that
an initial datum starting on the ($3$--dimensional) surfaces
$W_\e^{\s}$, $\s=\pm$, defined as
$$ \AA^\s(\f,\aa)=\AA'+\UU^\s_{\AA'}(\f,\aa) \; , \qquad I^\s(\f,\aa)
=\pm J_0g_0\,\sqrt{2(1-\cos\f)}+ V^\s_{\AA'}(\f,\aa) \; , \Eq(1.2)$$
evolves, when the time $t\to\pm\io$, tending to be confused with a quasi
periodic motion on a invariant torus $\TT(\AA')$, with rotation vector
$$ \oo'=(\o'_1,\o'_2) \; , \qquad 
\o'_1 \defi \hdp\O_{1}+\h J^{-1} A'_1 \; ,
\qquad \o'_2 \defi \hdm\O_{2} \; , \Eq(1.3)$$
and furthermore such asymptotic motion takes place with $\AA$ moving
quasi periodically {\it with average} $\AA'$.

{\it All this holds if $\oo'$ verifies the Diophantine condition}
$$|\oo'\cdot\nn|> C |\nn|^{-\t} \; , \qquad
\forall \nn\in\ZZZ^2\setminus\{\V0\} \; , \Eq(1.4)$$
for $C,\t>0$ (which may depend also on $\h$).
The values of $\e$ for which we
shall be able to prove the above will be so small that the part of the
stable and unstable manifolds with $|\f|< \fra32\p$ {\it can be
represented as a graph of $\AA,I$ over $\aa,\f$}.\annota1{Note that if
$\e=0$ they are graphs over $\aa,\f$ for $|\f|$ smaller than {\it any}
prefixed quantity $<2\p$.} Hence we look, since the beginning, for
invariant tori which have the latter property.

The approach to the invariant tori, of the points that lie on their
stable manifolds, will be exponential in the sense that their distances
$d(t)$ to the tori will be such that
$$ \lim_{t\to+\io} t^{-1}\log d(t)^{-1}=\lis g_0 \; , \qquad
\lis g_0\= \lis g_0(\e)\defi(1+\G(\e,g_0))\,g_0 \; , \Eq(1.5) $$
for a suitable analytic function $\G(\e,g_0)$, divisible by $\e$.  We
shall call $\lis g_0$ the {\it Lyapunov exponent} of the torus (it will
depend on $\e$ as well as on the considered torus, \ie on $\oo'$ and
on $\h$). The exponent relative to the approach to the same torus
along its unstable manifold (as $t\to-\io$) will be the same, by time
reversal symmetry defined below.

We fix throughout the paper $\t$ ($\t\ge1$) and {\it we shall mainly
study the dependence of $\e_0$, \ie our {\it estimate} for the
analyticity radius, as a function of $\h$: $\e_0=\e_0(C,\h)$.}

\*

\0{\bf 1.2.}
{\cs Remark.} The relation \equ(1.3) between the value of the average
action and the rotation vector is non trivial and it has been named in
[G1,G2] (where it was pointed out) by saying that the tori of
\equ(1.1) are ``torsion free'' or ``twistless''. It is a remarkable
symmetry property of \equ(1.1), see [G1,Ge2,GGM3].

\*

\0{\bf 1.3.}
If $\e=0$ the stable and unstable manifolds coincide (because the
pendulum separatrix is degenerate); it is a degeneracy that is lost when
$\e\ne0$ and generically the manifolds will have only pairwise
isolated trajectories in common, called {\it homoclinic trajectories}.

Nevertheless time reversal symmetry and parity symmetry\annota2{The
latter symmetry is due to the assumption of evenness of $f$.} hold for
\equ(1.1). If $S^t$ denotes the time evolution and the
involution map $i$ (composition of parity and time reversal) is
defined by $i(\f,\aa,I,\AA)=(2\p-\f,-\aa,I,\AA)$, then $iS^t=S^{-t}i$
and there are relations between the stable and unstable
manifolds that are preserved even for $\e\ne0$. Namely
$$ \eqalign{
\UU^+_{\AA'}(\f,\aa) = & \UU^-_{\AA'}(2\p-\f,-\aa) \; , \cr
V^+_{\AA'}(\f,\aa) = & V^-_{\AA'}(2\p-\f,-\aa) \; , \cr} \Eq(1.6) $$
where care must be exercised because the manifolds contain {\it two}
points over each $\aa,\f$.\annota3{This is in fact already so for
$\e=0$.} Hence if $\f\simeq \p$ the relations in \equ(1.6) concern
points that lie on different connected manifolds; to understand
what happens one should try a drawing taking into account that the
above representations are considered only for $|\f|<\fra32\p$.

Looking at the manifolds at $\f=\p$, {\it assuming their existence and
that they are graphs above $\aa,\f$ for $|\f|<\fra32\p$}, equations
\equ(1.6) imply that, fixed $\AA'$,
$$\QQ(\aa)\defi \UU^+_{\AA'}(\p,\aa)
-\UU^-_{\AA'}(\p,\aa) = -\QQ(-\aa) \; , \Eq(1.7)$$
so that $\QQ(\V0)=\V0$; but, in general, $\QQ(\aa)\ne \V0$ for $\aa\ne\V0$.

The function $\QQ(\aa)$ is called the {\it homoclinic splitting} (or
simply {\it splitting}) {\it vector} at $\f=\p$, and the
determinant of the matrix with entries $\dpr_{\a_i}Q_j(\V 0)$
(splitting matrix) is called the {\it splitting}.
One can more generally consider
$\Zz\=\Zz(\f,\aa)= (\UU^+_{\AA'} (\f,\aa)-\UU^-_{\AA'}(\f,\aa),
V^+_{\AA'}(\f,\aa)- V^-_{\AA'}(\f,\aa))$ which would be the splitting
vector at $\f$. Here and henceforth the vectors in $\RRR^\ell$ will be
denoted with an underlined letter (while the boldface is used for vectors in
$\RRR^{\ell-1}$); so far $\ell=3$, but shortly we shall consider
$\ell\ge 3$.  The function $\Zz$ can be written as the gradient
of a generating function $\F$, \ie $\Zz=(\dpr_\f \F, \dpr_\aa \F)$.
This is a result due to Eliasson who points out that it follows
immediately from the Lagrangian nature of the stable and unstable
manifolds. It is a further symmetry property.\annota4{It can
alternatively be easily seen from the explicit expressions for the
stable and unstable manifolds equations derived in [G1], which also
provide a general algorithm for constructing the function $\F$ as a
convergent series in $\e$ for $\e$ small; see [G3].}

The symmetry of \equ(1.1) (hence the consequent oddness of
$\QQ(\aa)$) implies that there is one trajectory which swings through
$\f=\p$ when $\aa$ is exactly $\V0$: it tends to the same invariant
torus as $t\to\pm\io$, provided the torus exists and its stable and
unstable manifolds are graphs over $\aa,\f$ over an interval of $\f$
greater than $|\f|<\p$.

In this paper we prove the following result.

\*

\0{\bf 1.4.}
{\cs Theorem.} {\it Given the Hamiltonian \equ(1.1), given constants
$s,\O>0$ and given $\h>0$ small enough, the following assertions
hold.\\
\pallino
There are invariant tori with rotation vectors $\oo'$ for all $\oo'$
verifying the Diophantine condition \equ(1.4) with constant
$C=C(\h)=\O e^{-s\h^{-1/2}}$ and
$|\o'_1|\in[\fra12\O_1\hdp,2\O_1\hdp]$.\\
\pallino
Such tori exist for $|\e|<\e_0=O(\h^2)$ and for $\h$ small enough.\\
\pallino
They can be parameterized by their average actions $\AA'$; the angular
velocity is then given by the rotation vector $\oo'\=(\O_1\hdp+\h
J_1^{-1}A'_1,\O_2\hdm)$ (\ie they are ``twistless'') and the Lyapunov
exponents have the form $\lis g_0= (1+\G(\e,g_0))\,g_0$, with
$\G(\e,g_0)$ analytic in $\e$ and divisible by $\e$.\\
\pallino
The parametric equations for such tori and for their stable and
unstable manifolds (``whis\-kers'') can be computed by a convergent
perturbation series in powers of $\e$ around the unperturbed tori with the same
rotation vector and their corresponding stable and unstable manifolds.\\
\pallino
At the homoclinic intersection with $\f=\p$ (existing by symmetry),
between the stable manifold and the unstable manifold of each torus,
the splitting is generically given by the Mel'nikov
integral which is of order $O(\e^2\h^{-\b}e^{-\fra\p2
g_0^{-1}\hdm})$, for $\e$ small enough, with $\b$ depending on the
degree $N_0$ in $\f$ of the perturbation $f$: one can take $\b=2N_0-1$
and the asymptotic formula holds if $|\e|< \h^{\z}$,
$\z=2(N_0+3)$ and $\h$ is small enough.}
\*

\0{\bf 1.5.}
{\cs Remark.} The novelty of the theorem is the ``sharp'' bound
$\e_0=O(\h^2)$. If we ``only'' require $\e_0=O(\h^{\fra92+})$ where
$\fra92+$ is any prefixed positive number $>\fra92$ the result is
proved in [GGM3] (see also [CG] or [Ge2]).  The improvement is made
possible by the {\it totally different technique} used (with respect
to [GGM3]): a technique that has interest in its own right and, we
think, beyond the result itself.
In fact the proof of the last assertion of the theorem is
the content of [GGM2], and the values of the constants $\b$ and
$\z$ are taken from Appendix A2 of [GGM2].

\*

\0{\bf 1.6.} Theorem 1.4 will be proved by a further extension of
Eliasson's method, [E,G2,Ge1,Ge2], for the KAM theorem. The following
discussion will show the correctness of the intuition that ``new''
small divisors appear in the perturbation expansion {\it at orders
spaced by} $O(\h^{-1})$. So that the coupling constant is effectively
$O(\e^{\h^{-1}})$ and the analyticity condition is expected to be
$\e^{\h^{-1}} C(\h)^{-q}$ small (for some $q>0$, determined as in the
discussion in Remark 5.16, item (4), below).  Hence the analyticity
condition will be $\e C(\h)^{-q\h}$ small rather than the far worse
$\e C(\h)^{-q}$ small, that is implied directly from lemma 1 in [CG]
(where $q=6$ is an estimate).

In the one degree of freedom case the corresponding problem is studied
in [N]: it is a problem that arises naturally in the context of
Nekhoroshev theory. In our case the rotation vector is not
one-dimensional, so that the cancellations between resonances typical
of small divisors problems, [E,G1,Ge1,Ge2], have to be exploited in
order to prove convergence of the perturbative series. The fact that
the two components of the rotation vector \equ(1.3) are so different
in scale has the consequence that small divisors can appear only at
high orders, so that the dependence of the radius convergence on the
Diophantine constant $C(\h)$ is highly improvable with respect the
``na\"{\i}ve'' one, as explained above: the proof of such an assertion
is the subject of the present paper (as, in the weaker form, already of
[GGM3]).
\*

\0{\bf 1.7.} The paper is organized as follows.
In \S 2,3,4 the formalism is concisely illustrated and the graphic
representations of the whiskers in terms of tree graphs is exhibited
(for systems more general than \equ(1.1); see \equ(2.1) below). The
analysis is brief but selfcontained, with references to [G1,Ge2] only
given for further insight and details. The basic formalism is in \S 2,
then we work out in \S 3 two specific examples to explain the origin
of the graphical interpretation, and in \S 4 we set up the general
Feynman rules for evaluating the equations of the whiskers
(and the splitting vector as a particular case) as a sum of
quantities that can be elementarily evaluated.  In \S 5 bounds are
derived, assuring the convergence of the perturbative series defining
the whiskers in the more general system in \equ(2.1) below and leading
to Theorem 1.4, when restricted to the Hamiltonian \equ(1.1).

The bounds are derived along the lines of [G1,Ge2]: the main part is
the derivation of the bounds for the part of the expansion
corresponding to what we call the contributions due to ``trees without
leaves'': this is done fully and self consistently in \S 5 and in the
related appendices. Once the bounds on the contributions from trees
without leaves are established, {\it which is the real difficulty},
the same analysis can be applied to bound the other
contributions. Since this is simply reduced, without any further
technical problems, to the case of contributions from the simpler
trees with no leaves we do not repeat this part of the discussion
which is done in [Ge2] following the corresponding analysis done in
[G1,Ge1].

To Appendix A1 we relegate some technical details, while Appendix A3
concerns the cancellation analysis of [Ge2], needed in order to treat
the small divisors problems, with more details with respect to the
quoted paper.  An original technical part is also in Appendix A2 and
deals with the improvement of the dependence of the convergence radius
on the Diophantine constant $C(\h)$.

We do not comment here on the obvious relevance of the above results for
the theory of Arnol'd diffusion: see [GGM3,GGM4].

\vskip1.truecm
\0{\titolo \S 2. Lindstedt series for whiskered tori.}
\numsec=2\numfor=1\*

\0We use the formalism of [Ge2]: it would be pointless to repeat
here the technical work required to motivate the necessity or
usefulness of the notations, and we cannot imagine that any reader may
have interest in the matter that follows unless he has some experience
with Eliasson's method, as exposed in [E] and complemented in
[G1,G2,Ge1,Ge2]. The references to [G1,Ge2] are given only to point at
places where further details on the motivations of the assertions can
be found.

The following analysis innovates [Ge2] in \S 5 because of the
extension of Siegel-Bryuno's bound described in Appendix A2 below:
this section and the next two provide a {\it self contained} description of
the graphical algorithm exploited in \S 5 and Appendix A2.

\*

\0{\bf 2.1.} In the following we shall consider a  Hamiltonian
(``Thirring model'') more general than the one in \equ(1.1), \ie a
Hamiltonian which couples a pendulum with $\ell-1$ rotators via a
perturbation $f_1$ which is always an {\it even trigonometric polynomial},
$$ \HH = \oo\cdot\AA  + \fra{1}{2J}\AA\cdot\AA
+ \fra{I^2}{2J_0} +J_0g_0^2\, f_0(\f) +\e J_0g_0^2\, f_1(\f,\aa)
+ J_0g_0^2\,\g(\e, g_0)\,f_0(\f) \; , \Eq(2.1) $$
where $(\aa,\AA)\in\TTT^{\ell-1}\times\RRR^{\ell-1}$,
$(\f,I)\in \TTT^1\times\RRR^1$, $J_0>0$, $J$ is a diagonal matrix,
with $0<\det J \le +\io$, and
$$ \eqalignno{
f_1(\f,\aa) = & \sum_{n\in\zzz \atop |n|\le N_0}
\sum_{\nn\in\zzz^{\ell-1} \atop |\nn|\le N}
f^1_{\nvec}\, e^{i(\nn\cdot\aa+n\f)} \; , \qquad
f^1_{\nvec}=f^1_{-\nvec} \; , &\eq(2.2) \cr
f_0(\f) = & \left( \cos\f-1 \right) =
\sum_{|n|= 1\atop \nn=\V0} f^0_\nvec e^{i(\nn\cdot\aa+n\f)}
\; , \cr}$$
with $\nvec=(\n_0,\nn)\=(n,\nn)\in \ZZZ\times\ZZZ^{\ell-1}$ and
$|\nn|=\sum_{j=1}^{\ell-1}|\n_j|$; we prefer to consider the
Hamiltonian \equ(2.1) with $\ell$ arbitrary because the Lindstedt
series analysis holds for any $\ell\ge 1$. So that the value $\ell=3$
and the existence of three time scales will be used only to obtain the
second bound in \equ(5.13) below.

The last term in \equ(2.1) could be put together with the free
pendulum potential $J_0 g_0^2\,(\cos\f-1)$ thus modifying the
``gravity acceleration'' $g_0^2$ into $(1+\g(\e,g_0))\,g_0^2$: the term with
$\g(\e,g_0)=\sum_{k=1}^\io\g_k(g_0) \e^k$ is added because we follow
here the approach in [Ge2].  We show that, given $s,\O,\h,\oo'$, with
$\h$ small enough and $\oo'$ verifying the Diophantine condition in
Theorem 1.4, then one can fix $\g(\e,g_0)$ so that, for
$|\e|<O(\h^2)$, there is an invariant torus with average (over time)
action $\AA'$, with the properties in Theorem 1.4 and with Lyapunov
exponent $g_0$ and rotation $\oo'\=\oo+J^{-1}\AA'$. In other words by
adding a {\it counterterm} to the Hamiltonian \equ(1.1) one gets a new
Hamiltonian system, \equ(2.1), with an invariant torus with rotation
$\oo'$ {\it and Lyapunov exponent exactly equal to the prefixed
$g_0$} (see also \S 2.7 below).

We further show that, fixed $\oo'$, $\g(\e,g_0)$ is jointly analytic
in $\e, g_0$, if $g_0$ varies near a prefixed $\lis g_0>0$. Going back
to the original Hamiltonian \equ(1.1) we {\it therefore} set $g_0^2=
\lis g_0^2\,(1+\g(\e,\lis g_0))$ and we can invert the latter relation
as $\lis g_0^2= (1+\G(\e,g_0))\, g_0^2$ for $\e$ small enough (this
will mean: for $|\e|<O(\h^2)$). Hence by
interpreting $g_0^2$ in \equ(1.1) as $\lis g_0^2\,(1+\g(\e,\lis
g_0))$, so that \equ(1.5) holds, we obtain Theorem 1.4 as a corollary
of the above statements.

Of course a similar proof could be done without first fixing the
Lyapunov exponent $\lis g_0$ and then inverting the relation between
the ``dressed exponent'' $\lis g_0$ and the ``bare'' one $g_0$. But it
is well known, from the analogous problem in renormalization theory,
that it is wiser technically and conceptually to work, in perturbation
theory, with prefixed physical quantities (\ie dressed ones). The idea
that perturbation theory would be simpler, in the technical estimates,
is the key idea beyond [G1, Ge1] that is introduced in [Ge2].

\*

\0{\bf 2.2.} From now on let us denote by $\aa$ the initial value of
the rotators angles (\ie at time $t=0$).  We define by
$X_j^{\s}(t;\aa)$, $j=0,\ldots,2\ell-1$, the values of the variables
at time $t$ that are reached from initial data
$X^\s(0;\aa)=(\p,\aa,I^\s(0;\V0),\AA^\s(0;\V0))$, with the given
$\aa$, with $\f=\p$ and with $I,\AA$ such that $X^\s(0;\aa)$ is on the
stable ($\s=+$) or unstable ($\s=-$) manifolds of the invariant torus
that we are searching for; the convention on the labels of $X$ is that
$$ \eqalign{
& X_0^\s=\f^\s \; ; \qquad X_j^\s=\a_j^\s \; , \qquad  \hbox{for }
1<j<\ell \; ; \cr
& X_\ell^\s=I^\s \; ; \qquad X_j^\s=A_j^\s \; , \qquad \hbox{for }
\ell<j<2\ell \; . \cr} \Eq(2.3) $$
All functions in \equ(2.3) depend on $t$ and $\aa$
(the symbols $I^{\s}(t;\aa)$
and $\AA^\s(t;\aa)$ should not be confused with $I^\s(\f,\aa)$ and
$\AA^{\s}(\f,\aa)$ defined in \equ(1.2): in the following no ambiguity
can arise as the quantities $I^\s$ and $\AA^\s$ {\it will be used always
with the meaning in} \equ(2.3) and as functions of $t,\aa$).

This is a parameterization of the stable and unstable manifolds in
terms of $\aa,t$ where $\aa$ is the value of the angular coordinates
at the moment in which $\f=\p$, and $t$ is the time elapsed since. The
parameterization is different from the one in terms of $\aa,\f$ in
\equ(1.2) unless, of course, it is $\f=\p$ and correspondingly
$t=0$. Hence the splitting vector \equ(1.7) at $\f=\p$ can also be written
$Q_j(\aa)=X^+_j(0;\aa)- X^-_j(0,\aa)$, $j=\ell+1,\ldots,2\ell-1$.  Note
that we do not need to consider explicitly the splitting in the
$I$--coordinates  because, by
energy conservation, they are functions of $\f,\aa,\AA^\pm$.

Let $\AA'$ be given and let $\oo'$ in \equ(1.3) be Diophantine with
constants $C=C(\h),\t>0$; see \equ(1.4).
We look for an invariant torus and for its stable
and unstable manifolds with the property that the quasi periodic
rotation on the torus takes place at velocity $\oo'$ and, {\it at the
same time}, the action variables oscillate with an average position $\AA'$.

\*

\pallino Before proceeding we remark that the above {\it two}
requirements may seem contradictory as there may seem to be
no reason for being able to
prescribe simultaneously the ``spectrum'' $\oo'$ and the ``average
action'' $\AA'$ of the invariant tori. In fact this property of ``{\it
twistless}'' motion on the tori or of ``{\it absence of torsion}'' is
very remarkable (see the Remark 1.2 and [G1]): it will appear as due to the
special symmetries of the system \equ(2.1) and to the separation of
the energy into a quadratic part involving actions only and an angular
part involving only the angles.

\*

Note also that we could confine ourselves to study the torus with
average position $\AA'=\V0$, as in [G1,Ge2], because any torus
can be reduced to that one through a trivial canonical
transformation (a translation in the action variables).
This explains why in the quoted papers only the torus
covered with rotation vector $\oo$ is explicitly considered:
however in the following we consider also $\AA'\neq\V0$,
as we are interested in showing the abundance of such tori
in phase space (see the Remark 1.5).

The quantity $X_j^{\s}(t;\aa)$ can be graphically represented as
sum of {\it values} which can be associated with tree graphs, that we
shall call ``Feynman graphs'' or ``trees'' {\it tout court}, see
Fig.\equ(2.4) below. The trees are partially ordered sets of points,
called {\it nodes}, connected by unit lines, called {\it branches},
and they are ``oriented'' towards a point called {\it root}, which is
reached by a single branch of the tree.  Given two nodes $v$ and $w$
of a tree, we say that $w$ precedes $v$ ($w\le v$) if there is a path
connecting $w$ to $v$, oriented from $w$ to $v$.  With an abuse of
notations we shall sometimes consider a tree as the collection of its
nodes, sometimes as the collection of its branches and sometimes as
the collection of both nodes and branches. The root {\it will not} be
considered a node.

A typical tree considered below can be drawn as in Fig.\equ(2.4):
the labels meaning and the caption of such a drawing
(which has to be interpreted as a mathematical formula) will be
elucidated in the coming sections.

\*
\eqfig{199.99919pt}{141.666092pt}{
\ins{-29.16655pt}{74.999695pt}{\rm root}
\ins{0.00000pt}{91.666298pt}{$j$}
\ins{49.99979pt}{70.833046pt}{$v_0$}
\ins{45.83314pt}{91.666298pt}{$\d_{v_0}$}
\ins{126.66615pt}{99.999596pt}{$v_1$}
\ins{120.83284pt}{124.999496pt}{$\d_{v_1}$}
\ins{91.66629pt}{41.666500pt}{$v_2$}
\ins{158.33270pt}{83.333000pt}{$v_3$}
\ins{191.66589pt}{133.332794pt}{$v_5$}
\ins{191.66589pt}{99.999596pt}{$v_6$}
\ins{191.66589pt}{70.833046pt}{$v_7$}
\ins{191.66589pt}{-8.333300pt}{$v_{11}$}
\ins{191.66589pt}{16.666599pt}{$v_{10}$}
\ins{166.66600pt}{54.166447pt}{$v_4$}
\ins{191.66589pt}{54.166447pt}{$v_8$}
\ins{191.66589pt}{37.499847pt}{$v_9$}
}{bggmfig0}{\hskip.6truecm\eq(2.4)}
\kern0.9cm
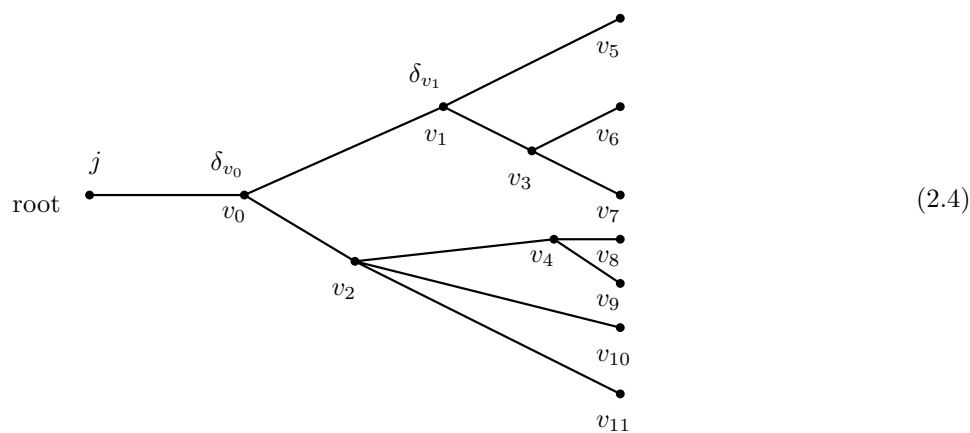
\didascalia{A tree $\th$ with $m=12$,
and some labels. The line numbers,
distinguishing the lines, and their orientation pointing at the root,
are not shown. The lines length should be the same but it is drawn of
arbitrary size. The nodes labels $\d_v$ are indicated only for two
nodes.}

The branch starting at the node $v$ and linking it to the uniquely
determined next node (or to the root), which we call $v'$, will be
denoted by $\l_{v}$: there is a unique correspondence between nodes
and branches starting at them.  We shall say that $\l_v$ exits from
$v$ and enters $v'$; given a node $v$ we shall say that a branch $\l$
{\it pertains} to $v$ if either $\l$ enters $v$ or $\l$ exits from
$v$; \eg in Fig.\equ(2.4) the line $v_1v_0\=\l_{v_1}$ ``exits'' $v_1$
and ``enters'' $v_0$, hence it pertains to both.

In [G1] two expansions are considered for the functions
$X^{\s}_j(t;\aa)$ representing the stable and unstable manifolds: one of
them is used to exhibit cancellations taking place at all orders in the
sums that express the coefficients of the power series in $\e$ of the
splitting vector, [G1,BCG,GGM2];
it is somewhat more involved than the other one that
is convenient to just discuss convergence of the perturbation series for
the splitting vector and that we shall use here.  This is the reason why (as
in [Ge2]) we shall not have trees whose lowest nodes carry a graphical
decoration called {\it form factor}, or {\it fruit} in [G1,GGM2].
Nevertheless some of the nodes will still have a particular
structure: to characterize them we introduce, below as in [Ge2], the
notion of ``{\it leaf}\/'', which is related to the notion of fruit in
[G1], from which it differs (and it, even, differs slightly from the
similar notion of leaf in [Ge2]), see below for the motivation of the
name.
\*

\0{\bf 2.3.} As mentioned the drawing Fig.\equ(2.4) has to be regarded
as a mathematical formula expressing a function of the labels and of
the topological structure of the trees. We now prepare the notation
for the definition of ``value'' of a tree (following [Ge2]) (see
[G1] for a simpler case): the derivation is not difficult but somewhat
long and unusual for the subject (the breakthrough work [E] still does
not seem to be well known in its technical aspects!). We discuss it in
detail not only for completeness but in the attempt to clarify a
construction that has generated quite a few new results starting from
the work of [E], see [G1,GGM2,BGGM].

Let us consider the unperturbed motion $ X^0(t)\=(\f^0(t),\aa+\oo'
t,I^0(t),\AA')$, where $(\f^0(t),I^0(t))$ is the separatrix motion,
generated by the pendulum in \equ(2.1) starting at $t=0$ in $\f=\p,\,
\AA=\AA',I=-2J_0 g_0$, so that $\f^0(t)=4 \arctan e^{-g_0t}$.  Let
$X^\s(t;\a)$, $\s={\rm sign}\,t=\pm$, be the evolution, under the flow
generated by
\equ(1.1), of the point on $W^\s_\e$ which at time $t=0$ is
$(\p,\aa,I^\s(\aa,\p),\AA^\s(\aa,\p))$, see \equ(1.2); let
$$X^\s(t)\=X^\s(t;\aa)\equiv \sum_{h\ge 0} X^{h\s}(t;\aa) \e^h=
\sum_{h\ge 0} X^{h\s}(t) \e^h,\qquad \s=\pm \; , \Eq(2.5)$$
be the power series in $\e$ of $X^\s$, (which we want to show to be
convergent for $\e$ small); note that $X^{0\s}\=X^0$ is the
unperturbed whisker. We shall often omit writing explicitly the $\aa$
variable among the arguments of various $\aa$-dependent functions, to
simplify the notations, and we shall regard the two functions
$X^{h\s}(t)$, as forming a single function $X^h(t)$, which is
$X^{h+}(t)$ if $\s=+,\, t>0$, and $X^{h-}(t)$ if $\s=-,\,t<0$.

Components of $X$ will be labeled $j$, $j=0,\ldots,2\ell-1$,
consistently with \equ(2.3), with the
convention that $X_0\defi X_-$ describes the coordinate $\f$,
$(X_j)_{j=1,\ldots,\ell-1}\defi\XX_\giu$ describes the $\aa$
coordinates, $X_\ell\defi X_+$ describes the $I$ coordinate and
$(X_j)_{j=\ell+1,\ldots,2\ell-1}\defi \XX_\su$
describes the $\AA$ coordinates,
$$ X \defi\, (X_j)_{j=0,\ldots,2\ell-1}\defi\,
(X_-,\XX_\giu,X_+,\XX_\su) \; , \Eq(2.6)$$
\ie we write first the angle and then the action components, first
the pendulum and then the rotators. The ${\bf \su}$ (``{\it up}'') and ${\bf
\giu}$ (``{\it down}'') labels recall that the components with labels ${\bf
\giu}$ ($0< j<\ell$) have ``lower'' index than the variables with
labels ${\bf \su}$ ($\ell<j$), which have a ``higher'' index (a
mnemonically useful fact, on first reading at least).

Inserting \equ(2.5) into the Hamilton equation associated with
\equ(2.1) we get that the coefficients $X^{h\s}(t)$, $h\ge1$, satisfy
the hierarchy of linear equations
$$ {d\over dt} X^{h\s}(t) = L(t) X^{h\s}(t)+F^{h\s}(t) \; ,\Eq(2.7)$$
with $F^{h\s}(t)$ a $2\ell$--vector and the $2\ell\times2\ell$--matrix
$L(t)$ is
$$\tst
L(t)=\pmatrix{
0                     &\V0         &J_0^{-1}        &\V0      \cr
\V0                   &0           &\V0             &J^{-1}   \cr
 g_0^2J_0 \cos\f^0(t) &\V0         &0               &\V0      \cr
\V0                   &0           &\V0             &0
\cr} \Eq(2.8)$$
For instance, $F^{1\s}(t)$ is a $2\ell$--vector with the first
$0,\ldots,\ell-1$ components vanishing (a consequence of the
assumption that the perturbation only depends on the angular
variables), with the $\ell$--th component equal to $-J_0g_0^2\dpr_\f
f_1(\f^0(t),\aa+\oo' t) +J_0g_0^2\g_1(g_0)\sin(\f^0(t))$ and with the
remaining components equal to $-J_0g_0^2\dpr_\aa f_1(\f^0(t),\aa+\oo' t)$.

In general $F^{h\s}$ depends upon $X^0,\ldots,X^{h-1\s}$ {\it but not on
$X^{h\s}$}.  The entries of the $(2\ell\times 2\ell)$ matrix $L$ have
different meaning according to their position: the $\V0$'s in the first
and third row are $(\ell-1)$--(row)--vectors, the $\V0$'s in the first
and third column are $(\ell-1)$--(column)--vectors, and the $0$'s and
$J^{-1}$ in the second and fourth column are $(\ell-1)\times
(\ell-1)$--matrices, while the $0$'s in the first and third columns are
scalars (as $J_0^{-1}$ is).  The perturbed motions will be described by
{\it dimensionless} quantities $\X,\F$:
$$\eqalign{
X^{h\s}_j = &\,\X^{h\s}_j, \qquad0\le j\le \ell-1,\qquad
X^{h\s}_j= \,J_0g_0\,\X^{h\s}_j,
\qquad  \ell\le j\le 2\ell-1 \; , \cr
\FF^{h\s}_\su=&\,J_0g_0^2\,\FFF^{h\s}_\su,\kern3cm F^{h\s}_+=J_0g_0^2
\,\F^{h\s}_+ \; , \cr}\Eq(2.9)$$
The simple form of the Hamiltonian equations for $\f,\aa$, namely
$\dot\f=J_0^{-1} I,\ \dot\aa=\oo+ J^{-1}\AA$ implies that
$\F^{h\s}_j=F^{h\s}_j\,\=\,0$, for $j=0,\ldots,\ell-1$.
For instance
$$ \F^{1\s}=\left( 0,\V0,-\dpr_\f f_1(\f^0(t),\aa+\oo' t)
+ \g_1(g_0)\sin(\f_0(t)),
-\dpr_\aa f_1(\f^0(t),\aa+\oo' t) \right) \; . \Eq(2.10) $$
Given the form of $L(t)$ and the vanishing of the first $\ell$ components
$F^{h\s}_-$, $\FF^{h\s}_\giu$ of $F^{h\s}$, for $h\ge 1$, the above
hierarchy of equations (determining the stable and unstable manifolds)
takes the form
$$ \eqalign{
& \fra1{g_0}
{d\over dt} \X^{h\s}_+= \cos\f^0\,\X^{h\s}_- + \F^{h\s}_+
\ ,\quad\quad\quad  \fra1{g_0}{d\over dt}
\XXX^{h\s}_\su=\FFF^{h\s}_\su \; , \cr
&  \fra1{g_0}{d\over dt} \X^{h\s}_- = \X^{h\s}_+ \ ,\quad\kern2.7cm
\fra1{g_0}{d\over dt} {\XXX}^{h\s}_\giu =J_0 J^{-1} {\XXX}^{h\s}_\su
\; . \cr} \Eq(2.11) $$
And, for all $h\ge1$, we can easily write (via Taylor expansion and
order matching) the following formula for $\F^{h\s}$ in terms of the
coefficients $\X^0,\ldots,\X^{h-1\s}$ and of the derivatives of $f_0$
and $f_1\=f$, see \equ(2.2).  The first $\ell$ components of $\F^{h\s}$
vanish, as said above, $\F_-^{h\s} \= 0,\, \FFF_\giu^{h\s} \= \V0$, and
$$ \eqalignno{
\FFF_\su^{h\s} = & -\sum_{|\mmm|\ge0} (\dpr_\aa f_1)_{\mmm}
(\f^0,\aa+\oo' t)
\sum_{(h^i_j)_{\mmm,h-1}} \prod_{i=0}^{\ell-1}\prod_{j=1}^{m_i}
\X^{h^i_j \s}_i \; ,\cr
\F_+^{h\s} \= & -\sum_{|\mmm|\ge 2} (\dpr_\f f_0(\f))_{\mmm} (\f^0)
\sum_{(h_j^0)_{\mmm,h}} \prod_{j=1}^{m_0} \X^{h_j^0\s}_- +&\eq(2.12) \cr&
- \sum_{p=1}^h \sum_{|\mmm|\ge 0} \g_p(g_0)\left(\dpr_\f
f_0(\f)\right)_{\mmm} (\f^0)
\sum_{(h_j^0)_{\mmm,h-p}} \prod_{j=1}^{m_0} \X^{h_j^0\s}_- + \cr
& - \sum_{|\mmm|\ge0} (\dpr_\f f_1)_{\mmm}(\f^0,\aa+\oo' t)
\sum_{(h^i_j)_{\mmm,h-1}} \prod_{i=0}^{\ell-1}\prod_{j=1}^{m_i}
\X^{h^i_j\s}_i \; , \cr} $$
where $(G)_{\mmm}(\cdot)$, with $G\in\{\dpr_\f f_0, \dpr_\aa f_1,
\dpr_\f f_1\}$,
and $(h^i_j)_{\mmm,q}$, with $h^i_j\ge 1$, are defined as
$$\eqalign{
(G)_{\mmm}(\cdot)\=&\Bigl(
{ \dpr^{m_0}_\f \dpr^{m_1}_{\a_1} \ldots \dpr^{m_{\ell-1}}_{\a_{\ell-1}}
\,G \over m_0!\,m_1!\,\ldots\,
m_{\ell-1}!} \Bigr)(\cdot) \; ,\cr
(h^i_j)_{\mmm,q}\=&(h^0_1,\ldots,h^0_{m_0},h^1_1,\ldots,h^1_{m_1},
\ldots,h^{\ell-1}_1,\ldots,h^{\ell-1}_{m_{\ell-1}})\qquad \hbox{ with }
\qquad \sum_{i=0}^{\ell-1}\sum_{j=1}^{m_i} h^i_j=q \; , \cr} \Eq(2.13) $$
and $m_i\ge0$, $\mmm=(m_0,\ldots, m_{\ell-1})$, $|\mmm|=\sum_{i=0}^\ell m_i$.
Note that the first two sums in the expression for $\F^{h\s}_+$ can only
involve vectors $\mmm$ with $m_j=0$ if $j\ge1$ (so that $|\mmm|=m_0$),
because the function $f_0$, see \equ(2.2),
depends only on $\f$ and not on $\aa$.
The evolution of $\X^h$ is determined by integrating
\equ(2.8), if the initial data are known.
The $h=1$ case requires a suitable interpretation of the symbols,
given explicitly by \equ(2.10).

Elementary quadrability of the free pendulum equations on the
separatrix leads to the following expression for the ``{\it Wronskian
matrix}'' $W(t)$ of the separatrix motion for the pendulum appearing
in \equ(2.1), with initial data at $t=0$ given by $\f=\p,I=-2g_0 J_0$,
\ie $\X^0_+=-2$. The matrix
$$ W(t)=\pmatrix{w_{00}(t)&w_{0\ell}(t)\cr
w_{\ell0}(t)&w_{\ell\ell}(t) \cr} \Eq(2.14) $$
is defined to be the solution of the linearization of the free pendulum
equation around the separatrix solution, with data $W(0)=1$ and with
$J_0=1$ ({\it because we use dimensionless solutions} $\X$, see
\equ(2.11)):
$$ W(t)=\pmatrix{
{1\over\cosh g_0t}&{\bar w(t)\over4}\cr
-{\sinh g_0t\over\cosh^2 g_0t}&
\left(1-{\bar w(t)\over4}{\sinh g_0t\over\cosh^2g_0t}\right)\cosh g_0t\cr},
\qquad \bar w(t)\={2g_0t+\sinh 2g_0t\over\cosh g_0t} \; . \Eq(2.15) $$
The evolution of the $I,\f$ components, \ie $\X^{h\s}_j$ with
$j=0,\ell$ (also identified with the components with subscripts $\pm$,
see \equ(2.6)) can be determined from $W(t)$, by integrating \equ(2.7)
for the $0$ and $\ell$ components, to be
$$\pmatrix{\X^{h\s}_-\cr \X^{h\s}_+\cr}= W(t)
\pmatrix{0\cr \X^{h\s}_+(0)\cr} +
W(t)\ig_0^{g_0t}{W\,}^{-1}(\t)\pmatrix{0\cr \F^{h\s}_+(\t)\cr}\ d\,g_0\t
\; . \Eq(2.16)$$
Thus, denoting by $w_{ij}$ ($i,j=0,\ell$) the entries of $W(t)$,
\equ(2.16) becomes, for $h\ge1$,
$$ \eqalign{
& \X^{h\s}_-(t) =
w_{0\ell}(t)\Big( \X^{h\s}_+(0)+ \ig_0^{g_0t}w_{00}(\t)
\F^{h\s}_+(\t)\,d\,g_0\t\Big)-w_{00}(t)\ig_0^{g_0t}w_{0\ell}(\t)
\F^{h\s}_+(\t)\,d\,g_0\t \; , \cr
& \X^{h\s}_+(t) =
w_{\ell\ell}(t)\Big(\X^{h\s}_+(0)+
\ig_0^{g_0t} w_{00}(\t) \F^{h\s}_+(\t)\,d\,g_0\t\Big)-w_{\ell0}(t)
\ig_0^{g_0t} w_{0\ell}(\t) \F^{h\s}_+(\t)\,d\,g_0\t \; , \cr} \Eq(2.17)$$
having used that $\X^{h,\s}_-(0)=0$ because the
initial datum for $\f$ is fixed and $\e$--independent.
Likewise integration of the equations \equ(2.11) for the $\su,\giu$
components yields, for $h\ge1$,
$$ \eqalign{
& \XXX_\giu^{h\s}(t) = {J^{-1} J_0}
\Big[ g_0t\Big(\XXX_\su^{h\s}(0)+
\ig_0^{g_0t}\FFF^{h\s}_\su(\t)\,d\,g_0\t \Big) -
\ig_0^{g_0t}g_0\t\,\FFF^{h\s}_\su(\t)\,d\,g_0\t \Big] \; , \cr
& \XXX_\su^{h\s}(t) = \Big(\XXX_\su^{h\s}(0)+\ig_0^{g_0t}
\FFF^{h\s}_\su(\t)\,d\,g_0\t\Big) \; , \cr} \Eq(2.18) $$
having used that the $\XXX^{h\s}_\giu (0)\=\V0$ because the initial datum
for $\aa$ is fixed and $\e$--independent. The equations
\equ(2.17), \equ(2.18) can be used
to find a reasonably simple algorithm to represent the whiskers
equations to all orders $h\ge1$ of the perturbation expansion.

\*

\0{\bf 2.4.} The initial data in \equ(2.17), \equ(2.18) have to be
{\it determined by imposing that the solutions (to all orders) become quasi
periodic} as $t\to\s\io$. This is quite easy and (as to be expected)
this condition is simply that $\X^{h\s}_+(0),\XXX_\su^{h\s}(0)$ are
determined by imposing that the integrals in parentheses become
integrals between $\s\io$ and $t$, \ie
$\X^{h\s}_+(0)=\ig_{\s\io}^0\ldots$ and
$\XXX_\su^{h\s}(0)=\ig_{\s\io}^0\ldots$; see below.

However the latter integrals are no longer
necessarily convergent properly (a few examples suffice to see this);
hence one has to go carefully through the process of imposing the
correct asymptotic behavior in order to see what is the meaning to be
given to such integrals $\ig_{\s\io}^t$. The
analysis can be found in [G1, Ge2]. The result is that all expressions
under integral sign can be written as sums of functions that are
rather special, namely
$$ M(t)=\s^\chi {(\s g_0 t)^j \over j!}
e^{i\oo'\cdot\nn t- p g_0\s t} \; , \Eq(2.19) $$
with $\chi,j,\nn,p$ integers and $p\ge -1$ (see below), so that one has
$$ \X^{h\s}(t)=\sum_{\nn\in\zzz^{\ell-1}}\sum_{p=-1}^{\io}
\tilde \X^{h\s}(\nn,p)
\,e^{i\oo'\cdot\nn t - pg_0\s t} \; , \qquad
\F^{h\s}(t)= \sum_{\nn\in\zzz^{\ell-1}}\sum_{p=-1}^{\io}
\tilde \F^{h\s}(\nn,p)
\,e^{i\oo'\cdot\nn t - pg_0\s t} \; , \Eq(2.20) $$
where we explicitly write down only the dependence on $\nn$ and $p$
(clearly also the fixed constants like $J,J_0,g_0,\ldots$ enter).

The series turn out to be convergent for $\s t>0$; however their sums
have {\it no singularity} at $t=0$ and can be anaytically continued
for $\s t<0$ (\ie $x\ge1$). More precisely the functions that one has to
integrate are contained in an {\it algebra} $\hat \MM$ on which the
integration operations that we need can be given a meaning.

\*

\0{\cs Definition} ([G1]).
{\it Let $\hat\MM$ be the space of the functions
of $t$ which can be represented, for some $k\ge 0$, as
$$M(t)=\sum_{j=0}^k{(\s t g_0)^j\over j!} M_j^\s(x,\oo t)\ ,\quad
x\=e^{-\s g_0t}\ ,\quad \s={\rm sign}\, t \; , \Eq(2.21)$$
with $M_j^\s(x,\pps)$ a trigonometric polynomial in $\pps$ with
coefficients holomorphic in the $x$-plane in the annulus $0<|x|<1$,
with possible singularities, outside the open unit disk, in a closed
cone centered at the origin, with axis of symmetry on the imaginary
axis and half opening $<\fra\p2$, and possible polar singularities at
$x=0$. The smallest cone containing the singularities will be called
the {\it singularity cone} of $M$.}

\*

The proper interpretation of the improper
integrals $\ig_{\s\io}^{g_0t} M(\t) d g_0 \t$,
which henceforth will be denoted by $\igb_{\s\io}^{g_0t} M(\t)
dg_0\t$, is simply the {\it residuum} at
$R=0$ of the analytic function
$$ \II_R M\defi\ig_{\s\io+i\theta}^{g_0t}e^{-Rg_0\s z} M(z)\,d\,g_0z \; ,
\Eq(2.22)  $$
(where $\theta$ is arbitrarily prefixed) which is
defined and holomorphic for $\Re R>0$ and large enough, \ie
$$\II M(t) \= \igb_{\s\io}^{g_0t} dg_0\t \, M(\t)
\defi \oint\fra{d R}{2\p i R} \,\II_R M(t) \; . \Eq(2.23)$$
By linear extension this defines the integration of function in $\hat
\MM$ for $|x|<1$. The analyticity in $x$ around $x=\pm1$ and the
remarks that $\fra{d}{d g_0 t}\II M(t)\= M(t)$, \ie $\II M(t)\=
\II M(t')+\ig_{g_0t'}^{g_0t} d\,g_0\t\, M(\t)$, so that $\II M(t)$ is
a special primitive of $M(t)$ (at fixed $\s$), allow us to
analytically continue the result of the integration to a function in
$\hat \MM$. The operator $\II$ maps the algebra $\hat \MM$ into itself
because one checks that on the monomial \equ(2.19) one has
$$\II M(t)=\cases{- g_0^{-1} \s^{\chi +1}e^{i\oo'\cdot\nn t-pg_0\s t}
\sum_{h=0}^j (g_0\s t)^{j-h} {1\over(j-h)!}
{1 \over(p- i \s g_0^{-1} \oo'\cdot\nn)^{h+1}}
\; , & if $|p|+|\nn|>0 \; , $\cr
g_0^{-1}\s^{\ch+1}\fra{(\s g_0 t)^{j+1}}{(j+1)!} \; ,
& otherwise $\; , $ \cr}\Eq(2.24)$$
showing, in particular, that the radius of convergence in $x$ of $\II
M$, for a general $M$, is the same as that of $M$. But in general the
singularities will not be polar, even when those of the
$M_j^\s$'s were such.

We shall see that the cases $|p|+|\nn|=0$ do not enter in the
discussion (a feature of the method of [Ge2]).  The complete
expression of $X^{h\s}(t)$ becomes
$$\eqalignno{
&\X^{h\s}_-(t) = w_{0\ell}(t)\II(w_{00}\F^{h\s}_+)(t)-w_{00}(t)\big(
\II(w_{0\ell}\F^{h\s}_+)(t)-\II(w_{0\ell}\F^{h\s}_+)(0^\s)\Big)\defi
\OO(\F^{h\s}_+)(t) \; , \cr
&
\XXX^{h\s}_\giu(t) = J^{-1}J_0 \Big(\II^2(\FFF^{h\s}_\su)(t)-\II^2(
\FFF^{h\s}_\su)(0^\s)\Big) \defi
\lis\II^2(\FFF^{h\s}_\su(t)) \; , & \eq(2.25)
\cr
& \X^{h\s}_+(t) = w_{\ell\ell}(t)\II(w_{00}\F^{h\s}_+)(t)-w_{\ell0}(t)
\Big(\II(w_{0\ell}\F^{h\s}_+)
(t)-\II(w_{0\ell}\F^{h\s}_+)(0^\s)\Big)\defi\OO_+(\F^{h\s}_+)(t)
\; , \cr
&
\XXX^{h\s}_\su(t) = \II(\FFF^{h\s}_\su)(t) \; , \cr} $$
where $\OO,\OO_+,\lis\II^2$ are implicitly defined here (and $\II^2$
is $\II$ applied twice); and
$\X^{h\s},\F^{h\s}\=(0,\V0,\F_+^{h\s},$ $\FFF^{h\s}_\su)$ are introduced in
\equ(2.9). While $\X^{h\s}$ has non zero components over both the
{\it angle} ($j=0,\ldots,\ell-1$) and over the {\it action}
($j=\ell,\ldots,2\ell-1$) components, the $\F^{h\s}$ has, as already
noted, only the action directions non zero; the notation $0^\s$ means
the limit as $t\to0$ from the left ($\s=-$) or from the right
($\s=+$), but below we shall drop the superscript on $0$ (always clear
from he context because it is the same as the superscript $\s$ of
the functions $\X^{h\s}$).  Furthermore, with
the definitions \equ(2.20) of $\tilde \FFF_\su^{h\s}(\nn,p)$ one
finds also the property (with the notations
in \equ(2.1))
$$ \tilde \FFF_{\su}^{h\s}(\V0,0) = \V0\; , \Eq(2.26)$$
for all $h\ge1$.

We shall repeatedly use that in order to compute $\X^{h\s}_j$ we only
need $\X^{h'\s}_{j'}$ with $0\le j'<\ell$ (\ie only
$\X^{h'\s}_+, \X^{h'\s}_\su$) and $h'<h$. This follows from
\equ(2.25) and \equ(2.12): whether we want to compute an ``{\it action
component}'' ($\X^{h\s}_j,\ j\ge\ell$) or an ``{\it angle component}''
($\X^{h\s}_j,\ j<\ell$) of $\X^{h\s}$, we only need the angle components
of lower orders, \ie $\X^{h'\s}_{j'}$ with $h'<h$ and $j'<\ell$.

\*
\0{\bf 2.5.} The linearity of the last of \equ(2.25), together with
\equ(2.26) and the $t$--dependence of $\FFF^{h\s}(t)$
in \equ(2.20), implies that the prefixed value $\AA'$ {\it has the
interpretation of average action} of the quasi periodic motion on the
invariant torus to which the trajectories that we study asymptote; see
the third statement in Theorem 1.4. This corresponds to the identity
of [CG] (see, in the latter reference, the first of (6.34) and its
proof in Appendix A12) that follows from the symplectic structure of
the equations of motion, according to a well known argument going back
to Poincar\'e, [P], discussed also in [E,CZ]. It is a property that
generated the qualification of ``{\it twistless tori'}'' given in [G1]
to such tori: the ``dispersion relation'' linking the frequencies to
the average actions {\it does not change} or {\it is not twisted} when
the perturbation is switched on. This is a property, established in
the present context in (33) of [Ge2], that can be ultimately traced
back to the fact that in the above models the twist condition is not
needed for establishing a KAM theorem.

\*

\0{\bf 2.6.} By combining \equ(2.25) and \equ(2.12), \equ(2.13) (and
recalling \equ(2.9)) the representation in terms of trees is immediate;
the integrals in \equ(2.25) and the lower order $X^h$ in \equ(2.12)
become {\it recursively} multiple (improper) integrals
over dummy ``time'' variables.

In this operation each function $(-\dpr_\aa f_1(\f^0(t),\aa+\oo'
t))_{\mmm}$ and $(-\dpr_\f f_0(\f^0(t)))_{\mmm}$ is expanded as a linear
combination of monomials $M(t)$ having the form
$\s^\chi (\s g_0t)^j(j!)^{-1} x^n e^{i\oo'\cdot\nn t}$
with $x=e^{-g_0 \s t}$; see \equ(2.19).

The form of \equ(2.12) shows that the integrations occur in a
hierarchical order: {\it hence one can describe them by a
tree}. The integrands can be identified by attaching to each node of
the tree suitably many labels.
We shall first illustrate the construction of the trees
via two examples (in \S 3 below): this can be useful in order to
understand the general case (see also [G1,Ge2]).

\*
\0{\bf 2.7.} We shall establish, also recursively, that $\X^{h\s}$ will
be expanded in monomial like \equ(2.19) with $j=0$ and $p\ge0$,
so that at $t\to\pm\io$ the quantities $\X^{h\s}$ will approach
exponentially fast quasi periodic functions describing the motion on
the invariant torus. The approach will be proportional to $e^{-
g_0|t|}$ or to a higher power of this quantity. This, together with
the remark that at order $0$ (\ie on the unperturbed motion) the
approach is precisely proportional to $e^{- g_0|t|}$ (in the $I,\f$
coordinates), will imply that at least for $j=0,\ell$ (and ``generically''
also for the other coordinates)
$$ \lim_{t\to\io} {1\over \s t} \log |\X^{\s}_j(t)|^{-1} = g_0
\; , \Eq(2.27) $$
\ie that the Lyapunov exponents of the torus are $\pm g_0$.

\*

Before stating the general graphical  rules to represent \equ(2.25) in
terms of explicitly performed integrals, we discuss in detail two examples:
understanding them facilitates enormously, we think, the understanding
of the general cases which will be exposed referring to the examples
to make it more concrete.
\ifnum\mgnf=0\pagina\fi
\vskip1.truecm
\0{\titolo \S 3. Two examples of the trees construction.}
\numsec=3\numfor=1\*

\0{\bf 3.1.}
We discuss how to make more explicit \equ(2.25) by performing two
``third order'' examples. The first order reduces trivially
to the first order formulae (Mel'nikov integral);
the second order is also a bit
too simple and is left to the reader: the first two orders
will be, of course, implicitly done below,
because to compute the third order one needs the first and second too.

To third order the last line in \equ(2.25) gives
$\XXX_{\su}^{3\s}(t)=\II(\FFF_{\su}^{3\s})(t)$, where
$\FFF_{\su}^{3\s}$ can be expressed through the first equation
in \equ(2.12), so that,
for $j=\ell+1,\ldots,2\ell-1$, one has
$$ \eqalign{
\F_j^{3\s} & = -{1\over2}\dpr_{\a_j}\dpr^2_\f f_1\,\X_-^{1\s}\X_-^{1\s}
-\dpr_{\a_j}\dpr_\f f_1\,\X_-^{2\s}
-\sum_{p=1}^{\ell-1}\dpr_{\a_j}\dpr_{\a_p}\dpr_\f f_1\,\X_p^{1\s}
\X_-^{1\s}+\cr &
-{1\over2}\sum_{p,q=1}^{\ell-1}\dpr_{\a_j}\dpr_{\a_p}
\dpr_{\a_q} f_1\,\X_p^{1\s}\X_q^{1\s}
-\sum_{p=1}^{\ell-1}\dpr_{\a_j}\dpr_{\a_p}
f_1\,\X_p^{2\s} \; , \cr} \Eq(3.1) $$
where $\X^{1\s}$ and $\X^{2\s}$ can be written
by using once more \equ(2.25) (the first two lines only as per the
general remark in the last paragraph of \S2.4).

We consider explicitly two contributions to $\XXX_\su^{3\s}(t)$.
Recalling that $\s=+$ corresponds to the stable manifold
and $\s=-$ to the unstable one, the first will be

$$\fra12 \igb_{\s\io}^{g_0t} dg_0\t_{v_0} (-\dpr_{\a_j\a_p\a_q}
f_1(\f^0(\t_{v_0}),\aa+\oo'\t_{v_0}) \,\,
\X^{1\s}_p(\t_{v_0})\,\X^{1\s}_q(\t_{v_0}) \; , \Eq(3.2)$$
arising from the fourth contribution in the r.h.s. of \equ(3.1).
The contribution \equ(3.2) can be written more explicitly,
by using again the expression for $\XXX^{1\s}_\giu$ in \equ(2.25), as
$$\fra12 \igb_{\s\io}^{g_0t} dg_0\t_{v_0} \left(-\dpr_{\a_j\a_p\a_q}
f_1\right) (\t_{v_0})\,\,\lis\II^2(-\dpr_{\a_p}f_1(\t_{v_1}))(\t_{v_0})
\,\,\lis\II^2(-\dpr_{\a_p}f_1(\t_{v_2}))(\t_{v_0}) \; , \Eq(3.3)$$
where the $\lis\II^2$ operations involve, see \equ(2.25), integrations
over variables that we can call $\t_{v_1}, \t_{v_2}$ and the
derivatives of $f_1$ are evaluated at
$(\f^0(\t_{v_n}),\aa+\oo'\t_{v_n})$, $n=0,1,2$. Such variables have
been indicated explicitly using the abbreviated notation $(\t_{v_n})$
and with a obvious abuses of notation (they should not appear at all,
except $\t_{v_0}$, being dummy).

The second example is obtained by considering the contribution with
$h^0_2=2$ from the first line of \equ(2.12), \ie the second
contribution in the r.h.s. of \equ(3.1),
$$ \igb_{\s\io}^{g_0t} \left( -\dpr_{\a_j\f}f_1 \right) (\t_{v_0})
\,\, \X^{2\s}_-(\t_{v_0})\, d\t_{v_0} \; , \Eq(3.4) $$
still imagining the derivatives of $f_1$ evaluated at
$(\f^0(\t_{v_0}),\aa+\oo'\t_{v_0})$. This contribution will be the sum
of several terms, because $\X^{2\s}_-(\t_{v_0})$ has to be expressed by
using \equ(2.25) and \equ(2.12). One of the (many) contributions will be
$$ \fra12\igb_{\s\io}^{g_0t} dg_0\t_{v_0} \left(
-\dpr_{\a_j\f}f_1\right) (\t_{v_0}) \,
\OO\Big(-\dpr^3_{\f}f_0 (\t_{v_1}) \,\OO\big(-\dpr_\f f_1(\t_{v_2})
\big)(\t_{v_1})\,
\OO\big(-\dpr_\f f_1(\t_{v_3}) \big)(\t_{v_1}) \Big)(\t_{v_0})
\; , \Eq(3.5)$$
where the $\OO$ operations involve, see \equ(2.25), integrations over
variables that we can call $\t_{v_1}, \t_{v_2},\t_{v_3}$ and the
derivatives of $f_0,f_1$ are evaluated at
$(\f^0(\t_{v_n}),\aa+\oo'\t_{v_n})$, $n=0,1,2,3$. Such variables have
been indicated explicitly with the same abuse of notation as above;
and the dependence on $\t$ of the derivatives of $f_0,f_1$ has again
been simply denoted by adding the symbol $(\t_{v_n})$ instead of the
full argument $(\f^0(\t_{v_n}),\aa+\oo'\t_{v_n})$.

A complete representation of the above two contributions to
$\X^{3\s}_j(t)$ is given, {\it with enormous notational
simplification}, by the following trees:

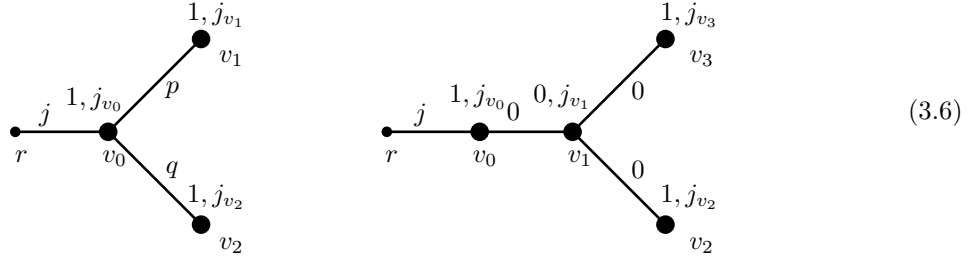
\eqfig{250pt}{100pt}{
\ins{9pt}{56pt}{$j $}
\ins{0pt}{38pt}{$r $}
\ins{33 pt}{38 pt}{$v_0 $}
\ins{57pt}{64pt}{$p $}
\ins{57pt}{34pt}{$q $}
\ins{233pt}{64pt}{$0 $}
\ins{233 pt}{34pt}{$0 $}
\ins{186pt}{56pt}{$0 $}
\ins{151pt}{56pt}{$j $}
\ins{77pt}{76pt}{$v_1 $}
\ins{77pt}{5pt}{$v_2 $}
\ins{140pt}{38pt}{$r $}
\ins{173pt}{38pt}{$v_0 $}
\ins{209pt}{38pt}{$v_1 $}
\ins{255pt}{5pt}{$v_2 $}
\ins{255pt}{76pt}{$v_3 $}
\ins{19pt}{62pt}{$1, j_{v_0}$}
\ins{65pt}{94pt}{$1, j_{v_1}$}
\ins{65pt}{25pt}{$1, j_{v_2}$}
\ins{164pt}{63pt}{$1, j_{v_0}$}
\ins{196pt}{63pt}{$0, j_{v_1}$}
\ins{244pt}{94pt}{$1, j_{v_3} $}
\ins{244pt}{25pt}{$1, j_{v_2} $}
}{albero1}{\eq(3.6)}
\*
\0where the labels on the nodes $v$ are denoted $\d_v,j_v$ and those
on the lines $\l_v$ are denoted $j_{\l_v}$.

The label $\d_v=0,1$ on the node $v$ indicates selection of
$f_{\d_v}$, \ie of $f_0$ or $f_1$, the label $j_v$ denotes a
derivative with respect to $\f$ if $j_{v}=\ell$ or with respect to
$\a_{j_v}$ if $j_v=\ell+1,\ldots,2\ell-1$. For the label
$j_{\l_v}$ associated with the branch $\l_v$ following $v$, one has
$j_{\l_v}=j_v-\ell$ for all $v$ except for the highest node $v_0$, for
which one has $j_{\l_{v_0}}=j_{v_0}$. In the examples above,
\equ(3.3) and \equ(3.5) correspond, respectively,
to the first figure in \equ(3.6) with $j_{v_0}=j,j_{v_1}=p+\ell,
j_{v_2}=q+\ell$ and to the second with
$j_{v_1}=j_{v_2}=j_{v_3}=\ell,j_{v_0}=j$, (hence
$j_{\l_{v_1}}=j_{\l_{v_2}}=j_{\l_{v_3}}=0$, $j_{v_0}=j$).  In the
examples the labels $p,q$ correspond to $\dpr_{\a_p},\dpr_{\a_q}$ in
\equ(3.3).

\*
\0{\bf 3.2.} {\cs Remark.}
The exception for the meaning of $j_{\l_{v_0}}$ is
convenient, in the above cases, as the
integration over $\t_{v_0}$ differs from the others: the inner
ones evaluate $\X^{h\s}_j$ for $j=0,\ldots,\ell-1$, because the
functions $f_0,f_1$ only depend on the angle variables (see the last
paragraph in \S 2.4); the last integral, however, evaluates in the
examples a component of $\XXX^{h\s}_\su$ (which is labeled
$j=\ell+1,\ldots,2\ell-1$), but, in general, $j$ can be any value
$j=0,\ldots,2\ell-1$. Note that this is not so for the inner labels
$j_\l$ which must be angle labels $j_\l=0,\ldots,\ell-1$.  So, in
general, we shall have that the value of a tree with $j_{\l_{v_0}}=j$
contributes to $\X^{h\s}_j$.

\vskip1.truecm
\0{\titolo \S 4. Trees and Feynman graphs approach to whiskers construction:
the general case.}
\numsec=4\numfor=1\*

\0We now proceed to describe the general case.

\*

\0{\bf 4.1.} To compute the splitting vector we only need to consider the
variable $t$ equal to $0$. However we shall be also interested in
$\X^{h\s}(\aa,t)$ for $\s t>0$, for instance in order to study how
fast the invariant torus is approached by the motions on its stable
and unstable manifolds (to obtain its Lyapunov exponent). Hence it
will be natural to attribute the label $t$ to the root: this will also
remind that the integral over $\t_{v_0}$ has to be performed between
$\s\io$ and $t$, (the value $\s=-$ corresponds to the unstable
manifold and the value $\s=+$ corresponds to the stable one). Since we
shall {\it never} consider the stable manifold for $t>0$ or the
unstable for $t<0$ the value of $\s$ will be the same as that of the
sign of $t$.

We shall be interested in computing not only $\XX_{\su}^\s(0;\aa)-\AA'$
(or $\XX^\s_{\su}(t;\aa)-\AA'$), as in [GGM2], but, more
generally, $X^\s(t;\aa)-X^0(t;\aa)$, with $\s=\hbox{sign}\,t$, (here
$X^{0}$ denotes the unperturbed motion).

In general the rules to express $X^\s(t;\aa)-X^0(t;\aa)$ as sum of
``values'' associated with trees will be described now, assuming that the
reader follows us by applying and checking them to the special cases
\equ(3.3), \equ(3.5), illustrated in \equ(3.6).

\*

The reader might be helped in following the construction of the
algorithm to express the stable and unstable manifolds below, by
keeping in mind that we simply decompose the (quite involved and
recursively defined by \equ(2.25),\equ(2.12)) expressions for the
whiskers, so far obtained, {\it further}.

The purpose being of reducing their evaluation to {\it very elementary
algebraic operations}: ultimately just products of simple factors
associated with the nodes (and their labels) of a tree, that we shall
call ``coupling constants'', and of factors associated with the
branches (and their labels), that we shall call ``propagators'', each of
which can be trivially evaluated and trivially bounded.

The reader familiar with Quantum Field Theory will realize the
striking analogy between the algorithms discussed below and the {\it
Feynman graphs}: in fact a ``tree'' will turn out as an analog to a
(loopless) Feynman graph and {\it very likely it is} a Feynman graph
of a suitable (non trivial) field theory. Our analysis amounts to a
renormalization group analysis of it and it partially extends, to the
case of the theory of the stable and unstable manifolds of hyperbolic
tori in nearly integrale systems, the field theoretic interpretation
already discussed in detail in previous works, see [GGM1] and appended
references, in the study of KAM tori.

\*

\pallino To each node we attach an {\it order label} $\d_v=0,1$,
see Fig.\equ(3.6), and a corresponding function $f_{\d_v}$: if a node
$v$ bears a label $\d_v=1$ the associated function is $f_1$ and if it
bears a label $\d_v=0$ it is $f_0$.

\*

\pallino To each node $v$ of a tree $\th$, see the Figure \equ(2.4) above,
we associate  an integration {\it time variable} $\t_v$ and an {\it
integration operation}, which corresponds to $\lis\II^2$ or $\OO$ if
the node {\it is not the highest node} $v_0$ and to
$\lis\II^2$ or $\OO$ or $\II$ or $\OO_+$
if the node $v$ {\it is the highest}, \ie $v=v_0$.
This is so because in the first case (a ``lower node'')
one must use the first two equations in \equ(2.25) because in
\equ(2.12) only angle components of $X^{h\s}$ appear, while in the
second case (that of the highest node) one can use all of
\equ(2.25) since we can evaluate either an angle coordinate
$\X^{h\s}_j(\aa,t)$, $j<\ell$, or an action coordinate, $j\ge\ell$.

When $v< v_0$ the choice between the two possibilities will be marked
by an {\it action label} $j_v$ associated with each node:
if $j_v=\ell$, $v<v_0$, then we choose $\OO$, if $j_v=\ell+1,\ldots,
2\ell-1$, $v<v_0$, we choose $\lis\II^2$.

\*

\pallino When $v$ is the highest node $v_0$, there are therefore
more possibilities: to distinguish between them we use the
{\it action label} $j_{v_0}$ and the {\it branch
label} $j_{\l_{v_0}}$, which can be equal either to $j_{v_0}$ or to
$j_{v_0}-\ell$.  So when $v=v_0$ and $j_{v_0}=\ell$, we choose $\OO$ if
$j_{\l_{v_0}}=j_{v_0}-\ell=0$ and $\OO_+$ if
$j_{\l_{v_0}}=j_{v_0}=\ell$, see \equ(2.25), while when $v=v_0$ and
$j_{v_0}>\ell$, we choose $\lis\II^2$ if $j_{\l_{v_0}}=j_{v_0}-\ell$ and
$\II$ if $j_{\l_{v_0}}=j_{v_0}$, see \equ(2.25).

As said in Remark 3.2, the meaning of the branch label
is that a tree with $j_{\l_{v_0}}=j$ is a graphic
representation of a ``contribution'' to $\X_j^{h\s}$. Therefore if
$j_{\l_{v_0}}\ge\ell$ we call the branch an {\it action branch} and if
$j_{\l_{v_0}}<\ell$ we call it an {\it angle branch}.

In the first of the figures in \equ(3.6) integrals with respect to
the nodes $v_1,v_2$ are of the type $\lis
\II^2$. In the second the integrals over the $\t_{v_n}$, $n=1,2,3$, are
all of the type $\OO$. In both cases the integrals over $\t_{v_0}$ are
of the form $\II$ because we fixed $j_{\l_{v_0}}=j>\ell$ to be an
action label.  We can associate a branch label $j_{\l_v}$ also to the
inner branches with $v<v_0$: however in this case one has necessarily
$j_{\l_v}=j_v-\ell$ because the inner branches necessarily represent
angle components $\X_j^{h\s}$ with $j<\ell$, see \equ(2.12). Hence no
information is carried by such labels that we define only for
uniformity of notation. The latter labels appear in \equ(3.6) as
$j,p,q$ in the first tree and as $j,0,0,0$ in the second one.  The
labels $j_{\l}$ corresponding to the lines pertaining to a node $v$
determine, as in the examples of \S3, which derivatives have to be taken
of the function $f_{\d_v}$ which is associated with $v$: each line $\l_v$
with label $j_{\l_v}$ corresponds to a derivative of $f_{\d_v}$ with
respect to $\f$ if $j_{\l_v}=0$ or to $\a_{j_{\l_v}}$ if $0<j_{\l_v}<\ell$.

\*

\pallino The integrations over the node times $\t_v$ must be thought
of as improper integrals, in the above sense, from $\s\io$ to
either $\t_{v'}$ or $0$ because \equ(2.25) contains various integrals
between such extremes. {\it It will be convenient to distinguish
between such terms}.

This can be easily done by adding, on each node, a new label $\r_v$
also equal to $0,1$: if $\r_v=1$ this means, naturally, that in the
evaluation of the integration operations relative to the node $v$ we
select the terms that correspond to integrations between $\s\io$ and
$\t_{v'}$ while if $\r_v=0$ we select the integrations between $\s\io$
and $0$. We shall imagine that also the highest node carries a label
$\r_{v_0}$ which is $1$ necessarily if we consider only $\XXX_\su(t)$
(because this implies that the function associated with the highest
node must appear differentiated with respect to a $\aa$--component,
see above and \equ(2.12)), but which could be $0$ for the other
components of $\Xi(t)$. Recall also that in this case $\t_{v'}\=t$,
see \equ(2.25).

\*

\pallino We remark that the hierarchical structure of the integrations
implies that if $\r_v=1$ and if $v'$ is the node immediately following
(in the direction of the root) $v$ along the tree then one has
$\t_{v}>\t_{v'}$ if $\s=+$ and $\t_{v}<\t_{v'}$ if $\s=-$, while
$\t_v,\t_{v'}$ have the same sign but are otherwise unrelated if
$\r_v=0$; see \equ(2.25) and check this in the examples.

Besides the labels already introduced also the labels $\r_v=0,1$, just
described but not shown in \equ(3.6), should be imagined carried by
each node.

\*

\pallino Given a tree labeled as above we pick up the nodes $v$
with $\r_v=0$ which are closest to the root, and consider the
subtrees having such nodes as highest nodes.
We call each such subtree, \ie each such node {\it together with the
subtrees ending in it} (and its labels), a {\it leaf}.\annota5{This
definition is slightly different from the one given in [Ge2], where
the leaf represents a collection of trees and, as explained below, is
related to a resummation operation (see also comments in \S 4.2, item
(v), below, and \equ(4.27)), that we do not consider here.}  The name
is natural if one imagines to enclose the part of the tree including
the node $v$ itself and half of the line $\l_v$ into a circle (or,
more pictorially, into a leaf shaped contour): hence, to whom tries
the drawing, it will look like the {\it venations} of a leaf and the
half line outside it will look like its {\it stalk}.

\*

\pallino
All nodes which do not belong to any leaves will be called
{\it free nodes}; they carry, by construction, a label $\r_v=1$, so that
the corresponding time variables are hierarchically ordered from the
lowest nodes up to the root: \ie if $w<v$ then $\t_w< \t_v$ if $\s=+$
and $\t_w>\t_v$ if $\s=-$. Given a tree $\th$ let us call $\th_f$ the
set of free nodes in $\th$, and call $\Th_L$ the set of highest nodes of the
leaves.
\*

\pallino Each
$f_{\d_v}$ function, associated with the node $v$ with order label
$\d_v$, can be decomposed into its Fourier harmonics. This can be done
graphically by adding to each node $v$ a {\it mode} label
$\nvec_v=(\n_{0v},\nn_v)\= (n_v,\nn_v)\in\ZZZ^{\ell}$, with
$|\nn_v|\le N$ and $|n_v|\le N_0$, that denotes the particular
harmonic selected for the node $v$. If $(j_v,\d_v,\r_v,\nvec_v)$ are
the labels of $v$ we will associate with $v$ the quantity
$f^{\d_v}_{\nvec_v}\,e^{i(\oo\cdot\nn_v
\t_{v}+n_v\f^0(\t_v))}$ multiplied by appropriate products of factors
$i n_v$ (one per $\f$--derivative) and $i\n_{vj}$ (one per
$\a_{j}$--derivative, $j=j_{\l_v}$).

If the mode labels $\nvec_v$ are specified for each $v$ we shall define
the {\it momentum} $\nn(v)$ ``flowing'' on a branch $\l_v$ as the sum
of all the angle mode components $\nn_w$ of the nodes $w$ {\it preceding}
the branch, with $v$ included,
$$ \nn(v) \defi\sum_{w\in\th,\ w\le v}\nn_w \; ; \Eq(4.1) $$
the momentum $\nn(v_0)$ flowing through the root branch will
be called the {\it total momentum} (of the tree).

We shall define also the {\it total free momentum} of the tree
as the sum of the mode labels of the free nodes: more generally,
for any free node $v$ we can define the {\it free momentum}
flowing through the branch $\l_v$ as
$$ \nn_0(v) = \sum_{w\in\th_f,\ w \le v} \nn_w \; . \Eq(4.2) $$
For instance in the above examples the two contributions \equ(3.3),
\equ(3.5) (represented by figure \equ(3.6)) are decomposed into sums of
several distinct contributions once the $\r_v$ and the mode labels
$\nvec_v$ are specified.

Likewise we can look at a leaf as a tree: the momentum $\nn'$
flowing through its stalk will then be called the internal {\it leaf
momentum}. Note that its value gives {\it no contribution} to the
total free momentum of the tree to which the leaf belongs.
\*

\pallino The free
momenta will turn out to describe the harmonics of the time dependent
quasi periodic motion around the invariant tori, while the Fourier
expansion modes of $X^{h\s}(t;\aa)$ as a function of
$\aa$ are related to the sum of the free momenta {\it and} of all the
internal leaf momenta. This is an important difference: it is a
property stressed in [G1] where it is referred as ``quasi flatness'',
source of the main difficulties and interest in the theory of
homoclinic splitting, see [G1,GGM2,GGM3,G3].

\*

\0{\bf 4.2.} The trees contributions of the
examples of \S 3 will be sums over the various labels of ``values'' of
trees decorated by more labels:
$$\eqalignno{ &\fra12\igb_{\s\io}^{g_0t} dg_0\t_{v_0}
(-i\n_{v_0j})(i\n_{v_0p})(i\n_{v_0q})\,f^1_{\nvec_{v_0}}\,
e^{i(\nn_{v_0}\cdot\oo'\t_{v_0}+ n_{v_0}\f^0(\t_{v_0}))}\cdot & \eq(4.3)
\cr &\quad\cdot\lis\II^2\big((-i\n_{v_1p})\,f^1_{\nvec_{v_1}}\,
e^{i(\nn_{v_1}\cdot\oo'\t_{v_1}+
n_{v_1}\f^0(\t_{v_1}))}\big)(\r_{v_1}\t_{v_0})\,
\lis\II^2\big((-i\n_{v_2q}) \,f^1_{\nvec_{v_2}}\,
e^{i(\nn_{v_2}\cdot\oo'\t_{v_2}+
n_{v_2}\f^0(\t_{v_2}))}\big)(\r_{v_2}\t_{v_0}) \; , \cr}$$
for \equ(3.3) and
$$\eqalignno{
&\fra12 \igb_{\s\io}^{g_0t} dg_0\t_{v_0}
(-i\n_{v_0j})(in_{v_0})\,f^1_{\nvec_{v_0}}\,
e^{i(\nn_{v_0}\cdot\oo' \t_{v_0}+ n_{v_0}\f^0(\t_{v_0}))}\,
\OO\Big((-in_{v_1})\,f^0_{\nvec_{v_1}}\, e^{i n_{v_1}\f^0(\t_{v_1})}
& \eq(4.4) \cr
& \qquad \OO\big((-i n_{v_2})\,f^1_{\nvec_{v_2}}\,
e^{i(\nn_{v_2}\cdot\oo' \t_{v_2}+ n_{v_2}\f^0(\t_{v_2}))} \big)
(\r_{v_2}\t_{v_1}) \,
\OO\big((-i n_{v_3})\,f^1_{\nvec_{v_3}}\,
e^{i(\nn_{v_3}\cdot\oo' \t_{v_3}+ n_{v_3} \f^0(\t_{v_3}))}\big)
(\r_{v_3}\t_{v_1}) \Big) \; , \cr} $$
for \equ(3.5), with the conventions following \equ(3.3)
about the dummy integration variables.

\*

\pallino The integration operations are still fairly involved, as it
can be seen from \equ(2.25) and from the expressions for $\lis \II^2$ and
$\OO$. With the above conventions for the dummy variables and noting that,
for any function $F$ in $\hat\MM$,
$$ \lis \II^2 \big( F(\t)\big)(t)=J^{-1}J_0\, \Big(
\II(g_0(t-\t)F(\t) )(t) - \II(g_0(t-\t)F(\t))(0) \Big) \; ,\Eq(4.5)$$
we see that the integration over the $\t_v$ has (by \equ(2.25)) one of
the two forms, when $\r_v=1$ and $v'$ is not the root (so that
$j_{\l_v}=j_{v}-\ell$),
$$ \eqalign{
(1)\quad& \II\big(
( w_{0\ell}(\t_{v'}) w_{00} (\t_v) -
w_{00}(\t_{v'}) w_{0\ell}(\t_v) )
e^{i(\oo'\cdot\nn_v\t_v+n_v\f^0(\t_v))}
G_v(\t_v)\big)(\t_{v'}), \qquad j_{\l_v}=0 \; , \cr
(2)\quad&\II\big(g_0(\t_{v'}-\t_v)\,
e^{i(\oo'\cdot\nn_v\t_v+n_v\f^0(\t_v))} G_v(\t_v)\big)(\t_{v'}),
\qquad \kern3.6truecm 0<j_{\l_v}<\ell  \; , \cr}\Eq(4.6)$$
where $G_v(\t_v)$ is a function that depends on the structure of the
tree formed by the nodes preceding $v$ and by the labels attached to
the nodes. If $\r_v=0$ it has one of the two forms
$$\eqalign{
(1)\quad & w_{00}(\t_{v'})\II\big(w_{0\ell}(\t_v)
e^{i(\oo'\cdot\nn_v\t_v+n_v\f^0(\t_v))} G_v(\t_v)\big)(0)\,,\qquad
j_{\l_v}=0 \; , \cr
(2)\quad&\II\big(g_0\t_v\,e^{i(\oo'\cdot\nn_v\t_v+n_v\f^0(\t_v))}
G_v(\t_v)\big)(0),\qquad \kern1.6 truecm
0<j_{\l_v}<\ell  \; . \cr}\Eq(4.7)$$
\*

\pallino When $v'$ is the root the operations involved in the evaluation of
the $\t_v$--integral are slightly different {\it if}
$j_{\l_{v_0}}=j_{v_0}$, \ie if we are considering contributions to the
action coordinates, (if $j_{\l_{v_0}}= j_{v_0}-\ell$ we still have
integrations of the form \equ(4.6) or \equ(4.7)). If $j_{\l_{v_0}}=
j_{v_0}$ the integrations are particularly simple if we are
interested in the evaluation of the splitting vector
\equ(1.7), that is $j_{v_0}>\ell$ and $t=0$;
in such a case the last two of \equ(2.25)
are relevant and setting $v=v_0$ the integration over $\t_{v_0}$ is
the value for $\t_{v'}$ of
$$\eqalign{
(1)\quad&\II\left(w_{00}(\t_v) e^{i(\oo'\cdot\nn_v\t_v+n_v\f^0(\t_v))}
G_v(\t_v)\right)(0) \; , \qquad j_{\l_v} = \ell \; , \cr
(2)\quad&\II\left(e^{i(\oo'\cdot\nn_v\t_v+n_v\f^0(\t_v))}
G_v(\t_v)\right)(0) \; ,
\qquad \kern1.2truecm j_{\l_v}>\ell \; . \cr} \Eq(4.8)$$
because, if $\t_{v'}=0$, one has $w_{\ell\ell}(0)=1$ and
$w_{\ell 0}(0)=0$; see \equ(2.15) and the last two of \equ(2.25).

More generally, if $\t_{v'_0}=t\neq 0$, setting $v=v_0$ and
$r=v'_0$, one defines for $\r_{v_0}=1$
$$\eqalign{
(1)\quad& \II\big(
( w_{0\ell}(\t_{v'}) w_{00} (\t_v) -
w_{00}(\t_{v'}) w_{0\ell}(\t_v) )
e^{i(\oo'\cdot\nn_v\t_v+n_v\f^0(\t_v))}
G_v(\t_v)\big)(\t_{v'}), \qquad j_{\l_v}=0 \; , \cr
(2)\quad&\II\big(g_0(\t_{v'}-\t_v)\,
e^{i(\oo'\cdot\nn_v\t_v+n_v\f^0(\t_v))} G_v(\t_v)\big)(\t_{v'}),
\qquad \kern3.6truecm 0< j_{\l_v}<\ell  \; , \cr
(3)\quad& \II\big(
(w_{\ell\ell}(\t_{v'}) w_{00} (\t_v) -
w_{\ell 0}(\t_{v'}) w_{0\ell} (\t_v) )
e^{i(\oo'\cdot\nn_v\t_v+n_v\f^0(\t_v))}
G_v(\t_v)\big)(\t_{v'}) \; , \qquad j_{\l_v}=\ell \; , \cr
(4)\quad&\II\big(
e^{i(\oo'\cdot\nn_v\t_v+n_v\f^0(\t_v))}
G_v(\t_v)\big)(\t_{v'}) \; , \qquad \kern5.3truecm j_{\l_v}>\ell \; ,
\cr} \Eq(4.9)$$
(see the last two relations in \equ(2.25)) and for $\r_{v_0}=0$
$$\eqalign{
(1)\quad & w_{00}(\t_{v'})\II\big(w_{0\ell}(\t_v)
e^{i(\oo'\cdot\nn_v\t_v+n_v\f^0(\t_v))} G_v(\t_v)\big)(0)\,,\qquad
j_{\l_v}=0 \; , \cr
(2)\quad&\II\big(g_0\t_v\,e^{i(\oo'\cdot\nn_v\t_v+n_v\f^0(\t_v))}
G_v(\t_v)\big)(0),\qquad \kern1.7truecm 0<j_{\l_v}< \ell  \; . \cr
(3) \quad & w_{\ell 0}(\t_{v'})\II\big(w_{0\ell}(\t_v)
e^{i(\oo'\cdot\nn_v\t_v+n_v\f^0(\t_v))} G_v(\t_v)\big)(0) \; ,
\qquad j_{\l_v}=\ell \; , \cr
(4) \quad & 0 \; , \qquad \kern7truecm j_{\l_v}>0
\; ; \cr}\Eq(4.10)$$
note that, for $\t_{v'}=t=0$ and $j_{\l_{v_0}}\ge \ell$,
\equ(4.9) and \equ(4.10), summed together, give \equ(4.8).

\*

\pallino Hence each node still describes a rather complicated set
of operations: it is, therefore, convenient to consider separately the
terms that appear in \equ(4.6)$\div$\equ(4.10). This can be done by
simply adding further labels at each node. To this end, looking at the
integrals in \equ(4.7) and \equ(4.10), at $\r_v=0$, and in
\equ(4.6) and \equ(4.9), at $\r_v=1$, we see that the
following kernels are involved in the integrals
$$ \eqalignno{
w^0_{j_{\l_v}}(\t_{v'},\t_v) & = \cases{
w_{00}(\t_{v'}) w_{0\ell}(\t_v) ,
& \kern2.8truecm$v>v_0\, , j_v=\ell \, $ $\to$ $j_{\l_v}=0\,,$ \cr
g_0\t_v , & \kern2.8truecm $v>v_0\, , j_v>\ell \,$
$\to$ $0<j_{\l_v}<\ell\,,$ \cr}
\cr
w^0_{j_{\l_{v_0}}}(t,\t_{v_0}) & = \cases{
w_{00}(t) w_{0\ell}(\t_{v_0}) ,
& \kern2.8truecm $j_{v_0}=\ell, \quad j_{\l_{v_0}}=0\,,$ \cr
g_0\t_{v_0} , & \kern2.8truecm $j_{v_0}>\ell ,
\quad 0<j_{\l_{v_0}}<\ell\,,$ \cr
w_{\ell0}(t) w_{0\ell}(\t_{v_0}) ,
& \kern3truecm$j_{v_0}=\ell,\quad j_{\l_{v_0}}=\ell$ \cr 0 , &
\kern3truecm$j_{v_0}>\ell,\quad j_{\l_{v_0}}>\ell \, , $ \cr}
& \eq(4.11) \cr
w^1_{j_{\l_v}}(\t_{v'},\t_v) & = \cases{
w_{0\ell}(\t_{v'}) w_{00}(\t_v) - w_{00}(\t_{v'}) w_{0\ell}(\t_v), &
$v>v_0\ , j_v=\ell \,$ $\to$ $j_{\l_v}=0\,,$ \cr g_0(\t_{v'}-\t_v), &
$v>v_0\ , j_v>\ell
\,$ $\to$ $0<j_{\l_v}<\ell\,,$ \cr}
\cr
w^1_{j_{\l_{v_0}}}(t,\t_{v_0}) & = \cases{
w_{0\ell}(t) w_{00}(\t_{v_0}) - w_{00}(t)
w_{0\ell}(\t_{v_0}), & $j_{v_0}=\ell , \quad j_{\l_{v_0}}=0\,,$ \cr
g_0(t-\t_{v_0}), & $j_{v_0}>\ell , \quad 0<j_{\l_{v_0}}<\ell\,,$ \cr
w_{\ell\ell}(t)w_{00}(\t_{v_0}) - w_{\ell0}(t) w_{0\ell}(\t_{v_0}), &
\kern0.3truecm$j_{v_0}=\ell, \, j_{\l_{v_0}}=\ell , $\cr 1 ,
&\kern0.3truecm $j_{v_0}>\ell, \,j_{\l_{v_0}}>\ell , $\cr} \cr}$$
respectively appearing in \equ(4.7) and \equ(4.10), at $\r_v=0$,
and in \equ(4.6) and \equ(4.9), at $\r_v=1$.

The function in \equ(4.11) involving the Wronskian matrix elements can
be computed from \equ(2.15) and one finds, for instance, that the function
in the seventh row on the r.h.s. is
$$
w_{0\ell}(\t_{v'}) w_{00}(\t_v) - w_{00}(\t_{v'}) w_{0\ell}(\t_v) =
\fra12 \left\{\fra{g_0(\t_{v'}-\t_{v})}{\cosh g_0\t_{v'}\,
\cosh g_0\t_{v}}+\fra{\sinh g_0\t_{v'}}
{\cosh g_0\t_{v}}-\fra{\sinh g_0\t_{v}}{\cosh g_0\t_{v'}}
\right\} \; ; \Eq(4.12)$$
hence if we consider \equ(4.6)$\div$\equ(4.10) we note that
the integrals over $\t_v$ involve functions that can be written, for
$\r=\r_v,\t=\t_v,\t'=\t_{v'}$ and for suitable coefficients
$c_j(\r,\a,v)$, ($\r=1$ if we consider \equ(4.6), \equ(4.9) and
$\r=0$ if we consider \equ(4.7), \equ(4.10)),
$$ \sum_{\a=-1}^2 T_\r^{(\a)}(\r\t',\t)\,Y^{(\a)}(\t',\t)
\, c_j(\r,\a,v) \; , \Eq(4.13)$$
where $Y^{(\a)}(\t',\t)$ are given, if $x=e^{-\s g_0 \t}$
and $x'=e^{-\s g_0 \t'}$, by
$$ \eqalignno{
Y^{(-1)}(\t',\t) =
&\fra12 {\sinh g_0\t\over\cosh g_0\t'} \,
\exp[in \f^0(\t)] =
\sum_{k'=1}^{\io}\sum_{k=-1}^{\io}
y_n^{(-1)}(k',k) {x'}^{k'}x^{k} \; ,\qquad k'\ {\rm odd}\,, \cr
Y^{(0)}(\t',\t) =
&\fra12 {\exp[in\f^0(\t)]\over\cosh g_0\t'\cosh g_0\t}
=\sum_{k'=1}^{\io}\sum_{k=1}^{\io}
y_n^{(0)}(k',k) x'^{k'}x^{k} \; , \qquad \kern1.2truecm k'\
{\rm odd}\,, &\eq(4.14)\cr
Y^{(1)}(\t',\t) =
&\fra12 {\sinh g_0\t'\over\cosh g_0\t} \,
\exp[in\f^0(\t)]= \sum_{k'=-1}^{\io}\sum_{k=1}^{\io}
y_n^{(1)}(k',k) {x'}^{k'}x^{k} \; , \qquad \kern.3truecm
k'\ {\rm odd} \, , \cr
Y^{(2)}(\t',\t) =
&\exp[in\f^0(\t)] = \sum_{k=0}^{\io} \tilde y_n^{(2)}(0,k) x^{k}
\; , \qquad \kern3.6truecm k'\=0\cr} $$
which define the coefficients $y_n^{(\a)}(k',k)$ for $\a=-1,0,1,2$ (it
is easily checked that $k'$ is {\it odd} in the first three relations)
and we set, for $\a=-1,0,1,2$,
$$ T^{(\a)}_\r(\r\t',\t) = \cases{
g_0(\t'-\t) & if $\a$ is either $0$ or $2$ and $\r=1\;$, \cr
g_0\t & if $\a$ is either $0$ or $2$ and $\r=0\;,$ \cr
1 & if $\a$ is either $-1$ or $1\;.$ \cr} \Eq(4.15) $$
{\it Likewise} we shall set, defining the coefficients $\tilde
y_n^{(\a)}(k',k)$, for $\a=-1,0,1$, and $\lis y_n^{(-1)}(k',k)$,
$$\eqalign{
&\tilde Y^{(\a)}(\t',\t)=-\tanh
g_0\t'\, Y^{(\a)}(\t',\t)\defi \sum_{k'=-\a}^{\io}\sum_{k=\a}^\io
\tilde y^{(\a)}_n(k',k) {x'}^{k'}x^{k}\; , \qquad\a=\pm1,k'={\rm odd}\;, \cr
&\tilde Y^{(0)}(\t',\t)=-\tanh
g_0\t'\, Y^{(0)}(\t',\t)\defi \sum_{k'=1}^{\io}\sum_{k=1}^\io
\tilde y^{(0)}_n(k',k) {x'}^{k'}x^{k} \;,\qquad\qquad k'\ {\rm odd}\, \cr
&\tilde Y^{(2)}(\t',\t)= Y^{(2}(\t',\t)\defi\sum_{k=1}^\io
\tilde y^{(2)}_n(0,k) x^{k} \;,\cr
&\lis Y^{(1)}(\t',\t)= {\cosh g_0\t'\over \cosh g_0\t}
\exp[in\f^0(\t)]\defi\sum_{k'=-1}^{\io}\sum_{k=1}^\io
\lis y^{(1)}_n(k',k)
{x'}^{k'}x^{k},\qquad\quad k'\ {\rm odd}\,,\cr
&\tilde T^{(0)}_1(\t',\t)=
g_0(\t'-\t),\qquad \tilde T^{(2)}_1(\t',\t)\=1,
\qquad \tilde T^{(0)}_0(0,\t)=T^{(0)}_0 \, ; \cr}
\Eq(4.16) $$
in all other cases the $T,\tilde T, \lis T$--functions will be defined
$1$ (no matter which is the value of the labels that we attribute to
them: this is done to uniformize the notation.

The label $k$ will be called the {\it incoming hyperbolic mode} and
$k'$ the {\it outgoing hyperbolic mode} for reasons that become clear
by contemplating \equ(4.19) below.

In terms of \equ(4.14)$\div$\equ(4.16) the functions
\equ(4.11) multiplied by $\exp[in_v\f^0(\t_v)]$ can be expressed
as in \equ(4.13),
thus defining implicitly the coefficients $c_j(\r,\a,v)$ in \equ(4.13):
$$ \eqalignno{
w^0_{j_{\l_v}}(\t_{v'},\t_v) \, \exp[in_v\f^0(\t_v)] & = \cases{
T^{(0)}_0(0,\t_v)\,Y^{(0)} (\t_{v'},\t_v) + Y^{(-1)} (\t_{v'},\t_v),
& $j_{\l_v}=j_v-\ell=0, $ \cr
T^{(2)}_0(0,\t_v)\, Y^{(2)} (\t_{v'},\t_v),
& $0<j_{\l_v}=j_v-\ell<\ell,$ \cr} \cr
w^0_{j_{\l_{v_0}}}(t,\t_{v_0}) \, \exp[in_{v_0}\f^0(\t_{v_0})] & = \cases{
T^{(0)}_0(0,\t_{v_0})\,Y^{(0)} (t,\t_{v_0}) + Y^{(-1)} (t,\t_{v_0}),
& $j_{\l_{v_0}}=j_{v_0}-\ell=0, $ \cr
T^{(2)}_0(0,\t_{v_0})\, Y^{(2)} (t,\t_{v_0}),
& $0<j_{\l_{v_0}}=j_{v_0}-\ell<\ell,$ \cr
\tilde T^{(0)}_0(0,\t_{v_0})\, \, \tilde Y^{(0)} (t,\t_{v_0}) +
\tilde Y^{(-1)} (t,\t_{v_0}),
& $j_{{v_0}}= j_{\l_{v_0}}=\ell,$ \cr
0 , & $j_{{v_0}}=j_{\l_{v_0}}>\ell,$ \cr} & \eq(4.17) \cr
w^1_{j_{\l_v}}(\t_{v'},\t_v) \, \exp[in_v\f^0(\t_v)] & = \cases{
T^{(0)}_1(\t_{v'},\t_v)\,Y^{(0)} (\t_{v'},\t_v)
+ Y^{(1)} (\t_{v'},\t_v) + & \cr \qquad - Y^{(-1)} (\t_{v'},\t_v),
& $j_{\l_v}=j_v-\ell=0,$ \cr
T^{(2)}_1(\t_{v'},\t_v)\,Y^{(2)} (\t_{v'},\t_v),
& $0<j_{\l_v}=j_v-\ell<\ell , $\cr} \cr
w^1_{j_{\l_{v_0}}}(t,\t_{v_0}) \, \exp[in_{v_0}\f^0(\t_{v_0})] & = \cases{
T^{(0)}_1(t,\t_{v_0})\,Y^{(0)} (t,\t_{v_0})
+ Y^{(1)} (t,\t_{v_0}) + & \cr \qquad - Y^{(-1)} (t,\t_{v_0}),
& $j_{\l_{v_0}}=j_{v_0}-\ell=0,$ \cr
T^{(2)}_1(t,\t_{v_0})\,Y^{(2)} (t,\t_{v_0}),
& $0<j_{\l_{v_0}}=j_{v_0}-\ell<\ell, $\cr
\tilde T^{(0)}_1(t,\t_{v_0})\,
\tilde Y^{(0)} (t,\t_{v_0})
+\tilde Y^{(1)} (t,\t_{v_0}) + &\cr \qquad -\tilde Y^{(-1)}
(t,\t_{v_0}) \big) +\lis Y^{(1)}(t,\t_v) , & $j_{v_0}=j_{\l_{v_0}}=\ell,$
\cr \tilde T^{(2)}_1(t,\t_{v_0}) \tilde Y^{(2)} (t,\t_{v_0}) ,
& $j_{v_0}=j_{\l_{v_0}}>\ell . $\cr} \cr} $$
One could avoid introducing the $\tilde T$ functions as they are
simply related to the $T$ functions or are just identically $1$:
however it is convenient to introduce them to make the above formulae
more symmetric and therefore easier to keep in mind while working with.

Finally we define the coefficients $\x_j(k',0)$ by the power series
expansion
$$ \eqalign{
{1 \over \cosh g_0\t{'}} & = \sum_{k'=1}^{\io} \x_\ell(k',0) x{'}^{k'}
\; , \qquad \ k'\ge1\,, \hbox{ odd } \; , \cr
1 & = \x_j(0,0) \; , \qquad j>\ell \; , \cr}\Eq(4.18) $$
where $x'=e^{-\s g_0 \t'}$ and $k'$ is odd, which occurs as
coefficient $w_{00}(\t')$ in \equ(4.7) (when $\r_v=0$, \ie
$v\in\Th_L$).

The above definitions (taken from (42) and (45) in [Ge2]) suffice to
discuss the whiskers (and therefore the splitting in the action variables).

\*

\pallino The \equ(4.13) allow us to introduce a
``relatively simple notation'': we can add to each node a {\it badge}
label $\a_v=(-1,0,1,2)$ that will distinguish which choice we make
between the possibilities in \equ(4.14) and \equ(4.16) and two
{\it hyperbolic mode} labels $k'_v,k_v$ which select which particular
term we choose in the sums in \equ(4.14) and \equ(4.16);
they are integer numbers $\ge-1$. We shall
not have to introduce labels to distinguish terms coming from the
expansions of $Y^{(\a)},\tilde Y^{(\a)}, \lis Y^{(\a)}$ bearing the
same badge $\a$ because one can check that the labels $\a_v$ together
with $j_v$ and $v$ itself uniquely determine which choice has to be made.

In terms of the latter labels we can define a {\it hyperbolic
momentum} of a line $\l_v$ as a label $p(v)\in \ZZZ$ which will be the
sum of all the hyperbolic modes of the nodes that precede $v$ {\it
plus} the incoming hyperbolic mode of the node $v$ itself: this
is the sum of the labels $k_w$ associated with all {\it free} nodes
$w\le v$, with $v$ included, and of the labels $k_w'$ associated with
all the {\it free} nodes $w<v$ or {\it highest} nodes of the leaves
$w<v$, with $v$ {\it not} included,
$$ p(v) = k_v+\sum_{w\in \th_f \atop w< v}\left( k_w+
k_w' \right) + \sum_{w\in\Th_L \atop w<v} k_{w}' \; .
\Eq(4.19) $$
{\it A very important property} is that $k_w+k_w'\ge0$, by \equ(4.14)
and \equ(4.16), and
$k'_w\ge0$ if $w\in \Th_L$, by \equ(4.18), so that $p(v)\ge-1$.
Furthermore if $p(v)=0$ then {\it either} $k_v=-1$ and $k_w+k'_w=0$
for all $w<v$ except one single node $\tilde w<v$ (which is either in
$\th_f$ or $\Th_L$) for which $k_{\tilde w}+k_{\tilde w}'=1$,
{\it or} $k_v=0$ and $k_w+k'_w=0$ for all $w<v$. If $p(v)=-1$
then $k_v=-1$ and $k_w+k'_w=0$ for all $w<v$.
\*

\pallino In the above analysis we have not taken explicitly
into account the possibility of contributions to $\F^{h\s}_+$ coming
from the third line in \equ(2.12), \ie counterterm contributions.
They are, of course, possible and they can be taken immediately into
account in
the graphical representation by considering the nodes with a label
$\d_v=0$ and adding to them a {\it counterterm label} $\k_v$, a non
negative integer. If $\k_v=0$ this will mean that the node represents
a contribution from the first line of the definition of $\F^{h\s}_+$,
\ie a contribution that is unrelated to the counterterms, while if
$\k_v\ge1$ the node represents a contribution from the term with $p=\k_v$
in the second line contribution to $\F^{h\s}_+$ in \equ(2.12).

\*

\0{\bf 4.3.}
The trees carry, at this point, quite a few decorating labels
and each tree together with all its labels will represent a ``very
simple'' contribution to the value of the $h$--th order coefficient in
the Taylor expansion in $\e$ (at fixed $\h$ of course) of the
$\X^{h\s}$ vector. Very simple means that the
improper integrals that
correspond to each term are very easy to evaluate explicitly and lead
to a result that can be expressed as a product of factors determined
by the labels of the tree and associated with the nodes and with
the lines, see \equ(4.30), below. We list here
the set of labels that have been introduced:
\*
\halign{\hfill $#$\ &\ #\hfill\cr
j_v &action labels\cr
j_{\l_v} &branch labels\cr
\d_v&order labels\cr
\r_v&leaf labels\cr
\nvec_v&mode labels\cr
\nn(v)&momentum in the branch $\l_v$ following $v$\cr
\nn_0(v)&free momentum in the branch $\l_v$ following $v$\cr
\a_v&badge labels\cr
(k'_v,k_v)&hyperbolic mode labels\cr
p(v)&hyperbolic momentum in the branch $\l_v$ following $v$\cr
\k_v& counterterm labels\cr
}
\*
There are some constraints between the labels, which follow
from the rules stated in \S 4.1 and \S 4.2 and from
the choice of the counterterms
(the latter will be discussed in \S 4.5 below):

\*
\pallino one has $j_{\l_v}=j_v-\ell$ if $v<v_0$ and $j_{\l_v}=j_v$ or
$j_{\l_v}=j_v-\ell$ if $v=v_0$ (see the third item in \S 4.1);

\pallino if $\r_v=0$ then $\a_v\neq 1$, (see \equ(4.17));

\pallino if $j_{\l_v}\ne0,\ell$, then $\a_v=2$,
otherwise if $j_{\l_v}=0,\ell$,
then $\a_v$ can be $-1,0,1$, (see \equ(4.17));

\pallino $\d_v=0$ implies $j_v=\ell$
(by the $\aa$--independence of $f_0$);

\pallino $k_v,k_v',p(v)\ge -1$, (see \equ(4.14), \equ(4.16)
and comment following \equ(4.19));

\pallino $(p(v),\nn_0(v))\neq(0,\V0)$, see Remark 4.6 below.
\*

In terms of such labels, given a decorated tree $\th_0$ with $m_0$ nodes
and with highest node $v_0$, we can define the {\it value} of a {\it
subtree} $\th$ with $m$ free nodes, highest node $w$
(preceding the highest node $v_0$ of $\th_0$: $w\le v_0$) and
label $j_{\l_w}=0,\ldots,\ell-1$, $\r_w=0,1$. It
will be given by the expression
$$ {\rm Val}(\th)= \Big[ \prod_{v\in\th_f \atop v\le w}
\igb_{\s\io}^{\r_v g_0\t_{v'}} dg_0\t_{v}\; \VV_v(\th) \Big]
\Big[ \prod_{v\in\Th_L} \LL_v(\th) \Big]
\Big[ \prod_{v\in\th_f \atop \d_v=0} \g_{\k_v}(g_0) \Big]
\; , \Eq(4.20)$$
where the integration is the improper integration $\II$ (in the sense of
\S 2.4), the tree $\th$ consists of a ``free'' $m$--nodes tree
$\th_f$ with leaves attached to a (possibly empty) subset of
the nodes of $\th_f$, and the following notation has been used.

\*

\0(i) The coefficients
$\VV_v(\th)$ and $\LL_v(\th)$ are described by the collection of
labels enumerated above. They can be written, respectively, as
$$ \VV_v(\th)=\bar F_{\nvec_v} \hat T^{(\a_v)}_{\r_v}(\r_v\t_{v'},\t_v)
\,e^{i\oo'\cdot\nn_v\t_v} x_v^{k_v}\prod_{w\in\th \atop w'=v} x_v^{k_w'}
\,(-1)^{\d_{\a_v,-1}}\,\hat y_{n_v}^{(\a_v)}(k_v',k_v) \;, \Eq(4.21) $$
and
$$ \LL_v(\th)=\x_{j_v}(k_v',0) 
L_{j_v\nn(v)}^{h_v\s}(\th) \; , \Eq(4.22) $$
where $x_v=\exp[-\s g_0\t_v]$ and $\hat y_{n_v}^{(\a_v)},\hat
T^{(\a_v)}_{\r_v}$ are (see \equ(4.17)) either\\
$\bullet$ $y_{n_v}^{(\a_v)}$, $T^{(\a_v)}_{\r_v}$, if
either $v<v_0$ or $v=v_0$ and $j_{\l_{v_0}}=j_{v_0}-\ell$, or\\
$\bullet$ $\tilde y_{n_v}^{(\a_v)}$ or $\lis y^{(1)}_{n_v}$ and
$\tilde T^{(\a_v)}_{\r_v}$ or $\lis T^{(\a_v)}_{\r_v}$,
if $v=v_0$ and $j_{\l_{v_0}}=j_{v_0}$.

Furthermore $\r_v=1$ if $v<w$, while $\r_w$ can be either $0$ or $1$;
$j_{\l_w}$ can any value $0,\ldots,2\ell-1$ if $w=v_0$, in any
other case $j_{\l_v}=0,\ldots,\ell-1$ (see above). In \equ(4.21)
%
$$ \bar F_{\nvec_v} = \Big( {J_0\over J}\Big)^{(1-\d_{j_v,\ell})
(1-\d_{j_v,j_{\l_v}})}f^{\d_v}_{\nvec_v}\Big[ (-i\n_v)_{j_v-\ell}
\prod_{w\in\th \atop w'=v} (i\n_v)_{j_w-\ell} \Big] \Eq(4.23) $$
depends on the labels $(\d,\nvec,j)$ of the node $v$ and of its
predecessors $w$'s (recall that by \equ(2.2) $\nvec_v=(n_v,\nn_v)$);
in \equ(4.22) the quantity
$L_{j_v\nn(v)}^{h_v\s}(\th)$ is called the ``{\it value of the
leaf}\/'' $v$ of order $h_v$ (see item (v) below for its definition).
The matrix $J$ is not, in general, a multiple of the identity and $J_0
J^{-1}$ will be interpreted as acting on the rotator components of
$\nvec_v$ (and it will be $1$ when raised to the power $0$).


\*

\0(ii) For the purposes of the cancellations analysis performed in
Appendix A3, the exact form of a few coefficients among the
$y_{n_v}^{(\a_v)}(k_v',k_v)$'s turns out to be essential, so that we
list them here:
$$ \matrix{
y_{n_v}^{(-1)}(2,-1)= 0 \; , \quad & \quad
y_{n_v}^{(-1)}(1,-1)= \s/2 \; , \quad & \quad
y_{n_v}^{(-1)}(1,0)= 2in_v \; , \cr
y_{n_v}^{(1)}(0,1)= 0 \; , \quad & \quad
y_{n_v}^{(1)}(-1,1)= \s/2 \; , \quad & \quad
y_{n_v}^{(1)}(-1,2)= 2in_v \; , \cr
y_{n_v}^{(2)}(1,0)=0 \; , \quad & \quad
y_{n_v}^{(2)}(0,0)=1 \; , \quad & \quad
y_{n_v}^{(2)}(0,1)=4in_v\s \; . \cr} \Eq(4.24) $$
The coefficients $\tilde y^{(-1)}(1,-1)$,
$\tilde y^{(-1)}(1,0)$,
$\tilde y^{(1)}(-1,1)$ and $\lis y^{(1)}(-1,2)$ are equal to
the corresponding (\ie with the same values of the labels $k',k$)
$y^{(\a)}(k',k)$ coefficients.

\*
\0(iii) The value of a leaf with highest node $v$ in \equ(4.22)
is {\it not} the same as the value $L_{j_v\nn(v)}^{h_v\s}$ in [Ge2]:
this is because of the above mentioned change in notation (see the
sixth item in \S 4.1)).  In [Ge2] leaf values are defined as sums of
the values of all leaves (in the sense we use now) with fixed order,
action label and total momentum. Then the leaf value considered here,
$L_{j_v\nn(v)}^{h_v\s}(\th)$, is a single contribution to the
$L_{j_v\nn(v)}^{h_v\s}$ of [Ge2], and depends {\it only} on the part
of the tree $\th$ consisting of the nodes $w\le v$; if we call $\th_v$
such a subtree, we can write ({\it temporarily, just for the purposes
of comparison}) the present definition of leaf value as
$L_{j_v\nn(v)}^{h_v\s}(\th)$ $\=$ $\bar L_{j_v\nn(v)}^{h_v\s}(\th_v)$
(as it depends only on the labels of the subtree $\th_v$). In order to
make a link between the different notations note that $L_{j\nn}^{h\s}$
in [Ge2] would be, with our present notations, just the sum
$$ L_{j\nn}^{h\s} \defi \sum_{\th_{v_0}\in\TT_{\nn,h} \atop j_{v_0}=j}
\bar L_{j\nn}^{h\s}(\th_{v_0}) \; , \Eq(4.25) $$
where $\th_{v_0}$ is the part of the tree $\th$ on which the leaf
value really depends.

Coming back to our notations we define the
{\it leaf value}
$L^{h\s}_{j\nn}(\th)$, with $j=j_{v_0}$, (where the the first and third of
\equ(4.11) should be used), to be the value of a tree $\th$ with
$j_{\l_{v_0}}=j_{v_0}-\ell$ and $\r_{v_0}=0$.

\*

\0(iv) By construction (see \equ(2.12) and corresponding comments), and
if $\Th_L$ is the set of highest nodes in the leaves, the total
perturbation order $k$ of $\th$ is
$$ k = \sum_{v\in\th_f\atop \d_v=0} \k_v + \sum_{v\in\th_f} \d_v
+ \sum_{v\in\Th_L} h_v =
\sum_{v\in\th \atop \d_v=0} \k_v + \sum_{v\in\th} \d_v
\; , \qquad m<2k\; . \qquad \Eq(4.26) $$
\*

\0(v) Both the counterterms and the leaf values of a given perturbation
order are recursively defined in terms of the same quantities with
lower orders.  In fact $\g_\k(g_0)$ admits a graphical representation
as sum of tree values defined as in \equ(4.20) with the difference
that the integration operation corresponding to the highest node of
the tree has to be suitably modified (see \equ(4.32) below).

If $\Val(\th)$ is defined as in \equ(4.20)
(and in item (iv) above) then, by construction, one has
$$ \X_{j}^{h\s}(t;\aa) =
\sum_{\nn\in\zzz^{\ell-1}}\X_{j\nn}^{h\s}(t) \,
e^{i\nn\cdot\aa} \; , \qquad \X_{j\nn}^{h\s}(t)
=  {1\over m_0!} \sum_{\th_0\in\TT_{\nn,h} \atop
j_{\l_{v_0}}=j } \Val(\th_0) \; , \Eq(4.27) $$
where $\TT_{\nn,h}$ is the collection of all trees with
total momentum $\nn$ and order $h$.
In \equ(4.30) $m_0!^{-1}$ is a combinatorial factor, which depends on
the way we count trees: the simplest is to think that the tree branches
of $\th$ are pairwise distinct and are distinguished by a label
$1,2,\ldots, m_0$, if $m_0$ is the number of nodes in the tree $\th$.
In the latter case, which corresponds to our choice, the factor is
simply $m_0!^{-1}$, see [G1,Ge2], provided we regard as identical two
trees that can be overlapped by pivoting the branches entering a node
(rigidly, together with the subtree attached to them) around any node:
as in [G1] we shall call {\it numbered trees} the trees so counted.

\*

\0{\bf 4.4.} {\cs Remark.}
Since the value of any leaf with highest node $v$ depends only on the
labels of the nodes $w\le v$, the equation
\equ(4.20) factorizes into a product of leaf values times
a product of counterterms (whose value, so far arbitrary, has still to
be specified and it will be, in the analysis between \equ(4.31) and
\equ(4.32) when intervening compatibility requirements will dictate its
value) times a factor
$$ \Big[ \prod_{v\in\th_f \atop v \le w}
\igb_{\s\io}^{\r_v g_0\t_{v'}} dg_0\t_{v}\; \VV_v(\th) \Big]
\Big[ \prod_{v\in\Th_L} \x_v(k_v',0) \Big] \; , \Eq(4.28) $$
which does not depend on the leaves.

\*

\0{\bf 4.5.} The extra effort with respect to the
approach without counterterms developed in [G1,Ge1],
gives here (as in [Ge2]) a reward: few
combinations of powers of the ``times'' $\t_v$ appear in the integrand
in \equ(4.21).  The time variables, by \equ(4.21) and
\equ(4.14)$\div$\equ(4.17), appear only via exponentials like
$$ e^{-\s(\t_v-\t_{v'})a} \qquad \hbox{ or } \qquad
(\t_{v'}-\t_v)\, e^{-\s(\t_{v}-\t_{v'})a } \; , \Eq(4.29) $$
for some complex $a=(g_0 p(v)-i\s\oo'\cdot\nn_0(v))$, yielding
respectively, upon integration, $a^{-1}$ or $a^{-2}$. Note also that
by the hierarchical structure of the trees one has
$\s\,(\t_{v'}-\t_v)\ge0$.

{\it This greatly simplifies the actual performance of the integration
operations} which, once one gets familiarity with the formalism, are
trivial. One can say that the absence of high powers of the $\t_v$'s is
due to having \ap fixed the Lyapunov exponent $g_0$ by means of the
counterterms (by contrast in [G1,Ge1] arbitrary powers of $\t_v$
appeared because $g_0$ is {\it not} fixed \ap).

\*

Of course the triviality of the integrations is entirely due to the
above {\it very fine} decomposition, into terms identified by labeled
trees, of the more compact (but ``difficult'' to integrate) integrands
appearing in \equ(4.6)$\div$\equ(4.12) and in the middle terms in \equ(4.14).

\*

Once all the integration operations will have been performed, the tree
value in \equ(4.20) will become a product of {\it ``factors''}, in
complete analogy with what one is accustomed to find when defining {\it
Feynman graphs} in Field Theory. The factors are associated with the
nodes $v$ and with the branches $\l_{v}$. The value of a tree $\th$
will then be {\it defined} as
$$ \eqalign{
{\rm Val}(\th) = &
e^{i\oo'\cdot\nn_0(v_0)t-\s g_0[k_{v_0}'+p(v_0)] t}
\Big[ \prod_{\l_v\in \th \atop v\in\th_f}
\Big( - \fra{\s g_0}{g_0p(v)-i\s\oo'\cdot\nn_0(v)} \Big)^{r_v} \Big]
\; , \cr & \Big[ \prod_{v\in\th_f}
\bar F_{\n_v}\,(-1)^{\d_{\a_v,-1}}\,y_{n_v}^{(\a_v)}(k_v',k_v) \Big]
\Big[ \prod_{v\in\Th_L} \x_{j_v}(k_v',0) L_{j_v\nn(v)}^{h_v\s}(\th)\Big]
\Big[ \prod_{v\in\th_f \atop \d_v=0} \g_{\k_v}(g_0) \Big] \cr} \Eq(4.30) $$
where $r_v$ is either $1$ or $2$, and the case
$(\nn_0(v),p(v))=(\V0,0)$ has to be {\it excluded} for any node
$v\in\th_f$. This is not to claim that no trees with
$\nn_0(v)=\V0,p(v)=0$ can be drawn by following the above rules: this is
a {\it further rule} to impose on the labels in order that the analysis
does not become contradictory requires fixing the function $\g(\e,g_0)$
conveniently: the consistence criterion determines $\g(\e,g_0)$ uniquely.
This rule is a natural extension of the corresponding rule holding in
perturbation theory of KAM tori, which was discussed by Lindstedt and
Newcomb for the lowest orders of the perturbation expansions and
which was proved to hold at all orders by Poincar\'e, [P];
see the last paragraph in \S 2.4 above and the Remark 4.6, (c), below.

The values of the numbers $r_v$ arise from the time variables integrals
via the mechanism just illustrated above (whereby one either gets
$a^{-1}$ or $a^{-2}$ from the integration of the functions \equ(4.29)).

The factors $-(\s g_0)^{r_v}[g_0p(v)-i\oo'\cdot\nn_0(v)]^{-r_v}$,
associated with the branches, will be called {\it propagators} or
{\it small divisors}. The first name arises from the possible
interpretation of the trees as Feynman graphs of a suitable field theory,
see [GGM1]; the second name corresponds to the usual name given in
Mechanics to such expressions generated by perturbation expansions.

It can be useful to write, if $v_0$ is the highest node of $\th$ and
$j_{\l_{v_0}}=0$,
$$ {\rm Val}(\th) = e^{i\oo'\cdot\nn_0(v_0)t-\s g_0[k_{v_0}'+p(v_0)]t}
\Big( - \fra{\s g_0}{g_0p(v_0)-
i\s\oo'\cdot\nn_0(v_0)} \Big)^{r_{v_0}} \; \overline{{\rm Val}}(\th)
\; , \Eq(4.31) $$
so defining the quantity $\overline{{\rm Val}}(\th)$ (this is a well
known kind of operation on Feynman graphs, which associates with a
graph another value gruesomely called the value of the {\it amputated}
graph -- amputated tree in our case).
Moreover we can define $\overline{{\rm Val}}(\th)$ also for
$(\nn_0(v_0),p(v_0))=(\V0,0)$ as no vanishing denominator appears in
its expression. It is however clear that nodes with
$(\nn_0(v),p(v_0))=(\V0,0)$ must not appear at all in the trees that we
consider, for \equ(4.30) to make sense as it is written. This implies,
not surprisingly, a consistence problem: namely one has to check
that the sum of all the $\lis{{\rm Val}} (\th)$ over trees of a given
order and with $(\nn_0(v_0),p(v_0))=(\V0,0)$ cancel so that lines
$\l_v$ with $(\nn_0(v),p(v))=(\V0,0)$ never appear,
neither for $v=v_0$ nor for $v<v_0$.

The cancellation is made possible because we still have freedom to fix
the counterterms and {\it their choice is in fact uniquely determined
by the conditions that they be such that the needed cancellation takes
place}.  The quantities $\overline{{\rm Val}}(\th)$ are convenient in
order to find and to express the counterterms
and also the ``resonance values'' introduced later (see Appendix A3).
One checks, see Appendix A1, that the counterterms can be explicitly
written, if $\TT_{\nn,h}$ is the collection of all trees with
total momentum $\nn$ and order $h$, as
$$ \g_\k(g_0) = - \fra12
\sum^*_{\th\in\TT_{\V0,\k},\a_{v_0}=-1 \atop p(v_0)=0,\,\nn_0(v_0)=\V0,
\, k_{v_0}'=1} \overline{{\rm Val}}(\th) \; , \Eq(4.32)$$
and the * means that the sum is further restricted so that the tree
contains no leaves. This choice being simply imposed
by the requirement that no contribution with
$(\nn_0(v_0),p(v_0))=(\V0,0)$ can arise for $\X_-^{h\s}(t)$; see
Appendix A1.

\*

\0{\bf 4.6.}
{\cs Remarks.}  (a) The presence of the counterterms will manifest
itself not only through the elimination of the trees whose value
would be meaningless if evaluated via \equ(4.30) but also, and mainly,
in the fact that the elements of the algebra $\hat \MM$ met in the
successive integrations have a special form (namely always like one of
the \equ(4.29)) which implies that the result of the improper integrals
is {\it not} as complicated as one could fear from \equ(2.24). This
leads to the simple expression \equ(4.32) (see Appendix A1 of [G1] for
what would otherwise happen without counterterms).

(b) From \equ(4.31) one sees that if $j_{\l_{v_0}}>\ell$ it is natural
to collect together the terms with $p(v_0)=0$: for them, since
$j_{\l_{v_0}}>\ell$, in \equ(4.30)
one must have $k'_{v_0}+p(v_0)=0$ by the last of
\equ(4.14). Note also that by \equ(2.26) the case
$(\nn_0(v),p(v_0))=(\V0,0)$ is excluded. If $j_{\l_{v_0}}\le \ell$ we,
likewise, collect the terms with $k'_{v_0}+p(v_0)=0$ and, for similar
reasons the term with $k'_{v_0}+p(v_0)=-1$ cannot be present
(see again \equ(4.14) and \equ(4.16), and use
$p(v_0)\ge -1$ supplemented by the relations between the labels
$p(v_0)$ and $k_{v_0}'$ which will be exhibited in \S 5.1).

Hence the cases with $p(v_0)+k'_{v_0}=-1$ are excluded by
construction\annota{6}{\rm The initial data $\X^{h\s}(0,\aa)$ were
determined precisely by imposing boundedness at $\s t=+\io$, \ie by
imposing the absence of divergent terms in the expansion in powers of
$x=e^{-g_0\s t}$ which would correspond to the terms with $p(v_0)=-1$.} and
we see that the sum of the values of the trees with
$p(v_0)+k_{v'_0}=0$ give us the equations for the actions and the
angles of the invariant torus to which the whiskers considered are
asymptotic: the terms with $p(v_0)+k_{v'_0}=0$ asymptote to quasi
periodic functions of $\oo t$ so that replacing $\oo t$ by $\pps\in
T^{\ell-1}$ one gets a parameterization of the points on the tori in
terms of a point $\pps\in T^{\ell-1}$ on a ``standard torus''.

And the terms with $k'_{v_0}+p(v_0)=1$ provide the leading
corrections. Since such terms are present already to order $0$ (as one
sees from the expression of the pendulum separatrix) the distance
between a point moving on the stable manifold of the torus and the
torus itself will be proportional to $x=e^{-g_0\s t}$ as $\s t\to\io$
so that $g_0$ has the interpretation of Lyapunov exponent of the
invariant torus; see \equ(2.27) in \S 2.27.

(c) Summarizing: {\it the case $(\nn_0(v),p(v))=(\V0,0)$ has to be
ruled out as a consequence of \equ(2.26) and of \equ(4.32),
respectively for the contributions to $\XXX_{\su}^{h\s}$ and to
$\X_+^{h\s}$ (see the last constraint listed at the beginning of \S
4.3). All cases with $k'_{v_0}+p(v_0)=-1$ are also excluded}.

\vskip1.truecm
\0{\titolo \S 5. Bounds.}
\numsec=5\numfor=1\*

\0{\bf 5.1.}
We now discuss how to bound the value of a tree or of a sum of a small
number of trees which we take for simplicity without leaves and
without counterterms.
The more general case will be eventually reduced, see
below, to the one we consider here.

We shall discuss first how to bound values of trees without leaves
and counterterms such that $p(v_0)=0$ if $v_0$ is the highest node;
hence we shall consider trees, always without leaves
and counterterms, with $p(v_0)= 0$. At the end we shall see
how the presence of leaves and counterterms modifies the analysis.

The following discussion is ``locally'' simple, but ``globally''
delicate and repeats that in [Ge2], \S4:
the conclusions are also summarized in the table 0,1,2,3 below.

{}From \equ(4.19) it follows that the hyperbolic momentum $p(v)$ is
$p(v)\ge -1$ and, as remarked after
\equ(4.19), $p(v)=0$ can occur only in special cases:
more precisely if $p(v)=0$, then $k_v$ is either $-1$ or $0$, and\\
(1) if $k_{v}=0$, all free nodes $w$ preceding $v$
(whether immediately or not) have $k_w'+k_w=0$, while\\
(2) if $k_v=-1$, all free nodes $w$ preceding $v$ have $k_w'+k_w=0$, {\it
except} for a single node $\tilde w<v$ such that $k_{\tilde
w}'+k_{\tilde w}=1$.\\
In the latter case we call $\PP$ the path of
nodes (\ie the totally ordered set of nodes) which connect $v$ to
$\tilde w$, both extremes included (see also [Ge2], \S 4).

\*

\pallino Supposing $p(v)=-1,0$ and recalling that $\a_v<2$ implies
$k_v'$ odd, the expansions \equ(4.14) impose that there are {\it very
few possible choices} of the values of the hyperbolic modes at $w\le
v$:\\ (1) if there is no path or there is a path linking $v$ to $\tilde
w$ but $w\neq \tilde w$ and $w\notin \PP$, then $k_w+k_{w'}=0$ and the
cases $\a_w=1$, $\a_w=-1$ and $\a_w=2$ require, respectively,
$(k_w',k_w)=(-1,1)$, $(k_w',k_w)=(1,-1)$ and $(k_w',k_w)=(0,0)$:
correspondingly $p(w)=1$, $p(w)=-1$ and $p(w)=0$. While, for $w\in\PP$,
the value of $p(w)$ ``increases by one unit'', \ie $p(w)=2$, $p(w)=0$
and $p(w)=1$ for $w\in\PP$;
\\
(2) if $w=\tilde w$ then $k_w+k_{w'}=1$ and the cases $\a_{\tilde w}=1$,
$\a_{\tilde w}=-1$ and $\a_{\tilde w}=2$ require, respectively,
$(k_{\tilde w}',k_{\tilde w})=(-1,2)$, $(k_{\tilde w}',k_{\tilde
w})=(1,0)$ and $(k_{\tilde w}',k_{\tilde w}) =(0,1)$ (correspondingly
$p(\tilde w)=2$, $p(\tilde w)=0$ and $p(\tilde w)=1$).

Note that in both cases $\a_w=0$ is not possible.

The above analysis covers both cases $v<v_0$ and $v=v_0$,
as the functions in \equ(4.14) and in \equ(4.16)
have the same dependence on $\t$ (hence on $k$).

\*

\pallino The latter properties
have to be considered as a further restriction to impose on the tree
labels, and play an essential r\^ole for the discussion of the
cancellations, [Ge2].
\*

\pallino Moreover if $p(v)=0$ then:

\0(1) if $k_v=0$ then $v$ can be preceded only by leaves with the highest
nodes $w$ having $j_w>\ell$, because $k_w'$ must be $0$ in
such a case, so that the second of \equ(4.18) applies;

\0(2) if $k_v=-1$, then all the leaves again must have the highest node
$w$ with $j_w>\ell$, except at most one leaf with highest node
$\tilde w$ with $j_{\tilde w}=\ell$ and $k_{\tilde w}'=1$.

\*
\0{\cs (Vertical) Table 0.} Possible cases when $p(v)=0,-1$.
\*
\halign{\strut\vrule\kern2truemm$#\quad$&
\vrule\kern2truemm$\quad#\quad$&
\vrule\kern2truemm$\quad#\quad$&
\vrule\kern2truemm$\quad#\quad$&
\vrule\kern2truemm$\quad#$\quad\hfill\vrule\cr
\noalign{\hrule}
p(v)    & k_v & k'_v           & \a_v  & j_{v}    \cr
-1      & -1  & {\rm odd}\ge 1 & -1    & \ell     \cr
0       & -1  & {\rm odd}\ge 1 & -1    & \ell     \cr
0       &  0  & {\ge1}         & -1    & \ell     \cr
0       &  0  & {\ge0}         & 2     & {>\ell}  \cr
\noalign{\hrule}
}

\*

\0{\cs (Horizontal) Table 1.} Cases $p(v)=0$, $w\notin \PP$.
\*
\halign{\strut\vrule\kern2truemm$#\quad$&
\vrule\kern2truemm$\quad#\quad$&
\vrule\kern2truemm$\quad#\quad$&
\vrule\kern2truemm$\quad#\quad$&
\vrule\kern2truemm$\quad#$\quad\hfill\vrule\cr
\noalign{\hrule}
\a_w       & -1     & 0                 &   1    &   2     \cr
(k'_w,k_w) & (1,-1) & \hbox{impossible} & (-1,1) & (0,0)   \cr
p(w)       & -1     & \hbox{impossible} &   1    &   0     \cr
j_w        & \ell   & \hbox{impossible} &   \ell & {>\ell} \cr
\noalign{\hrule}
}

\*

\0{\cs (Horizontal) Table 2.} Cases $p(v)=0$, $w\in \PP, w> \tilde w$.

\*

\halign{\strut\vrule\kern2truemm$#\quad$&
\vrule\kern2truemm$\quad#\quad$&
\vrule\kern2truemm$\quad#\quad$&
\vrule\kern2truemm$\quad#\quad$&
\vrule\kern2truemm$\quad#$\quad\hfill\vrule\cr
\noalign{\hrule}
\a_w   & -1  & 0 & 1 & 2 \cr
(k'_w,k_w)& (1,-1) & {\rm impossible} & (-1,1) & (0,0) \cr
p(w) & 0 & {\rm impossible} & 2 & 1 \cr
j_w & \ell & {\rm impossible} & \ell & {> \ell} \cr
\noalign{\hrule}
}

\*
\0{\cs (Horizontal) Table 3.} Cases $p(v)=0,\, w=\tilde w$.
\*
\halign{\strut\vrule\kern2truemm$\quad#\quad$&
\vrule\kern2truemm$\quad#\quad$&
\vrule\kern2truemm$\quad#\quad$&
\vrule\kern2truemm$\quad#\quad$&
\vrule\kern2truemm$\quad#$\quad\hfill\vrule\cr
\noalign{\hrule}
\a_w   & -1  & 0  & 1 & 2 \cr
(k'_w,k_w)& (1,0) & {\rm impossible} & (-1,2) & (0,1)\cr
p(w) & 0 & {\rm impossible} & 2 & 1 \cr
j_w & \ell & {\rm impossible} & \ell & {> \ell} \cr
\noalign{\hrule}
}

\*

\pallino We extend the definition of path also to the case $p(v)=0,
k_v=0$, by setting $\PP\defi\emptyset$ if $j_v>\ell$ and $\PP\defi v$ if
$j_v=\ell$, only for purposes of notational convenience (see \equ(5.3)
below). This is consistent with the above tables and does not change
them.

\*

\0{\bf 5.2.} {\cs Remark.}
Note that, if a tree (or subtree) $\th_0$ with highest node $v_0$
has total hyperbolic momentum $p(v_0)=0$, then there is one and
only one path $\PP$, and, if $\PP\neq\emptyset$, then $\PP$
connects the node $v_0$ to some node $\tilde w<v_0$.
In fact, if there is a path $\PP\neq\emptyset$, the node $\tilde w$
is so defined that $k_{\tilde w}+k_{\tilde w}'=1$: then if
$k_{v_0}=0$ there cannot be any of such nodes (and $\PP=v_0$
in such a case), while if $k_{v_0}=-1$ there must be one and only
one such node. This simply follows from the analysis in \S 5.1, by
noting that all nodes $w<v_0$ except $\tilde w$ must have $k_w+k_w'=0$.

\*

\0{\bf 5.3.}
The small divisors can be really ``small'' only when $p(v)=0$: if
$p(v)\neq 0$, they are bounded by $g_0^{-r_v}$, \ie by a quantity of
order $O(1)$. So one can consider all free nodes in the trees, among
the ones having $p(v)=0$, which are closest to the root. All free
nodes $v$ between them and the root have propagators which are not
small because $|p(v)|\ge1$ (see also [Ge2], page 298).

Given a tree $\th_0$ with $m$ free nodes,
from each subtree $\th$ ending in a node
$v_0$ with $p(v_0)=0$ (here $v_0$ is some node of $\th_0$: it becomes
the highest node of $\th$), one obtains contributions which can be
naturally collected together (recall Remark 4.4) into a contribution
to the tree value (see \equ(4.30)) consisting in a factor
$$ \prod_{v\le v_0} \bar F_{\nvec_v} G_v[\oo'\cdot\nn_0(v)]
\, y'_{v} \; , \Eq(5.1) $$
(here the product is over all free nodes preceding $v_0$)
times a product of
counterterms $\g_{\k_v}(g_0)$ for $v\in\th_f$ with $\d_v=0$, times a
products of factors $\x_{j_v}(k_v',0) L_{j_v\nn(v)}^{h_v\s}$, for each
$v\in\Th_L$; see \equ(4.30).  The vector $\nn_0(v)$ is the free momentum
(defined above; see \equ(4.2)) flowing through the branch $\l_v$, the
coefficients $y'_{v}$ are related to the expansions \equ(4.14) via
$$y'_{v} = \cases{
\fra12 \left[ y_{n_v}^{(1)}(-1,1)+y_{n_v}^{(-1)}(1,-1) \right] = \s/2
& if $v\in\PP$, $\a_v\in\{-1,1\} \; , $ \cr (-1)^{\d_{\a_v,-1}}\,
y_{n_v}^{(\a_v)}(k_{v'},k_v) & otherwise ; \cr} \Eq(5.2) $$
where $\PP$ denotes the path in $\th$ (there is always
one such path, possibly the empty set, because we suppose $p(v_0)=0$);
and $G_v[\oo'\cdot\nn_0(v)]$ is related to the propagator
of the branch $\l_v$, and it will have the form
$$ G_v[\oo'\cdot\nn_0(v)] = \cases{
g_0^2\left[ i\oo'\cdot\nn_0(v) \right]^{-2}
& if $v\notin\PP$, $j_v>\ell$, \cr
-\s g_0^2\left[ g_0^2+ ( \oo'\cdot\nn_0(v) )^2 \right]^{-1}
& if $v\notin\PP$, $j_v=\ell$, \cr
g_0^{2-\d_{j_v,\ell}}\left[-\s\,( g_0p(v)-i\s \oo'\cdot\nn_0(v))
\right]^{-(2-\d_{j_v,\ell})}
& if $v\in\PP$, $\a_v\neq-1$, $\to p(v)\ne0$ \cr
g_0\left[i\oo'\cdot\nn_0(v)\right]^{-1} &
if $v\in\PP$, $\a_v=-1$, \cr} \Eq(5.3)$$
because:

\0(a) The first line is such because if $j_v>\ell$ one has necessarily
$\a_v=2$, see Tables 0,1,2,3 and we have to integrate a function
$g_0(\t_{v'}-\t_v) e^{i n_v\f^0(\t_v)}$ so that $k_v\ge0$: hence
$k_v=p(v)=0$ and we have the second function in \equ(4.29) to integrate.

\0(b) The second line is such because if $v\not\in\PP,\, j_v=\ell$ we have
$w^1_\ell(\t_{v'},\t_v) e^{i n_{v}\f^0(\t_{v})}$ which is a sum of three
terms (see the third of \equ(4.17)): the first has $k_v+k_{v'}\ge2$ so
is excluded (recall that $p(v_0)=0$ and $v\le v_0$); while the second
only sees the contribution to $Y^{(1)}$ with $k_v'=-1,k_v=1$, see
\equ(4.14), and the third only contributes by the term with
$k_v'=1,k_v=-1$ in $Y^{(-1)}$. In the two cases one has $p(v)=1$ or
$p(v)=-1$ respectively; adding up together the latter two contributions
and using the first of
\equ(5.2) to compute the sum of the coefficients we get
$$ \fra{-\s g_0 y_{n_v}^{(1)}(-1,1) }{g_0-i\s\oo'\cdot\nn_0(v)}
- \fra{-\s g_0 y_{n_v}^{(-1)}(1,-1) }{-g_0-i\s\oo'\cdot\nn_0(v)}
= \fra\s2 \fra{-2\s g_0^2}{g_0^2
+(\oo'\cdot\nn_0(v))^2} \; , \Eq(5.4) $$
as it can be read from the coefficients in the intermediate
column of \equ(4.24) and from $p(v)=\a_v=\pm1$.

\0(c) The third line of \equ(5.3) is obtained by noting that, if
$v\in\PP$, $v> \tilde w$ one has $p(v)=1+k_v$, so that, if $j_v=\ell$
and $\a_v\ne-1$, then $p(v)>0$, see Table 2; if $v=\tilde w$ and
$\a_v\ne-1$, one has $p(v)\ne0$, see Table 3 (note that $\a_v\ne2,0$ so
that we have to consider the first integrand in \equ(4.29)).

If $j_v>\ell$ then $\a_v=2$, and, by the Tables 2,3, one has
$k_{v'}=0,k_v\ge0$ and $k_v+k_{v'}=1$, so that
$k_v=1$ and $p(v)=2$; while, if $v=\tilde w$, then $k_{v'}=0,
k_v\ge0$ and $k_{v}'+k_v=1$ imply $k_v=1$, so that $p(v)=1$.
So in both cases $p(v)\ge 1$.

\0(d) The fourth line is found by looking at the Tables 2,3 as follows: if
$\a_v=-1, v\in\PP, v>\tilde w$, one has  $k_v+k_v'=0$, hence
$k_v=-1,k_{v'}=1$ and $p(v)=0$; this happens only if $j_v=\ell$ so that
we have to consider the first integral in
\equ(4.29) and we get the fourth relation.

\*

This shows that the only trees that do not have a value tending to $0$
as $t\to\s\io$, \ie are those with $p(v_0)+k_{v_0}'=0$ (all the others
tend to $0$ as a power of $x=e^{-g_0\s t}$), have propagators that are
even functions of the momenta flowing in them. In fact the observation
on the absence of paths preceding $v_0$ implies that only the first
two propagators in \equ(5.3) appear in such trees. Since, as already
remarked, the trees with $p(v_0)+k_{v_0}'=0$ give the equations of the
tori this is an interesting check that the tori equations so obtained
at $t=+\io$ and $t=-\io$ do {\it coincide}. A similar analysis, and
check, holds for the cases $j_{\l_{v_0}}\le \ell$.

\*

\0{\bf 5.4.} {\cs Remark.}
Collecting together the contributions from $\a_v=-1$ and $\a_v=1$, for
$v\notin\PP$, is a convenient operation and has nothing to do with the
deeper resummations that imply the cancellations necessary for
convergence estimates: the systematic use
of this operation should be described by adding a label to the trees
on the nodes $v\notin \PP$ and replacing on the branches which give
rise to one of the two propagators in \equ(5.4) the $\a_v$ label by
the new label (\eg a $*$ label which indicates that we consider the
sum of the values of a tree with $\a_v=1$ and one with $\a_v=-1$).
We shall do this without explicitly mentioning the new label, to
simplify the notation. Moreover we can no more associate a label
$p(v)$ to a node of this kind, as two factors with different $p(v)$
label ($p(v)=\pm 1$ for $\a_v=\pm 1$) have been considered together;
nevertheless we shall modify slightly the definition of $p(v)$ by
setting $p(v)\defi 1$ in such a case (and letting it unchanged in all
the other cases).

We shall continue to call $G_v[\oo'\cdot\nn_0(v)]$ a {\it propagator}
as, for the purposes of the following analysis, only
such modified version of the original propagators
appearing in \equ(4.30) plays a r\^ole.

\*

\0{\bf 5.5.}
Furthermore we define the {\it degree} $D$ of a propagator to be $D=2$
if either $v\notin\PP$ or $v\in\PP,j_v>\ell$ (hence $\a_v\ne-1$), and
$D=1$ otherwise (the constraint, see \equ(4.3), $1\le r_v\le 2$
implies that the power to which the divisors appear raised is either
$1$ or $2$); by extension we shall say that a branch $\l$ has degree
$D_{\l}=D$ if the corresponding propagator has degree $D$.

The coefficients $\bar F_{\nvec_v}$ and $y'_{v}$
in \equ(5.1) satisfy the bounds
$$ |y_{v}'|\le 4N \; , \qquad \prod_{v\le v_0}
|\bar F_{\nvec_v}|\le (\CC N^2)^m \; , \Eq(5.5) $$
for some constant $\CC$ depending on the perturbation $f_1$ in
\equ(2.1); see \equ(2.6), \equ(2.13) and \equ(2.18).
For instance one can take
$$ \CC=\max\{|J^{-1}|J_0,1\} \max_{|n|\le N_0 ,
\, |\nn|\le N} |f_\nvec| \; ; \Eq(5.6) $$
see \equ(4.23), where $|J^{-1}|$ is the maximum of the matrix elements
of the (diagonal) matrix $J^{-1}$.

To bound the product in \equ(5.1), we shall consider simultaneously the
cases $k_{v_0}=0,-1$; if $k_{v_0}=0$ the path $\PP$ is supposed to be
reduced to a single node, $v_0$, or to the empty set, $\emptyset$,
depending on the value of $j_{v_0}$, (respectively $j_{v_0}=\ell$, and
$j_{v_0}>\ell$, see above).

What follows below and in Appendix A2 really goes beyond [Ge2],
although it constitutes a natural extension of it. From now
now let us consider the case $\ell=3$ and the Hamiltonian \equ(1.1).

We shall assume first a condition on the rotation vectors stronger than
the Diophantine one, as done in [G1,GG,Ge2], \ie we suppose that they
satisfy a {\it strong Diophantine condition}
$$ \eqalignno{
(1) & \quad C_0 | \oo_0 \cdot \nn| \ge |\nn|^{-\t} \; , \quad\quad
\V0 \neq \nn \in \ZZZ^{2} , \qquad C_0^{-1}=\hdm C(\h) \; ,
&\eq(5.7) \cr
(2) & \quad
\min_{0\ge p\ge n} \Big| C_0 |\oo_0 \cdot \nn| -
2^p \Big| \ge 2^{n+1} \; ,  \quad\hbox{if} \quad n \le 0,
\; \; 0 < |\nn| \le (2^{n+3})^{-1/\t} , \cr} $$
%
where $n, p \in \ZZZ$, $n\le 0$, and
$$ \oo_0\= \h^{-1/2}(\O_1+\hdp J^{-1}A_1)^{-1}\oo'=
\big(1,\h^{-1}(\O_1+\hdp J^{-1}A_1)^{-1}\O_2\big) \; , \Eq(5.8) $$
so that $\oo'\cdot\nn=\h^{1/2}(\O_1+\hdp J^{-1}A_1)\oo_0\cdot\nn$.
We suppose also that $A_1\in[-\hdm R,\hdm R]$, with $R\le J\O_1/2$,
so that $\h^{1/2}(\O_1+\hdp J^{-1}A_1)\ge \h^{1/2}\O_1/2$.

If we write $\oo'=(\hdp\O_1+\h J^{-1}A'_1,\hdm\O_2)$
then the measure of the set of
$A'_1$'s such that $\oo_0$ verifies the strong Diophantine condition
\equ(5.7) has measure of size $O(C_0^{-1}\h^{-3/2})$.

By reasoning as in [GG], once the case of strong Diophantine vectors
has been understood, it can be extended to cover also the case of the
usual (weaker) Diophantine condition (expressed by (1) in \equ(5.7)
above). Alternatively one could follow the approach in [GM] avoiding
completely considering condition (2) in \equ(5.7) and assuming only
the ``usual'' condition (1) in \equ(5.7).  We shall not perform such
an analysis (which can be easily adapted from the quoted papers), and
we shall confine ourselves to the case of strongly Diophantine
vectors.\annota{7}{\rm Basically the argument is the following: the
analysis that we present does not change if $2^p,2^n$ are replaced by
exponentials in another base $q$ (larger than $1$) or even if they are
replaced by $\g(p),\g(n)$, where $\g(p)/q^p\tende{p\to-\io}1$, and if
in the second of \equ(5.7) we substitute $2^p, 2^{n+1}, 2^{n+3}$ by,
respectively, $\g(p), \g(n+1), \g(n+3)$. One then proves a simple
arithmetic lemma (see [GG]), whereby it follows that, if the
first of \equ(5.7) is verified and if $\g(p)$ is suitably chosen,
then the second holds with $\g(p), \g(n+1), \g(n+3)$
replacing $2^p, 2^{n+1}, 2^{n+3}$.}

Keeping in mind that $C_0=\hdm e^{+s \hdm}$ is enormous we shall say that
$$\eqalign{
(1)\ & \  G_v[\oo'\cdot\nn_0(v)] \qquad\hbox{is on scale 1,
if $C_0|\oo_0 \cdot \nn_0(v) | > C_0/4 $, or if $p(v)\ne0$;}\cr
(2)\ & \   G_v[\oo'\cdot\nn_0(v)] \qquad\hbox{is on scale 0,
if $1/2<C_0|\oo_0 \cdot \nn_0(v) | \le C_0/4$;}\cr
(3)\ & \   G_v[\oo'\cdot\nn_0(v)] \qquad\hbox{is on scale $n \le -1$,
if $2^{n-1}<C_0|\oo_0\cdot\nn_0(v)|\le 2^n$.}\cr}\Eq(5.9)$$
\*

\0{\bf 5.6.}
{\cs Remark.} Note that in the above definition of scale the second
and third cases can arise only if $p(v)=0$. The propagators on scale 1
can be bounded by $4^2$ if $p(v)=0$ and by $1$ if $p(v)\ne0$.
Note also that the definition of the scales $n=0$
and $n=1$ is different from [G1,Ge2]: {\it this is an important
modification}, exploited in Appendix A2, useful in order to take
advantage from the existence of different scale times.

\*

\0{\bf 5.7.}
As it is well known, \equ(5.1) cannot usefully be bounded by just taking
the absolute value of each factor and bounding the denominators by using
the Diophantine condition. This is true not only if one wants to get the
improved bounds that we are studying, but also if one, more modestly,
wants to show convergence for $\e$ small enough: this is a problem
usually referred to as a ``{\it small divisors problem}''.

Useful bounds are nevertheless possible, as shown first in similar cases
in [E], because one can collect the contributions from the various trees
into pairwise disjoint (``{\it non overlapping}\/'') classes whose
values add up to a quantity that verifies much better bounds than the
individual elements of the same class. Each class $\FFFF(\th)$ will be
determined by one of its elements $\th$ called a {\it
representative}. This means that there are important cancellations
within each class.

The classes can be constructed by collecting trees which have the same
{\it resonance structures}. The key notion of resonance
is recalled below and the description of the classes will follow it.

\*

\0{\cs Definition}.
{\it A {\it ``cluster\/''} $T$ of scale $n_T$
will be a maximal connected set of branches with scales
$n\ge n_T$ and with at least one branch of scale $n_T$.}

\*

A free node $v$ will be defined to be internal to $T$, $v\in T$, if at
least one of the branches leading to it or coming from it, \ie {\it
pertaining to $v$} (as defined in \S 4), belongs to $T$; a leaf with
highest node $v'$ will be defined to be internal to the cluster $T$ if
$v'\in T$.

A branch $\l_v$ is called {\it external} to $T$ if it does not belong
to $T$ but it pertains to a node $v$ internal to $T$, and it said to
be entering $T$ if the node $v'$ following it is in $T$, exiting from $T$
if $v\in T$, (note that an external branch of $T$ is not any branch
outside $T$). We define the {\it degree} $D_T$ of a cluster $T$ to be the
degree of its exiting branch, and the {\it order} $k_T$ of $T$ to be given
by the same formula as \equ(4.26), with the extra constraint that the
nodes are internal to $T$.

\*

\0{\cs Definition.} {\it A
{\it ``resonance\/''} $V$ will be a cluster with only two external
branches $\l_{v_0}$ and $\l_{v_1}$ carrying the same free momentum,
$\nn_0(v_0)=\nn_0(v_1)$ and with order ``not too high'', \ie
$$k_V<\max\{ N^{-1}2^{-(n_{\l_{v_0}}+3)/\t},(\g N\h)^{-1} \},\qquad
\g\defi\fra{4\O_1}{\O_2} \; . \Eq(5.10)$$
The branch exiting from a resonance will be called a {\it resonant
branch}, and the scale $n_{\l_{v_0}}$ of the two branches entering and
exiting the resonance will be called the resonance-scale. The degree of
the propagator of the exiting branch will be called the degree $D_V$ of
the resonance.}

\*

Even though a node $v$ either with $\d_v=0$, $\k_v\ge 1$ or with
$\d_v=1$, $\nn_v=\V0$ is not a cluster in the above sense (because it
does not consist of branches) we shall nevertheless regard it as a
cluster when there are only one incoming branch and one exiting branch
of equal scale.  {\it Therefore we shall also regard it as a
resonance}, if $\k_v<\max\{N^{-1}2^{-(n_{\l_{v_0}}+3)/\t},(\g
N\h)^{-1}\}$ when $\d_v=0,k_v\ge1$ and if
$1<\max\{N^{-1}2^{-(n_{\l_{v_0}}+3)/\t},(\g N\h)^{-1}\}$ when
$\d_v=1,\nn_v=\V0$; note that the restriction that if $\d_v=0=\k_v$
there are at least two branches entering $v$ implies that no node with
$\d_v=0=\k_v$ can be a resonance.
\*

\0{\cs Definition.}
{\it A resonance will be called {\it ``strong\/''} if $p(v_0)=p(v_1)=0$.}

\*

All resonances on scale $\le 0$ are strong (see Remark 5.8).

\*

\0{\bf 5.8.} It is important to note that a strong resonance of degree
$2$ is necessarily such that {\it also} the degree of the entering
branch {\it must} be $2$. No branch inside it can be of order $1$
and no path can precede $v_0$.
This is so because $D_{\l_{v_0}}=2$
implies $j_{v_0}>\ell$ (see the first of
\equ(5.3)) and $p(v_0)=0$ implies
that $\a_{v_0}=2, k_{v_0}=0, k'_{v_0}=0$ so any path preceding $v_0$
would necessarily imply the contradiction $p(v_0)=1$. Also if
$D_{\l_{v_1}}=1$, $D_{\l_{v_0}}=2$ one must have $j_{v_1}=\ell$ hence
$k_{v_1}=0$ (otherwise $p(v_0)>0$) so that $k'_{v_1}=0$: {\it but}
$\a_{v_1}<2$ and $k'_{v_1}$ must be odd. The cases $D_{\l_{v_0}}=1$,
$D_{\l_{v_1}}=1,2$ are both allowed.

\*

\0{\bf 5.9.}
Given a tree $\th$, let $V$ be a resonance (if there are any) with
entering branch $\l_{v_1}$ of degree $D_{\l_{v_1}}=2$.  Then consider
the family of all trees which can be obtained from $\th$ by detaching
the part of the tree having $\l_{v_1}$ as root branch and reattaching
it to all the remaining nodes {\it internal to $V$ but external to the
resonances contained inside the cluster $V$} (if any); to the just
defined set of trees we add all the trees obtained by reversing
simultaneously the signs of the latter modes of the nodes
(this can be done as the sum of the mode vectors $\nn_w$ of such nodes,
$w\in V$, vanishes).  The set of all the so obtained trees will be
denoted $\FFFF_V(\th)$.

The definition of resonance and the strong Diophantine condition
insures that all the trees so constructed have a well defined value
(\ie no division by zero occurs in evaluating it with the above
rules); see the Remark 5.10, (1), below.

If the entering branch $\l_{v_1}$ of the resonance
has degree $D_{\l_{v_1}}=1$ then also the exiting branch $\l_{v_0}$
has degree $D_{\l_{v_0}}=1$, and we
collect together with the considered tree also the tree which is
obtained from $\th$ through the following operation.  Replace the
resonance $V$ with {\it a single node} $v$ carrying labels $\d_v=0$
and $\k_v=k_V$, if $k_V$ is the order of the resonance.  The set of
all the so obtained trees will be denoted by $\FFFF_V(\th)$: the
definition of the class $\FFFF_V(\th)$ will therefore depend on the
degree of the branch entering $V$.

Then repeat the above operations for all resonances in $\th$.
Thus a class $\FFFF(\th)$
has been constructed and the number of elements of $\FFFF(\th)$ is
bounded by the product $\prod_V 2 \NN_V$ of the numbers $\NN_V$ of
branches in each resonance $V$ {\it which are not} contained inside
inner resonances. The latter product is bounded by
$\exp \sum_V 2\NN_V\le \exp 2m $; the $\FFFF(\th)$ can be obtained
starting from any of its elements (which therefore we shall call
{\it representatives} of the class): this is again a {\it consequence of
the strong Diophantine condition}, see [Ge2].

\*

\0{\bf 5.10.} {\cs Remarks.}
(1) The strong Diophantine condition plays a r\^ole here that should be
stressed.  In fact one checks that because of it the scale of a branch
inside a resonance {\it cannot} change too much, as one considers the
different members of a given family. Not enough to change the sets of
branches that belong to a given resonance and insures that the different
families of trees {\it do not overlap}: for this reason the strong
Diophantine condition leads to a simplification of the analysis (the
simplification in the simpler case of the KAM theory). A simplification
that is however not major (as explained informally in [G1] and as shown
in [GG], see footnote 7 above).

(2) To see how the above difficulty is bypassed by using the
alternative approach of [GM1,GM2], we refer to the conclusive comments
in [GM1], \S 3.

\*

\0{\bf 5.11.} Consider trees with $p(v_0)=0$, if $v_0$ is the
highest node of the tree; then the expression of each tree value
contains a product like \equ(5.1). As mentioned in the introduction {\it
we consider only trees without leaves.}

Since the leaf values factorize with respect the product \equ(5.1),
they can be dealt with separately, and no overlap arises with the
cancellation mechanisms acting on the product
\equ(5.1): so that leaves can be easily taken into
account; see \S A3.3 in Appendix A3 (see also [G1,Ge1,Ge2]).

The counterterms can also be explicitly expanded in terms of tree
values, according to \equ(3.2), which again we can imagine to have no
leaves, (see however the comments in \S A3.3 below).

The cancellation mechanisms described in [Ge1,Ge2] (and recalled in
Appendix A3) lead to the bound (on a given family $\FFFF(\th)$
described above, in \S 5.9), see \equ(5.1), \equ(4.30), \equ(4.23)
$$ \eqalignno{
\left( {1\over \h^{1/2}} \right)^{2m}
 & \Big[ (4N^3\CC')^{m} 2^{4m}e^{2m}
\prod_{n\le0} \big( C_0^{2N_n^2} 2^{-2nN^2_n} \big)
\big( C_0^{N_n^1} 2^{-nN^1_n} \big) \Big] \; \cdot & \eq(5.11)\cr
& \cdot \Big[\prod_{n\le 0}\;\prod_{T,\,n_T=n}
\prod_{i=1}^{m^1_T(n)}\,2^{(n-n_{i}+3)}
\prod_{i=1}^{m^2_T(n)}\,2^{2(n-n_{i}+3)}
\Big] \; , \cr} $$
where

\0$\bullet$ $\CC'=\max\{(2g_0/\O_1)^2,4^2\}\CC$, with $\CC$ the
dimensionless constant defined in \equ(5.6);

\0$\bullet$ $m$ is the number of nodes $v\ge v_0$;

\0$\bullet$ $N^j_n$ is the number of propagators on scale $n$ and of degree
$j$ in $\th$, which can be written as
$$ N^j_n=\bar N^j_n+\sum_{T \atop n_T=n, D_T=j}
(-1) + \sum_{T \atop n_T=n} m^j_T(n) \; , \Eq(5.12) $$
where $m^j_T(n)$ is the number of resonances on scale $n$ and degree
$j$ (\ie with entering branch having a propagator of degree $j$)
contained inside the cluster $T$;

\pallino the terms $\bar N^j_n$, $j=1,2$, which count the number of
propagators {\it which do not correspond to resonant branches} plus
the number of clusters on scale $n$ and of degree $j$ in $\th$,
satisfy the bounds
$$ \sum_{j=1}^2 \bar N^j_n \le 4 m N 2^{(n+3)/\t},\qquad
\sum_{n=-\io}^{0} \sum_{j=1}^2 \bar N^j_n \le
4 m \g N \h \; , \Eq(5.13) $$
(with $\g=4\O_1/\O_2$) which are proven in Appendix A2;

\0$\bullet$ the first square bracket in \equ(5.11) is the bound on
the product of individual elements in the family $\FFFF(\th)$ times the
bound on their number $\prod_V 2\NN_V< e^{2m}$, see above.

\0$\bullet$ the second square bracket term is the part coming from the maximum
principle, (in the form of Schwarz's lemma), applied to bound the
sums of the tree values (``{\it resummations}'') over the classes
$\FFFF(\th)$
introduced above: this is a {\it non trivial product of small factors}
that arise from the
cancellations associated with the resummations, see Appendix A3.  In
\equ(5.11) $n_i$ is the scale of the cluster $V_i$ which is the $i$--th
resonance inside $T$, as in [Ge2];

\pallino the $\eta^{-m/2}$
arises as a lower bound on the small divisors of the form $\oo'\cdot\nn$
on scale $n=1$ (for $n=1$ we use the better bound $|\oo_0\cdot\nn|\ge
2^2\hdp$).
\*

\0{\bf 5.12.}
{\cs Remark.} The first bound \equ(5.13) holds for all $n$ and for all
Hamiltonians of the form \equ(2.1). On the contrary the second bound in
\equ(5.13) will follow from the fact that the rotation vector $\oo_0$
has the form \equ(5.8), with $\h$ small, and will be used to control
the (huge) factors $C_0$ in \equ(5.11).

\*

\0{\bf 5.13.}
Hence by substituting \equ(5.12) and the first of \equ(5.13) into
\equ(5.11) we see that, for $j=1,2$, the $m^j_T(n)$ is taken away by
the first factor in $\,2^{jn} 2^{-jn_{i}}$, while the remaining
$\,2^{-jn_i}$ are compensated by the $-1$ before the $+m^j_T(n)$ in
\equ(5.11) taken from the factors with $T=V_i$ (note that there are
always enough $-1$'s), and therefore
\equ(5.11) is bounded by
$$ \eqalign{
& \left( {2\over \h^{1/2} } \right)^{2m} (4N^3\CC')^{m}
e^{m} 2^{4m}2^{8m} C_0^{8m\g N\h}
\prod_{n=-\io}^{0} 2^{-8 m N n 2^{(n+3)/\t} } \; , \cr} \Eq(5.14) $$
because the product of the factors $C_0$ in \equ(5.11) can be bounded by
using the second of \equ(5.13), since the product does not contain the
$n=1$ factor).  The last product in \equ(5.14) is bounded by
%
$$ \prod_{n=-\io}^{0} 2^{-8mNn 2^{(n+3)/\t}} \le \exp \Big[ 8mN2^{3/\t}
\log2 \sum_{p=1}^{\io} p 2^{-p/\t} \Big] \; , \Eq(5.15) $$
hence, by adding the remark that the perturbation degree $k$ and the
number of tree nodes $m$ are related by $m<2k$, a bound on the
sum over all the subtrees of order $k$ with $p(v_0)=0$, $\nn(v_0)=\nn$
(recalling that the number of trees with $m$ nodes is $<4^m m!$) is
$$ \D_k\defi \Big| \fra1{|\FFFF(\th)|}
\sum_{\th'\in\FFFF(\th)} \prod_{v\in\th'}
\bar F_{\nvec_v} G_v[\oo'\cdot\nn_0(v)]\,\tilde y_{v}\Big|
\le B_0^{2k} \h^{-2k}\; , \Eq(5.16)$$
for some positive constant $B_0$.
The normalization constant $|\FFFF(\th)|$ is
introduced in order to avoid overcountings: in fact if
$\th'\in\FFFF(\th)$ then $\th\in\FFFF(\th')$, so that,
without dividing by $|\FFF(\th)|$ in \equ(5.16), each tree would
be counted $|\FFFF(\th)|$ times.

If $C_0^{-1}=\hdm C(\h)$ is chosen
as in the statement of Theorem 1.4, an explicit calculation gives
the bound on \equ(5.11) of the form $(\hdm)^{4k} B_0^k$, $k\ge1$, and
$$ B_0 = 2^{18}(4N^3\CC') \, \exp \Big[  2+4\g N\h\log\h+ 8s\g N\hdp
+ 8N 2^{3/\t}\log 2 \sum_{p=1}^{\io} p 2^{-p/\t} \Big] \; , \Eq(5.17) $$
which is bounded uniformly in $\h$ (for $\h\le1$).

\*

\0{\bf 5.14.} In the previous section trees with $p(v_0)=0$
have been considered; in particular only the contributions
\equ(5.1) arising from the value \equ(4.30), once the
corresponding tree has been deprived of leaves and counterterms,
have been bounded and the bound \equ(5.16) has been obtained
through a suitable resummation operation.
In such a case the sum over the labels
$(k_v',k_v)$ is trivial because the condition $p(v_0)=0$ imposes that
only a few values (up to three per node) can be assumed by the
hyperbolic mode labels; also the sum over the mode labels $\nvec_v$
cannot create any problems. In fact for any node $v$ one has
$|\nvec_v|\le N$ and $|n_v|\le N_0$ (see the eighth item in \S 4.1).

The cases $p(v_0)\ne0$ as well as those involving graphs containing
leaves or counterterms can be treated in the same manner as already
done in [G1,Ge2]. We provide, in Appendix A4, a quick description
of the construction of the analyticity bound $\e_0=D^{-1}$ with
$$e_0^{-1}=D=\left[ B 2^6\ell (2N+1)^{2\ell-1}(2N_0+1) \right]^2 \; ,
\qquad B=\max(B_0\h^{-1},B_1)\Eq(5.18)$$
and $B_1$ is a suitable numerical constant.

The part of Theorem 1.4 not concerning the connection between the
average action $\AA'$ and the rotation vector $\oo'$ nor the splitting
size follows.

\*

\0{\bf 5.15.}
Determining the exact splitting size (\ie the leading behavior
asymptotically as $\h\to0$ with $\e< B\h^2$) is {\it not} trivial
because of the existence of major cancellations in the evaluation of
the determinant of the splitting matrix; however the analysis in
[GGM2] dealt with this question in detail: in the latter paper
remarkable cancellations are exhibited and an exact formula for the
splitting angles is derived (see (7.19) of [GGM2]).

One gets the results in the last item of Theorem 1.4 simply if
[GGM2] and the first part of Theorem 1.4 (to estimate the
remainders) are used: then the claimed bounds on the splitting follow
immediately (see Remark 1.5). In [GGM3] an improvement of lemma 1 and
lemma 1' of [CG] was used instead to control the density of
tori in phase space (the lemmata in [CG] were,
as such, useless already in the case in [GGM2]
because they would require that $\e$ be far smaller than the $\e_0$ of
Theorem 1.4); see [GGM3], where this is discussed in detail and differs
from our case only because it relied on a theorem weaker than
Theorem 1.4 above (as the radius of convergence estimate there is
proportional to $\h$ to the power $\fra92+$ rather than our $2$).

\*

\0{\bf 5.16.} {\cs Remarks.} (1)
The bound \equ(5.16) and the discussion in \S 5.14
imply the convergence of the perturbative
expansions for the parametric equations of the invariant tori (for the
Hamiltonian \equ(1.1)), if $|\e|<\e_0=O(\h^2)$. This bound on the
convergence radius should be compared with the value given by
[GGM3], which, for $\NN=O(\hdm)$, gives $\e_0=O(\h^{\fra92}/
\log^2\h^{-1})$. {\it As usual the Lindstedt series gives a much
better estimate than the classical method} (\ie an exponent $2$ versus
$\sim 4.5$). We do not see immediately how to improve substantially the
classical estimate without important changes in the architecture of the
proof of [GGM3], although this should be possible; on the other hand,
from the above analysis, $\e_0=O(\h^2)$ might be close to an optimal
result. If so it should be no surprise that our analysis is so delicate.

(2) In the Hamiltonian \equ(1.1),\equ(2.1) the polynomial dependence
of the interaction on the rotators angles has very likely a purely
technical motivation (as it simplifies the analysis) and could
probably be relaxed into a more general analytical dependence, as in
[BCG]. On the contrary the hypothesis that the perturbation is a
trigonometric polynomial of degree $N_0$ in $\f$ is fundamental to
get the correct asymptotic behavior, in order to apply the results in
[GGM2], where the dominance of Mel'nikov integral is proven {\it
provided the perturbation is polynomially small in a power of
$\h^{N_0}$} (so that the results of [GGM2] become meaningless for
$N_0\to\io$).

(3) A bound of the form \equ(5.16)
holds under the weaker condition that $C(\h)\le e^{-s\h^{-a}}$, with
$a\le 1$ (see \equ(5.17)).

(4) If $q$ is defined as in \S1.6 so that $|\e| C(\h)^{q\h}< 1$ implies
analyticity in $\e$, the above analysis gives that $q$ can be taken
$q=8\g N$.

In general all the bounds found so far are not uniform in $N$; in order
to deal with the analytical case in the frame of the exploited formalism
one should bound the small divisors by using the results of [GM2] or
Eliasson-Siegel's bound (see for instance [BGGM]), and use explicitly as
in [BCG] the exponential decay in $\nn$ of the Fourier coefficients
$f^1_{\nvec}$.

(5) Note that we have convergence for $|\e|<O(\h^2)$, while the
asymptoticity of the splitting estimate follows only for $|\e|<O(\h^{\z})$
(with $\z=2(N_0+3)$), which {\it much smaller} for $\h\to0$. Then the
question is: what will be the asymptotics for $\e$ small enough to be
in the convergence domain but too large to be in the domain of the
asymptotic result?  There is some evidence that there is a critical
value $T_c$ such that, if $\e=\h^T$, then for $T> T_c$ the asymptotic
formula that we can prove only for $T> \z$ holds, but for
$T=T_c$ it is modified remaining qualitatively of the same size
$O(e^{-\fra12\hdm})$ and for $T<T_c$ it becomes qualitatively
different. The analogy with the critical point scaling phenomena seems
to be substantial and, keeping in mind that the above theory can be
interpreted as a field theory, see [GGM1], one would say that the
region $T>T_c$ is described by a trivial fixed point; a non trivial
fixed point describes the case $T=T_c$ and another ``low temperature''
fixed point describes the cases $T<T_c$.  Evidence in this direction
comes also from the theory of the standard map and its developments,
[La,Gel2,Gel1].

\vskip1.truecm
\0{\titolo Acknowledgments.}
\*

\0{\it This work is part of the research program of the European Network
on ``Stability and Universality in Classical Mechanics", \#
ERBCHRXCT940460. Supported in part by CNR-GNFM and Rutgers
University.}

\vskip1.truecm
\0{\titolo Appendix A1. Counterterms}
\numsec=1\numfor=1\*

\0{\bf A1.1}
To order $h$ one can write, by using \equ(2.12), \equ(2.25) and \equ(4.11),
$$ \eqalignno{
\X_-^{h\s}(t) & = \igb_{\s\io}^t\, d\t\,
w_{0}^1(t,\t)
\left( \Phi_+^{h\s}(\t;\g_1(g_0),\ldots,\g_{h-1}(g_0)) +
\g_h(g_0) \sin \f^0(\t) \right) &\eqa(A1.1) \cr
& + \igb_{\s\io}^0 d\t \, w_{0}^0(t,\t)
\left( \Phi_+^{h\s}(\t;\g_1(g_0),\ldots,\g_{h-1}(g_0)) +
\g_h(g_0) \sin \f^0(\t) \right) \; , \cr} $$
where $\Phi_+^{h\s}(t;\g_1(g_0),\ldots,\g_{h-1}(g_0))$
takes into account all the contributions to
$\Phi_+^{h\s}=(J_0g_0^2)^{-1} F^{h\s}_+$
except the only one explicitly depending on $\g_h(g_0)$, which is given
by $\g_h(g_0)\sin\f^0(\t)$.

We shall impose, recursively, that the contributions to
the integrands in \equ(A1.1) arising from amputated trees
(see comments after \equ(4.31)) with $(\nn_0(v_0),p(v_0))=(\V0,0)$,
{\it without leaves and without end nodes bearing a counterterm label}
(see Remark A1.2 below),
compensate exactly the contributions due to the trees with a single
node representing a counterterm of order $h$ (\ie the terms
with $\g_h(g_0)$ in \equ(A1.1)). The first described
type of contributions can be written
$$ \WW_1 = w_{\ell}^1(t,\t) \,
\Phi_+^{h\s}(\t;\g_1(g_0),\ldots,\g_{h-1}(g_0)) \=
\sum_{\nn\in\zzz^{\ell-1}}
\sum_{p=-1}^{\io} \tilde \WW_1(\nn,p)\,
e^{i\oo'\cdot\nn t-pg_0\s t} \Eqa(A1.2) $$
in the first integral in \equ(A1.1) and
$$ \WW_0 = w_{\ell}^0 (t,\t) \,
\Phi_+^{h\s}(\t;\g_1(g_0),\ldots,\g_{h-1}(g_0)) \=
\sum_{p=-1}^{\io} \sum_{\nn\in\zzz^{\ell-1}} \tilde \WW_0 (\nn,p)\,
e^{i\oo'\cdot\nn t-pg_0\s t} \Eqa(A1.3) $$
in the second one. Imposing that such contributions are canceled by
the contributions with $p=0$ arising from
$w_{\ell}^1(t,\t)\g_h(g_0)\sin\f^0(\t)$ and
$w_{\ell}^0(t,\t)\g_h(g_0)\sin\f^0(\t)$ gives our prescription
on how to fix $\g_h(g_0)$.

Since two integrals are involved (one for $\r_{v_0}=1$ and one for
$\r_{v_0}=0$), the first time dependent and the second
time independent, two conditions may seem to be required: however note
that $\sin \f^0(t)=2\sinh g_0 t/\cosh^2 g_0 t $ is expanded in odd
powers of $x=e^{-\s g_0 t}$, hence the only terms appearing in
\equ(A1.2) and \equ(A1.3) which can contribute to $p=0$ are, in both
cases, those involving $y^{(-1)}(k_{v_0}',k_{v_0})$, with
$k_{v_0}=-1$.  This means that the contributions with $p=0$ arising
from \equ(A1.2) and \equ(A1.3) are equal, so that no compatibility
problem arises.

Note that an expression analogous to \equ(A1.1)
is obtained also for $\X_+^{h\s}(t)$; however the terms
with $p=0$ have the same form as in \equ(A1.1) (see \equ(4.17) for
$j_{\l_{v_0}}=\ell$ and for $j_{\l_{v_0}}=0$), so that if no term with
$(\nn_0(v_0),p(v_0)=(\V0,0))$ contributes to $\X_-^{h\s}(t)$, then also
no term with $(\nn_0(v_0),p(v_0))=(\V0,0)$ contributes
to $\X_+^{h\s}(t)$.

\*

\0{\bf A1.2.} {\cs Remark.}
It may seem strange that we exclude from the definition of the
counterterms trees with leaves: in fact one can imagine to realize
trees with $(\nn(v_0),p(v_0))=(\V0,0)$ also by trees which contain leaves
with a stalk bearing a label $j_w>\ell$ or $j_w=\ell$ and internal
momenta $(\nn',p')$. Such terms would give rise to $\aa$--dependent
counterterms which of course are not allowed: however it turns out
that the sum over all contributions to tree values of trees with
$(\nn(v_0),p(v_0))=(\V0,0)$ from such trees cancel {\it exactly}:
this is explained, together with the other cancellations built in
our algorithm, in Appendix A3 (see \S A3.5 in particular).

\*

\0{\bf A1.3.}
Let us consider the first integral in \equ(A1.1).  Corresponding to the
node $v_0$ of each tree whose value contributes to $\X_-^{h\s}(t)$ there
is a coefficient $\tilde y_{n_{v_0}}(k_{v_0}',k_{v_0})$, see \equ(4.14),
\equ(4.17).  Then from \equ(A1.1) and the just formulated
condition to impose we obtain
$$ \sum_{\th\in\TT_{\V0,h} , \a_{v_0}=-1 \atop p(v_0)=0 ,k_{v_0}'=1}
\overline{{\rm Val}}(\th) + \g_h(g_0) \left.
w_{\ell}^1(t,\t)\,
\sin\f_0(\t) \right|_{k'=1,p=0} = 0 \; , \Eqa(A1.4) $$
where the sum is over the set $\TT_{\V0,h}$ of
all trees of order $h$ and momentum $\nn(v_0)=\V0$,
with $\nn_0(v_0)=\V0$ (see Remark A1.2),
$p(v_0)=0$, $j_{v_0}=\ell$ and $k_{v_0}'=1$; hence if $p(v_0)=0,
j_{v_0}=\ell$, one must have $k_{v_0}=-1$, hence $\a_{v_0}=-1$ and
$k'_{v_0}=1$ which is a possible case indeed.

A trivial calculation (just take into account that
$y_{n_v}^{(-1)}(1,-1)$ $=$ $\s/2$ and
$\sin\f^0(\t)=4\s x + O(x^3)$) gives
$$ \left. w_{\ell}^1(t,\t)\,
\sin\f_0(\t) \right|_{k'=1,p=0} = 2 \; , \Eqa(A1.5) $$
so that \equ(4.32) follows; the above follows [Ge2], page 287.

\*


\vskip1.truecm
\0{\titolo Appendix A2. (Improved) resonant Siegel-Bryuno's bound}
\numsec=2\numfor=1\*

\0{\bf A2.1.} We follow the idea of P\"oschel, [P\"o] (see also [G1,
GG,Ge2]).  In the discussion, we focus on the scale labels, so
that it is quite irrelevant which value the $p(v)$'s, $v \in \th$,
assume, and therefore which resonances are strong and which are not.

Calling $N^*_n(\th)$ the number of non resonant branches
carrying a scale label $\le n$, in a tree $\th$ with $m$ nodes,
we shall prove first that
$$ N^*_n(\th) \le 2m E_n - 1 \; , \qquad
E_n \defi N2^{(3+n)/\t},\qquad n\le1\; , \Eqa(A2.1) $$
provided that $N^*_n(\th)>0$, and
$$ N^*_{0}(\th) \le 2m \g N \h - 1 \; , \qquad
\g \defi 4\O_1/\O_2 \; , \Eqa(A2.2) $$
if $N_{0}^*(\th)>0$.

Define, as in \S 5.7, $\oo_0=(1,\h^{-1}(\O_1+\hdp J^{-1}A_1)^{-1}\O_2)$.
Then $C_0|\oo_0\cdot\nn|>|\nn|^{-\t}$ for all
$\V0\neq\nn\in\ZZZ^{\ell-1}$; see \equ(5.7).
Assume also $\h$ so small that $C_0\ge 2$,
(this is not restrictive as we are interested in $\h\to 0$).

Set $E_n\= N2^{(n+3)/\t}$ as in \equ(A2.1). Note that if $m\le
E_n^{-1}$ one has $N_n^*(\th)=0$. In fact $m\le E_n^{-1}$ implies
that, for all $v\in\th$, $|\nn_0(v)| \le N E_n^{-1}$, \ie
$C_0|\oo_0\cdot\nn_0(v)|\ge (N^{-1}E_n)^{\t}$ $=$ $2^{n+3}$, so that
there are {\rm no} clusters $T$ with $n_T=n$.  Note also that if
$m\le (\g N \h)^{-1}$, with $\g=4\O_1/\O_2$, then $N_{0}^*=0$, as
$|\oo_0\cdot\nn_0(v)|\ge 1$ for all $v\in\th$ in such a case.

\*

\0{\bf A2.2.} Let us prove first the inequality \equ(A2.1).
If $\th$ has the root branch either with scale $>n$,
or with scale $\le n$ and resonant,
then calling $\th_1,\th_2,\ldots,\th_k$
the subtrees of $\th$ ending into the highest node $v_0$ of $\th$
and with $m_j>E_n^{-1}$ nodes, $j=1,\ldots,k$,
one has $N_n^*(\th)=N_n^*(\th_1)+\ldots+N_n^*(\th_k)$ and the statement is
inductively implied from its validity for $m'<m$ and from
the just proved fact that $N_n^*(\th)=0$ if $m\le E_n^{-1}$.

If the root branch is on scale $\le n$ and non resonant,
one has $N^*_n(\th)\le 1+\sum_{i=1}^k N_n^*(\th_i)$:
if $k=0$ the statement is trivial, if $k\ge2$ the statement is again
inductively implied by its validity for $m'<m$.
If $k=1$ one has $N^*_n(\th)\le 1+2 m_1 E_n-1$, hence
we once more have a trivial case unless the order $m_1$ of
$\th_1$ is $m_1 > m- (2E_n)^{-1}$: but in the
latter case we shall show that the root branch of $\th_1$ has scale $>n$.

Accepting the last statement (which will be proved below), one will
obtain $N_n^*(\th)=1+N_n^*(\th_1)= 1+N_n^*(\th'_1)+\ldots+N_n^*
(\th'_{k'})$, where $\th'_j$'s are the $k'$ subtrees ending into the
highest node of $\th'_1$ with orders $m'_j>E_n^{-1}$,
$j=1,\ldots,k'$. Going once more through the analysis the only non trivial
case is if $k'=1$ with the root branch of $\th'_1$ non resonant;
and in such case
$N_n^*(\th'_1)=N_n^*(\th^{\prime \prime}_1) + \ldots +
N_n(\th^{\prime \prime}_{k^{\prime \prime}})$, \etc., until we reach a
trivial case or a tree of order $\le m-(2E_n)^{-1}$.

It remains to check that if $m-m_1<(2E_n)^{-1}$ then the root branch
of $\th_1$ has scale $>n$. Let us proceed by {\it reductio ad absurdum}.
Suppose that the root branch of $\th_1$ is on scale $\le n$. Then
$C_0|\oo_0\cdot\nn_0(v_0)|\le\,2^n$ and $C_0|\oo_0\cdot\nn_0(v_1)|\le
\,2^n$, if $v_1$ is the highest node of $\th_1$.
Hence $C_0|\oo_0\cdot(\nn_0(v)-\nn_0(v_1))|< 2^{n+1}$ (equality
would imply violation of the strong Diophantine property, \equ(5.7)), and
the Diophantine condition implies that
$$ |\nn_0(v_0)-\nn_0(v_1)|> 2^{-(n+1)/\t} \= \d \; , \Eqa(A2.3) $$
because $\nn_0(v_0)\neq\nn_0(v_1)$ (the root branch of $\th$ being
supposed non resonant).
But $m-m_1<(2E_n)^{-1}$, so that $|\nn_0(v_0)-\nn_0(v_1)|<
(2E_n)^{-1}N < 2^{-1}2^{-(n+3)/\t}$ $=$ $2^{-(1+2/\t)}\d<\d$,
which contradicts inequality \equ(A2.3).

\*

\0{\bf A2.3.} Let us prove now \equ(A2.2).
If $\th$ has the root branch either with scale $1$,
or with scale $\le 0$ and resonant,
then calling $\th_1,\th_2,\ldots,\th_k$
the subtrees of $\th$ ending into the highest node $v_0$ of $\th$
and with $m_j>(\g N\h)^{-1}$ nodes, $j=1,\ldots,k$,
one has $N_{0}^*(\th)=N_{0}^*(\th_1)+\ldots+N_{0}^*(\th_k)$
and the statement is inductively implied from its validity for $m'<m$
and from the fact that $N_{0}^*(\th)=0$ if $m\le (\g N \h)^{-1}$.

If the root branch is on scale $\le 0$ and non resonant,
one has $N^*_{0}(\th)\le 1+\sum_{i=1}^k N_{0}^*(\th_i)$:
if $k=0$ the statement is trivial, if $k\ge2$ the statement is again
inductively implied by its validity for any $m'<m$.
If $k=1$ we once more have a trivial case unless the order $m_1$ of
$\th_1$ is $m_1>m- (2\g N\h)^{-1}$: but in the
latter case the root branch of $\th_1$ has scale $1$.

Accepting the last statement (which will be proved below),
one will obtain $N_{0}^*(\th)=1+
N_{0}^*(\th_1)= 1+N_{0}^*(\th'_1)+\ldots+N_{0}^*(\th'_{k'})$,
where $\th'_j$'s are the $k'$ subtrees ending into the highest
node of $\th'_1$ with orders $m'_j>(2\g N\h)^{-1}$.
Going once more through the analysis the only non trivial
case is if $k'=1$ and in that case
$N_{0}^*(\th'_1)=N_{0}^*(\th^{\prime \prime}_1) + \ldots +
N_{0}(\th^{\prime \prime}_{k^{\prime \prime}})$, \etc., until
we reach a trivial case or a tree of order $\le m-(2\g N\h)^{-1}$.

It remains to check that, if $m-m_1<(2\g N\h)^{-1}$,
then the root branch of $\th_1$ has scale $1$.
Suppose that the root branch of $\th_1$ is on scale $\le 0$. Then
$p(v_1)\neq0$ and $|\oo_0\cdot\nn_0(v_0)|\le1/4$,
$|\oo_0\cdot\nn_0(v_1)|
\le 1/4$, if $v_1$ is the highest node of $\th_1$, \ie
$$ |\oo_0\cdot(\nn_0(v_0)-\nn_0(v_1))| \le 1/2 \; . \Eqa(A2.4) $$
As the root branch of $\th$ is supposed non resonant, then $m-m_1<(2\g
N\h)^{-1}$ implies that $0$ $<$ $|\nn_0(v_0)-\nn_0(v_1)|$ $<$ $(2\g
N\h)^{-1}N = (2\g\h)^{-1}$, so that one would have
$|\oo_0\cdot(\nn(v_0)-\nn(v_1))|\ge 1$, which is contradictory with the
inequality \equ(A2.4).
%

\*

\0{\bf A2.4.} A similar induction can be used to prove that if
the number of branches on scale $n$ is $N_n(\th)>0$ then
the number $p_n(\th)$ of clusters of scale $n$ verifies the bound
$$ p_n(\th) \le 2m N 2^{(n+3)/\t}-1 \; . \Eqa(A2.5) $$
In fact this is true for $m\le E_n^{-1}$, if
$E_n$ is defined as in \S A2.1. Otherwise,
if the highest tree node $v_0$ is not in a cluster on scale $n$,
one calls $\th_1,\ldots,\th_k$ the subtrees ending into $v_0$, and
one has $p_n(\th)=p_n(\th_1)+\ldots+p_n(\th_k)$,
so that the statement follows by induction.
If $v_0$ is in a cluster $V$ of scale $n$, and $\th_1$, $\ldots$, $\th_k$
are the subtrees entering the cluster containing $v_0$ and with
orders $m_j> E_n^{-1}$, one will find
$p_n(\th)=1+p_n(\th_1)+\ldots+p_n(\th_k)$.
Again we can assume that $k=1$, the other cases being trivial.
But in such case there will be only one branch entering the cluster $V$
and it will have a propagator of
scale $\le n-1$. Therefore the cluster $V$ must contain at least
$E_n^{-1}$ nodes. This means that $m_1\le m-(2E_n)^{-1}$.

Finally, the bound
$$ \sum_{n=-\io}^0 p_n(\th) \le 2m\g N\h-1 \Eqa(A2.6) $$
is a trivial consequence of \equ(A2.2).

\*

\0{\bf A2.5.} Let $\bar N^*_n\le N^*_n$ be the number of non
resonant branches on scale $n$.
Then if $\bar N_n$ is the number of non resonant branches {\it plus}
the number of clusters on scale $n$, $\bar N^*_n$ verifies the bounds
$$ \bar N_n^* = \big( \bar N_n^* + p_n \big) - p_n \= \bar N_n-p_n
\le 4m N 2^{(n+3)/\t}-2- \sum_{T \atop n_T=n} (1)
\le 4m N 2^{(n+3)/\t}+\sum_{T\atop n_T=n} (-1) \; . \Eqa(A2.7) $$
This proves that \equ(A2.1) and \equ(A2.5) imply
an inequality analogous to the first of \equ(5.13);
likewise one derives an inequality similar to the second of \equ(5.13)
by combining \equ(A2.2) and \equ(A2.6).

\vskip1.truecm
\0{\titolo Appendix A3. Cancellations between resonances}
\numsec=3\numfor=1\*

\0In this appendix we recall briefly the cancellation mechanisms of
[Ge2]. We provide this as a guide to the reader and as a tune up of a
fine points of the analysis of [Ge2] (the analysis in A3.2 is given
here in full details while in [Ge2] it was left out).

\*

\0{\bf A3.1.}
Consider a tree $\th$ with a strong resonance $V$ of order $k_V$.
Let $\l_{v_0}$ and $\l_{v_1}$ be, respectively, the exiting
and entering branches of $V$.
There are two possibilities: either the degree of the propagator
corresponding to the branch exiting from $V$ is $D_{\l_{v_0}}=2$
or it is $D_{\l_{v_0}}=1$ (equivalently the degree of the
resonance is either $D_V=2$ or $D_V=1$).

Let us discuss first the case in which the degree $D_V$
of the resonance is $D_V=2$. Then $j_{v_0}>\ell$ (see \equ(5.3)
and the comments after the definition of strong resonance in \S 5.9)
and, by following the notations of \S 5.1, we shall say that
$\PP=\emptyset$, \ie there is no path $\PP$ ending into $v_0$.
It follows, from the properties of $\PP$ discussed at the beginning of
\S 5.1 above, that $p(v_1)=0$ implies $j\=j_{v_1}>\ell$
and $D_{\l_{v_1}}=2$ (see again \equ(5.3)).

Consider all the trees belonging to the class $\FFFF(\th)$ which
are obtained from $\th$ by detaching the subtree
having as branch root the entering branch $\l_{v_1}$ of the resonance
and attaching it to all the remaining nodes of $V$
(see the definition of the class $\FFFF_V(\th)$ in \S 5.9).
As a consequence of such an operation\\
$\bullet$ some of the branches
internal to the resonance have changed the
free momentum
by an amount $\nn_0(v_1)$, and
\\
$\bullet$ if $w$ is the node inside $V$ to which
the branch $\l_{v_1}$ is attached and $j_{v_1}-\ell>0$, then
$\bar F_{\nvec_w}$ (see \equ(4.23)),
has the form of an even function of $\nvec_w$ times
a factor $(i\n_{wj})$.

We shall call {\it resonance value} $\RR_V$
the product of factors appearing in
the definition of tree value and relative only to the nodes and branches
internal to the resonance $V$:
$$ \RR_V = \bar F_{\nvec_{v_0}} y_{v_0}'
\prod_{v\in V \atop v <v_0} \bar F_{\nvec_v} y_{v}'
G_v[\oo'\cdot\nn_0(v)] \; , \Eqa(A3.1) $$
and shall consider the resonance value as a function of the
quantity $\m\=\oo_0\cdot\nn_0(v_1)$.

Then for $\m=0$ a quantity proportional to
$\sum_{w\in V}\nn_{wj}$ is constructed, but such a quantity
is vanishing by definition of resonance, as $j=j_{v_1}-\ell>0$.

If we sum also on an overall change of signs of the mode labels of the
nodes internal to the resonance (by following the definition of the
class $\FFFF(\th)$ given in \S 5.9), we obtain a zero contribution also to
first order in $\m$ (here the even parity of the perturbation $f$ is
essential, see [G1,Ge2]).

This can be seen by using the explicit form of the functions in
\equ(4.21), \ie the coefficients listed in \equ(4.24).
Noting that in the present case {\it there cannot be any $\PP$ inside
$V$} the only propagators we can associate with the branches internal to
$V$ have the form of the two first terms of \equ(5.3), so that, for
$\m=0$, {\it they are even functions of the mode labels}.  Moreover in
such a case the analysis in \S 5.1 shows that $\a_v=-1$, $\a_v=1$ and
$\a_v=2$ imply, respectively, $k_v'=-k_v=1$, $k_v'=-k_v=-1$ and
$k_v'=-k_v=0$ (the case $\a_v=0$ is not possible here): then no $n_v$
labels appear in the coefficients $y_{n_v}^{(\a_v)}(k_v',k_v)$
corresponding to the nodes $v\le v_0$ (see the list of coefficients in
\equ(4.24)).  Therefore all the dependence on the $n_v$ labels is
through the factors $\bar F_{\nvec_v}$ in \equ(4.23). This yields that
there is an even number of the $n_v$ (if there are any) corresponding to
the nodes $v\in V$: two for each branch $\l_v$ with $j_v=\ell$, by
taking into account that $j_{v_0},j_{v_1}>\ell$, so that no change is
produced by the sign reversal (since, by the parity properties of the
Hamiltonians \equ(1.1) and \equ(2.1), one has also
$f^{\d_v}_{\nvec_v}=f^{\d_v}_{-\nvec_v}$). This means that the resonance
value is an even function of $\m$.

\*

\0{\bf A3.2.}
Let us now consider the case in which the strong resonance is of degree
$D_V=1$ and the tree $\th$ has no leaves inside $V$. In such a case
$\a_{v_0}=-1$ and $j_{v_0}=\ell$, hence $D_{\l_{v_0}}=1$ (see
\equ(5.3)): then a first order zero in $\m$ will be enough. Moreover
there is a $\PP$ inside the resonance: we shall distinguish between
the cases $v_1\notin\PP$ and $v_1\in\PP$.

Let us consider first the case $v_1\notin\PP$ (in particular this is
the case when $\PP=v_0$, $k_{v_0}=0$, provided $k_V\ge 2$). In such a
case $j_{v_1}>\ell$ and we can reason as above to obtain a first order
zero.  Note that in such a case there would be no cancellations
between tree values of trees obtained by the sign reversal operation.

On the contrary, if $v_1\in\PP$, then $k_{v_0}=-1$,
and one has also $\a_{v_1}=-1$ and $j_{v_1}=\ell$.
In this case consider together with
the tree $\th$ also the tree $\th'$ obtained from
$\th$ by performing the following operation
(recall the definition of $\FFFF_V(\th)$):
replace the resonance $V$ with a single node $v$
carrying labels $\d_v=0$ and $\k_v=k_V$
(if $k_V$ is the order of the resonance $V$),
then express the counterterm $\g_{\k_v}(g_0)$ associated with
the node $v$ in terms of trees.
If $\th_1$ is the subtree having $\l_{v_1}$ as root branch,
then the values of the two considered trees
$\th$ and $\th'$ can be written, respectively,
as ${\rm Val}(\th)=A(\th)\RR_V{\rm Val}(\th_1)$
and ${\rm Val}(\th')=A(\th)[\g_{k_V}(g_0)\s/2]{\rm Val}(\th_1)$,
where $\s/2=y_{v}^{(-1)}(1,-1)$ and $A(\th)$ takes into account the
factors corresponding to all nodes {\it not} preceding $v_0$,
and has the same value for both $\th$ and $\th'$.

The resonance value $\RR_V$, for $\m=0$, can be written as
$$ \RR_V = \overline{{\rm Val}}(\th_0)\, in_{v_1'} \; , \qquad
\hbox{ for some } \th_0\in\TT_{\V0,k} \hbox{ with } p(v_0)=0
\hbox{, } k_{v_0}=-1 \; , \Eqa(A3.2)$$
see the definitions \equ(4.30) and \equ(4.31)
of tree value and the definition \equ(A3.1) of resonance value:
remember that we are considering resonances $V$ with degree
$D_V=1$, so that $k_{v_0}=-1$ and, as a consequence, $k_{v_0}'\ge 1$;
see \equ(4.14). The counterterm $\g_\k(g_0)$ can be represented
in terms of trees as in \equ(4.32); note that, if
the tree contributing to $\g_\k(g_0)$ has $k_{v_0}=-1$,
the condition $\a_{v_0}=-1$ implies that such a tree
has a node $w>v_0$ with $k_w+k_w'=1$,
while all the other nodes $v\neq w$ have $k_v+k_v'=0$.

Among the contributions in \equ(4.32) to $\g_{k_V}(g_0)$
there will be a quantity $\overline{{\rm Val}}(\th_2)$, where $\th_2$
will have the same topological form of $\th_0$ in \equ(A3.2)
with the node $w$ such that $k_w+k_w'=1$ corresponding
to the node $v_1'\in V$; then we denote both nodes by $w$.

Then $\overline{{\rm Val}}(\th_0)$ will be related to
$\overline{{\rm Val}}(\th_2)$ by
$$ \overline{{\rm Val}}(\th_2) = -\left[
{y_{n_v}^{(\a_v)}(k_w',k_w)|_{k_w'+k_w=1}\over
y_{n_v}^{(\a_v)}(k_w',k_w)|_{k_w'+k_w=0}} \right]
\overline{{\rm Val}}(\th_0) \; , \Eqa(A3.3) $$
so that a look at the coefficients listed in and after \equ(4.24) shows
that the factor in square brackets in \equ(A3.3) (when it is not
vanishing) is equal to $4in_w\s$.  The quantity $\overline{{\rm
Val}}(\th_2)$, in order to contribute to $\g_{k_V}(g_0)\s/2$, has to be
multiplied by a factor $-4\s$ extra with respect to $\overline{{\rm
Val}}(\th_0)$, which, on the other hand, has to be multiplied by $in_w$
in order to contribute to the resonance value $\RR_V$ (see \equ(4.32).

Then, for $\m=0$, by summing the values
of the two considered contributions one obtain
$$ A(\th) \Big[ \overline{{\rm Val}}(\th_0)\,in_w-{1\over4\s}
\overline{{\rm Val}}(\th_2)\Big]{\rm Val}(\th_1) \; , \Eqa(A3.4) $$
which is zero by \equ(A3.3), so that a first order zero is obtained.

\*

\0{\bf A3.3.} If there are leaves, nothing changes in the discussion
of \S A3.1, as $k_{v_0}=0$ implies that only leaves $w$ with
$j_w>\ell$ are possible, and $\x_w(k_{w}',0)\=1$ in such a case
(see \equ(4.18)).

In \S A3.2, when discussing the case $v_1\in\PP$, one has to take care of
the case in which there is a leaf with highest node
$\tilde w$ with $k_{\tilde w}'=1$
(such a leaf will be at the end of the path $\PP$).
In fact the resonances having as entering branch a branch of the
path $\PP$ cannot have any leaves with $k_w'=1$, while
when considering the graphical representation for $\g_\k(g_0)$, there
will be also contributions arising from trees containing a leaf:
such contributions will be either of the form \equ(4.32) with
$\overline{{\rm Val}}(\th_2)=
\overline{{\rm Val}}(\th_2)in_{v_1'}\x_{v_1}(1,0)
L^{h_1\s}_{\ell\nn(v_1)}(\th_2)$, or of the form $\g_\k(g_0)=
\g_{\k-h_1}(g_0)\,
in_{v_1'}\x_{v_1}(1,0) L^{h_1\s}_{\ell\nn(v_1)}(\th_2)$,
where $h_1\ge 1$, and $\th_2'$ is a suitable tree of order $k-h_1$.
Then one realizes that the two contributions cancel exactly,
so that no new case has to be discussed with respect to the analysis
of \S A3.2.

\*

\0{\bf A3.4.} The above discussion completes the proof of
approximate cancellations of resonance values (\ie of cancellations to
first and second order, according to the degree of the resonant
branch).  The existence of cancellations, approximate to the first or
second order, is all is needed to obtain the bound \equ(5.16): the
analysis continues exactly as in [Ge2] and is based on simple
analyticity arguments that allow us to exploit, via the maximum
principle, the fact that in a resonance with momentum $\nn$
the functions of $\m=\oo_0\cdot\nn$ that have been considered above have
a zero in $\m$ of order $1$ or $2$.

A complete analysis showing that the higher orders contributions (\ie
the part which does not cancel) can be performed as in [Ge2], Appendix
B, and the final result is given by the bound
\equ(5.16) in \S 5.11.

\*

\0{\bf A3.5.} We shall show now that all contributions
with $(\nn_0(v_0),p(v_0))=(\V0,0)$ involved in the
definition of the counterterms (see Appendix A1)
must have automatically also $\nn(v_0)=\V0$.
The analysis performed in \S 5.1 shows that in order
to have $p(v_0)=0$ (for $j_{v_0}=\ell$), there can be
any number of leaves with highest nodes $w$ such that
$j_w>\ell$ and only one leaf $w$ with $j_w=\ell$
(contributing, respectively, $k_{w}'=0$ and $k_{w}'=1$
to $p(v_0)$).

Each time a leaf with $j_w>\ell$ appears, if we sum together the
values of all trees obtained by detaching the leaf with its stalk,
then reattaching it to all the other nodes of $\th_f$, we obtain
a vanishing contribution: simply by the cancellation mechanism
described in \S A3.1 (assuring there the first order zero),
{\it which, now, is an exact cancellation as the leaf does
not contribute to the free momenta of the branches of $\th_f$,
so that it does not modify the propagators.}
So we can suppose that no leaf with $j_{w}>\ell$ is possible in trees
involved in the determination of the counterterms.

In the same way, if we have a tree $\th$ having a leaf with $j_w=\ell$
and $h_w=h-h_1$ (for some $h_1$),
we can reason as in \S A3.2 and consider, together with
$\th$, also the tree formed by only one free node,
carrying a counterterm label $h_1$ and bearing the same leaf as $\th$.
The same cancellation mechanism described in \S A3.2 apply now:
again the only difference is that now the
cancellation is exact (by the same reason as before).

This shows that no tree with leaves can contribute to
$(\nn_0(v_0),p(v_0))=(\V0,0)$, so that for such trees
one has $\nn(v_0)=\nn_0(v_0)=\V0$. This, together
with the analysis in \S A1.1, proves \equ(A1.4) in \S A1.2.

\*
\ifnum\mgnf=0\pagina\fi
\vskip1.truecm
\0{\titolo Appendix A4. Graphs with non zero total hyperbolic
momentum, with leaves or with counterterms}
\numsec=4\numfor=1\*

\0{\bf A4.1}
Consider first the cases $p(v_0)\neq 0$. In this case we consider the
nodes $w<v_0$ with $p(w)=0$ and which are the nearest to $v_0$: by
construction all nodes $z$ between $v_0$ and the just singled out nodes
have $p(z)\neq 0$. Let us denote by $\tilde\th$ the set of such nodes
$z$ and $k_1$ the sum of their order labels.
The subtrees having as root branches the branches
exiting from the nodes $w$ can be considered as trees of the kind of the
previous sections (\ie with $p(w)=0$), so that the integrations
corresponding to their nodes can be performed and discussed as before
and a bound $B_0^{2k_0}\h^{-2k_0}$ follows, if $k_0=k-k_1$.

All the other nodes (in $\tilde\th$) have $p(z)\neq 0$,
so that no real small divisor appears
(\ie the propagators are trivially bounded by 1). The {\it only
delicate point to discuss} concerns the sum over the hyperbolic
mode labels, but this can be done as in [Ge2], p. 292, (or in [G1],
item 7 in Appendix A1), to which we refer for details beyond the
summary that follows.

By noting that the Laurent expansion of each function
(of $x$ and $x'$) appearing in \equ(4.14) and \equ(4.16)
starts from $k\ge -1$ and $k'\ge =1$, we can denote by $M_1$
the maximum of all such functions (multiplied by $1/x$ and $1/x'$
respectively when $k=-1$ and $k'=-1$) in a disk of radius $\l=1/2$.
If $M_2$ is the coefficient of the term with $k_v=-1$ or $k_v'=-1$,
set $M=\max\{M_1,M_2\}$.

Consider the tree value \equ(4.20). If $\s t<1$, the first integral
(corresponding to the highest node $v_0$) can be split into the sum of
two integral, the first one from $\s\io$ to $\s1$ (here the value
$1$ is an arbitrarily chosen positive number) and the second one from
$\s1$ to $t$. Let us denote by $I_{m}(\th)$ the first integral and
$J_m(\th)$ the second one, if $m$ is the number of nodes in $\th$.

For the nodes $v\in\tilde\th$, one has\\
(1) for each node the associated propagator is bounded by 1
(as $p(v)\neq 0$),\\
(2) $\prod_{v\in\tilde\th}|\bar F_{\n_v}|\le (CN^2)^{2k_1}$,\\
(3) for each node $v$ one has $|y_{n_v}^{(\a_v)}(k_v,k_v')|
\le M\,2^{k_v+k_v'}$,\\
(4) the last integration (on $\t_{v_0}$) produces a factor
$\exp[-k_{v_0}'+p(v_0)]=\exp[-\sum_{v\in\tilde\th}(k_v+k_v')]$.

Then the sum over the hyperbolic mode labels can be performed and
gives, for each node $v\in\tilde\th$, a factor $A^2$, where
$$ A=\sum_{k=-1}^{\io} \left( {2\over e } \right)^k =
{e^2 \over 2(e-2)} \; . \Eqa(A4.1) $$
The contribution to $I_m(\th)$ arising from $\tilde\th$ a bound
$B_1^{2k_1}$ is obtained, with $ B_1 = A^2 \CC N^2 M$, so that, for
$I_m(\th)$ a bound
$$ B^{2k} , \qquad B\ge \max\{B_0\h^{-1},B_1\} \Eqa(A4.2) $$
is obtained (see \equ(5.6) for the meaning of $\CC$).

By taking into account the integral $J_m(\th)$,
one can perform a splitting of the integration domain
for the integrals corresponding to the nodes immediately
preceding $v_0$ (now for each such nodes the second integral
is from $\s1$ to $\t_{v_0}$), and, iterating such a splitting,
one finds that $\Val(\th)$ can be written as sum of at most
$2^m$ terms each of which has the form (for some integer $p$)
$$ \left[ \prod_{v\in \th^*} \int_{\s1}^{\t_{v'}}
d\t_{v} \left( \ldots
\right) \right] I_{m_1}(\th_1)\ldots I_{m_p}(\th_p) \; , \Eqa(A4.3) $$
where $\th_1,\ldots,\th_p$ are disjoint subtress of $\th$
and $\th^*$ is the set of the $m_0$ nodes in $\th$ not belonging to
any such subtrees and $m_1+\ldots+m_p+m_0=m$.
The dots between the parentheses denote the product of
the functions in
$$ \prod_{w\in\th} Y^{(\a_w)}(\t_w',\t_w)\Eqa(A4.4) $$
which depend on $\t_v$, for a given $v\in\th^*$, and therefore is a
quantity bounded by $1$ (see \equ(4.14) and \equ(4.16)).
Note that the functions $Y$ have {\it no singularity} as functions
of their arguments $x=e^{\s g \t},x'=e^{\s g \t'}$, when $x,x'=1$
(or $\t,\t'=0$), even though the values $x,x'=1$ lie on the convergence
circle (the singularities being at $x,x'=\pm i$).
Furthermore each integration from $\s1$ to
$\t_{v'}$, once the integrand has been bounded, gives 1, while the
integrals $I_{m_j}(\th_j)$, $j=1,\ldots,p$ can be bounded as before.

Of course, if $\s t >1$, the discussion is easier
as no splitting of the integration domains is needed.

So we can conclude that a final bound $(2B)^{2k}$
is obtained for $\Val(\th)$; so far neither leaves
nor counterterms have been considered.

\*

\0{\bf A4.2}
Introducing the leaves and the counterterms, one sees (recall
Remark 4.4) that the value of any tree $\th$ can be always be
written as the product of a factor like \equ(4.28) times
the product of the counterterms and of the leaf values;
each counterterm can be decomposed in turn as sum of values of
amputated trees (see \equ(4.32)). As each leaf and each
amputated tree can contain other leaves and counterterms
we can iterate such a decomposition procedure, until, at the end,
the value of the tree $\th$, with highest node $v_0$,
turns out to be given by the product
of factorizing terms which\\
{(1)} either are of the form \equ(4.28),
with $\r_{w}=0$ for any subtree with highest node $w<v_0$ and
with $\r_{w}=0,1$ if $w=v_0$,\\
\0{(2)} or differ from \equ(4.28)
simply because no integration is performed corresponding
to the highest node.

The terms as in item (1) correspond to subtrees contributing
to leaf values (for $w<v_0$) and to $\Val(\th)$ itself (for $w=v_0$),
while the terms as in item (2) correspond to amputated trees
contributing to counterterms.
Then a natural decomposition of the tree $\th$ into
subtrees $\tilde \th$ (amputated or not) follows: each of such subtree
contains neither counterterms nor leaves (by construction).
Furthermore the subtrees contributing to leaves are linked
to nodes of some other subtrees through their stalks, while
the amputated subtrees are not linked to any node
(as there is no branch exiting from the highest node).
To keep memory of the node to which the counterterm label
is attached we can draw a hatched line connecting the amputated
subtree to such a node.

So for each subtree $\tilde \th$ (amputated or not) one can reason as
above and a bound $B^{m_0}$ is obtained, if $B$ is the same constant
as before and if $m_0\le 2k_0$ is the number of free nodes of the
subtree.  For all of them the resummation described in \S 5.9 has to
be performed, to bound the values of the subtrees $\th_v$, with
$p(v)=0$, $v\in\tilde \th$: such a resummation is taken into account
by the constant $B$.  By collecting together all bounds one obtains,
for the (normalized) sum of the values of the all trees $\th'$
generated by the resummations corresponding to the families
$\FFFF(\th_v)$, a bound $B^{2k}$, if $k$ is the order of $\th$.

Therefore we are left with the sum of all possible ways
to arrange leaves and counterterms. The choice of
the leaves is uniquely determined by the assignments
of the labels $\r_v$, $v \in \th$, so that it gives a factor $2^{m}$
(recall that the number of nodes $m$ is such that $m<2k$).

In the same way one can deal with the counterterms: simply
one has to distinguish between solid and hatched lines,
so that another factor $2^{m}$ is produced.

\*

\0{\bf A4.3}
Then the sum over all the other labels can be
performed, in the same way as for the contributions without
leaves and without counterterms (see the beginning of this subsection).
The sum over the hyperbolic modes has been taken
into account by the constant $B$ (see \equ(A4.2)); moreover\\
{$\bullet$} the sum over the mode
labels is bounded by $(2N+1)^{m(\ell-1)}(2N_0+1)^m$,\\
{$\bullet$} the sum over the angle labels is bounded by $\ell^m$,\\
{$\bullet$} the sum over the order labels is bounded by $2^m$,\\
{$\bullet$} the sum over the badge labels is bounded by $2^m$.

Therefore, by taking into account that the momenta and the hyperbolic
momenta are uniquely determined by the mode labels and, respectively,
the hyperbolic mode labels and that the sums over the leaf labels and
counterterms labels have been already considered, we are left with the
sum of (unlabeled) numbered trees (see comments after \equ(4.27)): but
these are no more than $2^{2m}m!$, so that, both for
$X_{j\nn}^{k\s}(t)$ and $\g_{k}(g_0)$, a final bound $D^k$ is
obtained, for some constant $D$: in terms of $B$ the constant $D$ is
given by $D=B\,2^6\ell\,(2N+1)^{(\ell-1)}(2N_0+1)$, \ie \equ(5.18). In
particular one has that $D$ is proportional to $\h^{-2}$, as $B$ is
so.

Note that, as a matter of fact, we have bounded $\X_{j\nn}^{h\s}(t)$ in
\equ(4.27) by neglecting the constraint on $j$ and $\nn$. Therefore,
by making use of the fact that the Fourier coefficients with $|\nn|>h
(N+N_0)$ vanish at order $h$ as a consequence of the trigonometric
assumption on the perturbation $f_1$, see \equ(2.2), the bound
$(2B)^{2k}$ is a bound both for the Fourier coefficients of
$X^\s(t;\aa)$ and for the function $X^\s(t;\aa)$ itself.

\vskip1.truecm
\0{\titolo References}
\*\*

\item{[BCG]} Benettin, G., Carati, A., Gallavotti, G.: {\it A rigorous
implementation of the Jeans--Landau--Teller approximation for
adiabatic invariants}, Nonlinearity {\bf 10} (1997), 479--507.

\item{[BGGM]} Bonetto, F., Gallavotti, G., Gentile, G., Mastropietro, V.:
{\it Lindstedt series, ultraviolet divergences and Moser's Theorem},
Annali della Scuola Normale Superiore di Pisa {\bf 26}, 545--593 (1998).

\item{[CG] } Chierchia, L., Gallavotti, G.:
{\it Drift and diffusion in phase space},
Annales de l'Institut Henri Poincar\'e, B {\bf 60} (1994), 1--144; see
also the {\it Erratum},  B {\bf 68} (1998), 135.

\item{[CZ] } Chierchia, L., Zehnder, E.,:
{\it Asymptotic expansions of quasi-periodic motions},
Annali della Scuola Normale Superiore di Pisa
Cl. Sci. (4) {\bf 16} (1989), 245--258.

\item{[E] } Eliasson, L.H.:
{\it Absolutely convergent series expansions for quasi-periodic motions},
Mathematical Physics Electronic Journal {\bf 2} (1996),
http:// mpej.unige.ch.

\item{[G1] } Gallavotti, G.:
{\it Twistless KAM tori, quasi flat homoclinic intersections, and
other cancellations in the perturbation series of certain completely
integrable Hamiltonian systems. A review}, Reviews on Mathematical
Physics {\bf 6} (1994), 343--411.

\item{[G2] } Gallavotti, G.: {\it Twistless KAM tori},
Communications in Mathematical Physics {\bf 164} (1994), 145--156.

\item{[G3] } Gallavotti, G.: {\it Reminiscences on science at I.H.E.S.
A problem on homoclinic theory and a brief review}, Publications
Scientifiques de l'IHES, Special Volume for the 40-th anniversary
(1998), ???--???, [preprint in chao-dyn 9804043].

\item{[GG] } Gallavotti, G. , Gentile, G.:
{\it Majorant series convergence for twistless
KAM tori}, Ergodic Theory and Dynamical Systems {\bf 15} (1995), 1--69.

\item{[GGM1] } Gallavotti, G. , Gentile, G., Mastropietro, V.:
{\it Field theory and KAM tori}, p. 1--9, Mathematical Physics
Electronic Journal {\bf 1} (1995), http:// mpej.unige.ch.

\item{[GGM2] } Gallavotti, G. , Gentile, G., Mastropietro, V.:
{\it Separatrix splitting for systems with three degrees of freedom}, to
appear in Communications in Mathematical Physics, 1999, [preprint with
title {\it Pendullum: separatrix splitting} in chao-dyn 9709004].

\item{[GGM3] } Gallavotti, G., Gentile, G., Mastropietro, V.:
{\it Hamilton-Jacobi equation, heteroclinic chains and Arnol'd
diffusion in three time scales systems},
submitted for publication [Preprint in chao-dyn \#9801004].

\item{[GGM4] } Gallavotti, G., Gentile, G., Mastropietro, V.:
{\it Melnikov's approximation dominance. Some examples},
to appear in Reviews in Mathematical Physics, 1999,
[preprint in chao-dyn \#9804043].

\item{[Ge1] } Gentile, G: {\it A proof of existence of whiskered tori with
quasi flat homoclinic intersections in a class of almost integrable
hamiltonian systems}, Forum Mathematicum {\bf 7} (1995), 709--753.

\item{[Ge2] } Gentile, G.: {\it Whiskered tori with prefixed
frequencies and Lyapunov spectrum}, Dynamics and Stability of Systems
{\bf 10} (1995), 269--308.

\item{[Gel1] } Gelfreich, V.: {\it Melnikov method and exponentially
small splitting of separatrices}, Physica D {\bf101} (1997), 227--248.

\item{[Gel2] } Gelfreich, V.: {\it Reference systems for
splitting of separatrices}, Nonlinearity {\bf10} (1997), 175--193.

\item{[GM1] } Gentile, G., Mastropietro, V.:
{\it KAM theorem revisited}, Physica D {\bf 90} (1996), 225--234.

\item{[GM2] } Gentile, G., Mastropietro, V.:
{\it Methods for the analysis of the Lindstedt series for KAM tori
and renormalizability in Classical mechanics.
A review with some applications},
Reviews in Mathematical Physics {\bf 8} (1996), 393--444.

\item{[N] } Neishtadt, A.I.:
{\it The separation of motions in systems with rapidly rotating phase},
Journal of Applied Mathematics and Mechanics {\bf 48} (1984), 133--139.

\item{[HMS] } Holmes, P., Marsden, J., Scheurle, J: {\it Exponentially
small splittings of separatrices with applications to KAM theory and
degenerate bifurcations}, Contemporary Mathematics {\bf 81} (1989), 213--244.

\item{[La] } Lazutkin, V.F.,, {\it
Splitting of separatrices for the Chirikov's standard map}, mp$\_$arc@
math.utexas.edu, \#98--421, (annotated english translation of a 1984
paper).

\item{[P] } Poincar\'e, H.:
{\it Les m\'ethode nouvelles de la M\'ecanique C\'eleste}, Vol. III,
Gauthier--Villard, Paris, 1899.

\item{[P\"o] } P\"oschel, J.: {\it Invariant manifolds of complex analytic
mappings}, Les Houches, XLIII (1984), Vol. II, p. 949-- 964, Eds. K.
Osterwalder \& R. Stora, North Holland, 1986.
\*

\FINE

\ciao